\documentclass[useAMS,usenatbib]{mn2e}

\usepackage{amssymb, amsmath, amsfonts}

\usepackage{graphicx, shadow}

\usepackage{epstopdf}

\usepackage{rotating}
\usepackage{wasysym}
\usepackage{verbatim}
\usepackage{natbib}

\newcommand{\sunrise}{\textsc{Sunrise}}

\newcommand{\hst}{{\sl HST}}

\newcommand{\illustris}{{Illustris}}

\newcommand{\gmtwenty}{$G$-$M_{20}$}

\newcommand{\fgmtwenty}{$F(G,M_{20})$}

\newcommand{\atlas}{\textsc{Atlas}$^{\rm 3D}$}
\newcommand{\sauron}{\textsc{Sauron}}

\newcommand{\sersic}{S\'{e}rsic}

\renewcommand{\vec}[1]{ {\bmath #1} }

\title[Simulated Galaxy Morphologies at $z=0$]{Galaxy Morphology and Star Formation in the Illustris Simulation at $z=0$}

\author[G.F.\ Snyder et al.]{Gregory F. Snyder$^1$, Paul Torrey$^{2,3}$, Jennifer M.\ Lotz$^1$, Shy Genel$^{4,5}$\thanks{Hubble Fellow},  \newauthor
Cameron K.\ McBride$^4$, Mark Vogelsberger$^2$, Annalisa Pillepich$^4$, Dylan Nelson$^4$,  \newauthor
Laura V.\ Sales$^4$, Debora Sijacki$^6$, Lars Hernquist$^4$, Volker Springel$^{7,8}$ \\
$^1$ Space Telescope Science Institute, 3700 San Martin Dr, Baltimore, MD 21218 \\
$^2$ Department of Physics, Kavli Institute for Astrophysics \& Space Research, Massachusetts Institute of Technology, Cambridge, MA, 02139, USA \\
$^3$ TAPIR, Mailcode 350-17, California Institute of Technology, Pasadena, CA 91125, USA \\
$^4$ Harvard-Smithsonian Center for Astrophysics, 60 Garden Street, Cambridge, MA, 02138, USA \\
$^5$ Department of Astronomy, Columbia University, 550 West 120th Street, New York, NY, 10027, USA \\
$^6$ Institute of Astronomy and Kavli Institute for Cosmology, University of Cambridge, Madingley Road, Cambridge CB3 0HA, UK\\
$^7$ Heidelberg Institute for Theoretical Studies, Schloss-Wolfsbrunnenweg 35, 69118 Heidelberg, Germany \\
$^8$ Zentrum f\"{u}r Astronomie der Universit\"{a}t Heidelberg, ARI, M\"{o}nchhofstr. 12-14, 69120 Heidelberg, Germany}
\begin{document}


\label{firstpage}

\maketitle

\begin{abstract}

We study how optical galaxy morphology depends on mass and star formation rate (SFR) in the \illustris\ Simulation.  To do so, we measure automated  galaxy structures in $10808$ simulated galaxies at $z=0$ with stellar masses $10^{9.7} < M_*/M_{\odot} < 10^{12.3}$. We add observational realism to idealized synthetic images and measure non-parametric statistics in rest-frame optical and near-IR images from four directions.  We find that \illustris\ creates a morphologically diverse galaxy population, occupying the observed bulge strength locus and reproducing median morphology trends versus stellar mass, SFR, and compactness.  Morphology correlates realistically with rotation, following classification schemes put forth by kinematic surveys.  Type fractions as a function of environment agree roughly with data.  These results imply that connections among mass, star formation, and galaxy structure arise naturally from models matching global star formation and halo occupation functions when simulated with accurate methods.  This raises a question of how to construct experiments on galaxy surveys to better distinguish between models.  We predict that at fixed halo mass near $10^{12} M_{\odot}$, disc-dominated galaxies have higher stellar mass than bulge-dominated ones, a possible consequence of the \illustris\ feedback model.  While \illustris\ galaxies at $M_* \sim 10^{11} M_{\odot}$ have a reasonable size distribution, those at $M_* \sim 10^{10} M_{\odot}$ have half-light radii larger than observed by a factor of two. Furthermore, at $M_* \sim 10^{10.5}$--$10^{11} M_{\odot}$, a relevant fraction of \illustris\ galaxies have distinct ``ring-like'' features, such that the bright pixels have an unusually wide spatial extent.

\end{abstract}


\section{Introduction} \label{s:intro}

Cosmological simulations of galaxy formation are poised to address
long-standing challenges in extragalactic astronomy. They may achieve this
not only by the ordinary means of testing theories of galaxy astrophysics with
numerical experiments, but also by enabling the direct modeling of
observations.  In observed distant galaxies, often our only dynamical information is a
snapshot of the stellar orbits traced by light, called the galaxy's
morphology.  Despite copious data, the origins of the distribution of galaxy
morphologies remain debated. The presence of a bulge-like light profile
correlates tightly with a lack of star formation \citep[e.g.,][]{Kauffmann2003, Bell2012, Woo2015}, but we know neither the
complete physics of quenching nor the assembly histories of bulges as a
function of halo mass or environment.  Although rare, galaxy
mergers can play an important role in massive galaxies \citep[e.g.,][]{Bell2006,lotz08_hst,Lotz2011}.  However,
their rates and consequences are largely unknown at $z > 1$. 

Analyses of galaxy morphologies typically span two broad properties: galaxy structure and disturbed appearance.  Important methods for conducting these analyses are visual inspection and automated algorithms, both of which have matured as the quantity of high resolution galaxy data increases.  However, the results of such analyses are for the most part empirically motivated: they determine whether a galaxy appears to be bulge- or disc-dominated and whether it appears to be experiencing a merger, interaction, or disturbance. We often define these rigidly, using wide bins that may be ineffective for learning how galaxies formed.  

In principle, simulations can link observed
morphology with underlying physical processes in powerful ways.  The principal idea is to apply
stellar population synthesis techniques
\citep[e.g.,][]{tinsley68,Tinsley1976,Tinsley1978,Gunn1981} to galaxy
evolution models.  For example, analytic and semi-analytic models make
unified predictions for observed fluxes, dynamics, and structure based on
simple physical prescriptions \citep[e.g.,][]{Henriques2015,Lu2014}.  Indeed, these calculations now form the basis for ``mock observatories'' in which we predict galaxy properties, including morphology, in realistic synthetic imaging surveys and release these to researchers through advanced databases \citep[e.g.,][]{Overzier2012,Bernyk2014}.  However, by design they do not trace the full dynamics of matter in galaxies, especially rare and subtle features such as tidal tails, disturbed morphologies, and rapid morphological transformations.  In contrast, using hydrodynamical simulations with sufficiently high spatial resolution and enough physics to predict chemically enriched stars, gas, and dust, we can assign spectral energy distributions (SEDs) to their components and predict how they would be observed in imaging surveys.  Moreover, because such simulations model directly the gravity of baryons and dark matter, they also capture processes such as local collapse and angular momentum transport which are essential for setting galaxy morphologies.

Until recently, the direct simulation approach has been limited either to very small samples of cosmologically assembling galaxies or to hand-crafted models of individual galaxies or groups.  However, significant progress has been made toward achieving direct, statistically relevant predictions of galaxy formation.  By including significant feedback from supernovae, massive star formation, and supermassive black holes, the realism of resulting galaxy models has continued to improve \citep[e.g.,][]{Crain2009,Guedes2011,Ceverino2014, Cen2014}. These advances owe in part to preceding generations of semi-analytic modeling which outlined the broad requirements for the impacts of feedback on massive galaxies \citep[e.g.,][]{croton06,Somerville2008}.  Moreover, refinements in numerical methods have enabled the field to conduct very large hydrodynamical simulations containing tens of thousands of galaxies whose internal dynamics are at least partially resolved \citep[e.g.,][]{Schaye2010,Khandai2015, Dubois2014, Vogelsberger2014a, Schaye2014}.  Since they predict the distribution and appearance of rare events, and directly link the physics of star formation quenching with the evolution of galaxy structure, these simulations increase the information learned from galaxy surveys.  We can, in principle, discard empirically motivated and coarsely binned classification schemes in favor of explicit ones.  

As a first step, in this paper we study the morphology of 10808 galaxies at $z=0$ from the \illustris\ Project.  We measure each galaxy from four viewing directions in four filters, for a total of $\sim 1.7\times 10^5$ synthetic images.  By doing so, we seek to enable direct comparison between simulated and observed galaxy populations and to prepare for future studies exploiting these models to interpret survey data. In Section~\ref{s:images}, we describe our methods for creating realistic synthetic observations from ideal images.  In Section~\ref{s:morphologies}, we determine how these galaxy images compare with observations using non-parametric image diagnostics, and show how \illustris\ galaxy models occupy the space of stellar mass, SFR, and morphology.  Section~\ref{s:rotation} links optical morphology with the kinematic structure of the simulated galaxies.  In Section~\ref{s:otherscience}, we explore correlations of morphology with galaxy size, the mass of supermassive black holes, and halo mass.  We will make our non-parametric morphology measurements available as online supplementary tables (see Appendix~\ref{a:morphs}).  We have released our synthetic images\footnote{www.illustris-project.org}, and publicly releasing our code for adding realism to model images\footnote{https://bitbucket.org/ptorrey/sunpy}.

For all calculations, we adopt the cosmology used for the \illustris\ simulation, a $\Lambda$CDM cosmology with $\Omega_m = 0.2726$, $\Omega_{\Lambda} = 0.7274$, $\Omega_b = 0.0456$, $\sigma_8 = 0.809$, $n_s = 0.963$, and $H_0 = 100 h\rm\ km s^{-1}$ with $h = 0.704$, consistent with the WMAP-9 measurements \citep{Hinshaw2013}.  

\section{Mock Image Analysis} \label{s:images}

In this section we briefly outline our methods to create realistic mock images and measure morphologies in observational units.  

\subsection{Simulations and Ideal Images} \label{ss:imagepipeline}

The \illustris\ Project consists of hydrodynamical simulations of galaxy
formation in a periodic volume of size $(106.5\rm\ Mpc)^3$ \citep{Vogelsberger2014b,Vogelsberger2014a,Genel2014}.  This volume was
simulated with and without baryons, and at several resolution levels; unless
otherwise noted, here we present results from the simulation with the
highest resolution ($2\times1820^3$), which resolves baryonic matter with mass $1.26\times 10^{6} M_{\odot}$ and employs a plummer equivalent gravitational softening length of 710 pc at $z=0$.  Using the Arepo code for gravity
and gas dynamics \citep{Springel2010}, these simulations evolve simultaneously the gas, stars, and dark matter components from cosmological initial conditions \citep{Vogelsberger2012,Keres2012,Sijacki2012}.  The
\illustris\ simulation galaxy physics models consist of primordial
and metal line cooling, star formation, gas recycling,
metal enrichment, supermassive black hole (SMBH) growth, and gas heating by
feedback from supernovae and SMBHs \citep{Vogelsberger2013}. Parameters
were chosen to match the $z=0$ stellar mass and halo occupation functions,
plus the cosmic history of SFR density. \citet{Torrey2014} showed that this
parameterization reproduces the observed stellar mass function, SFR-Mass main
sequence, and the Tully-Fisher relation, from $z=3$ to $z=0$. \citet{Sijacki2015} described the evolution of SMBHs in \illustris\ and their co-evolution with the galaxy population.  { The \illustris\ simulations are publicly available at {www.illustris-project.org/data} \citep{Nelson2015a}.}

We generate noiseless high-resolution images with an approach described by
\citet[][hereafter T15]{Torrey2015} built around the \sunrise\footnote{
  \sunrise\ is freely available at https://bitbucket.org/lutorm/sunrise } code
\citep{jonsson06,jonsson09,Jonsson:2010gpu}.  T15 published idealized images to enable studies focused on data from different observatories with the same simulated dataset. For example, \citet{Wellons2015} used such tools to demonstrate how high-redshift compact massive galaxies would be observed with the \emph{Hubble Space Telescope}. Similarly, \citet{Moody2014} created and analyzed mock \hst\ images from very high-resolution cosmological simulations of clumpy disc galaxies \citep{Ceverino2014}.  We describe our methods for adding realism and our example model observatory settings tuned for low-redshift \illustris\ galaxies in Section~\ref{ss:realism}.  

To create the images, we begin by segmenting the
simulation output into separate files, one for each galaxy (subhalo) identified with the SUBFIND code \citep{Springel2001}, which uses a friends-of-friends algorithm \citep[e.g.,][]{Davis1985} with a linking length of 0.2 times the mean particle separation to identify bound dark matter halos, and then applies a secondary linking stage to associate baryonic mass to these halos \citep{Dolag2009}. In this paper, we analyze only the mass associated with individual subhalos, ignoring nearby subhalos such as other members of the parent halo. This allows us to separate galaxies into images and cleanly measure structural parameters.  However, with this approach we are unable to directly study signatures of galaxy mergers such as multiple nuclei, since the halo finder may separate objects that would be combined in real image-based segmentation maps.  Moreover, the procedure used here occasionally leaves spurious faint surface brightness features from satellite galaxies; these do not affect the global optical morphology of the bright pixels, but could affect diagnostics of faint or subtle stellar halo features.  For studying galaxy merger signatures with simulations like \illustris\ \citep[e.g.,][]{Rodriguez-Gomez2015}, we will use images containing all nearby mass.  

With \sunrise, we assign to each star particle a spectral energy distribution
(SED) based on its mass, age, and metallicity using stellar population models
by \citet{bc03} with a \citet{chabrier03} initial mass function. Each star
particle radiates from a region with a size directly proportional to the
radius that encloses its $16$ nearest neighbors: the emission region is
smaller (larger) in areas of high (low) stellar density.  This adaptive
smoothing technique leads to much more realistic surface brightness maps in
regions of low stellar density; see T15 for full details.  {This step preserves the light distribution in bright regions of well resolved galaxies, and prior to any analysis, we convolve with a point-spread function (PSF) larger than the typical spacing between star particles near the centers of galaxies.  Therefore, the results of our study do not depend strongly on this assignment.  However, certain measures of clumpiness, such as Gini alone, are sensitive to individual bright star particles and therefore care must be taken when interpreting the original high-resolution images regardless of the light assignment method.  In Appendix A, we used a representative subset of galaxies to compare our choice of star particle sizes with the most common alternative, a constant radius equal to stars' gravitational softening length, and find that the morphology diagnostics measured here are identical to within $1\%$.}

We then project these SEDs with \sunrise\ without performing dust radiative transfer (RT) through the ISM.  We skip the RT stage in order to avoid over-modeling galaxies with ISM resolution insufficient to gain substantial accuracy by performing the RT {, and because the primary focus of the present work is on the morphologies of massive ($M_* > 10^{9.5} M_{\odot}$) galaxies at $z=0$, which we do not believe to be especially senstive to dust attenuation, on average. Therefore, in this study we have used \sunrise\ only as a tool to assign stellar population models to star particles, project these quantities in several directions, and create synthetic images in arbitrary filters.  While this may be slightly more computationally intensive than required for the present work, we viewed the advantages of starting with a widely used open-source code as the backbone of our analysis pipeline, with the option of using dust RT in future studies or in higher resolution simulations, to outweigh the disadvantages. }

To assist our analysis, we also project surface density maps of physical quantities onto the same pixels, using the options provided by \sunrise.
These maps include SFR, stellar mass, gas mass, stellar metallicity, gas
metallicity, luminosity- and mass- weighted stellar age, and gas temperature.
In this paper, we neglect the important effect that dust attenuation will have
on our synthetic images. However, we have performed extensive tests derived from
these 2-D mass maps, from which we derived a resolved slab dust
model by expanding on techniques to estimate dust on semi-analytic models
\citep[e.g.,][]{DeLucia2007,Kitzbichler2007,Guo2009, Guo2011, Guo2012a}. We found that the primary results and correlations presented in this paper are qualitatively unaffected by dust modeled in this way, so we have chosen to reserve a comprehensive exploration of this topic to a future paper.

The unattenuated rays travel until they exit the grid or enter one of four viewing apertures (``cameras''). The output of this step is the SED at each of 512$\times$512 pixels in each camera.  From these data cubes, \sunrise\ creates raw mock images by integrating the (optionally redshifted) SED in each pixel over a set of common astronomical filters, from the UV through IR.  In this paper, we perform this filter synthesis in the rest frame of each galaxy.  The spatial extent of each image is set to ten times the 3D stellar half-mass radius, and therefore the physical pixel scale varies.  We use pixel sizes of roughly $100$--$300\rm\ pc$, sufficiently small to simulate SDSS, Pan-STARRS, and LSST images of sources at $z \gtrsim 0.02$ regardless of the limited resolution of the simulation, which is a separate constraint that we consider below. 

\begin{figure*}
\begin{center}
\includegraphics[width=5.0in]{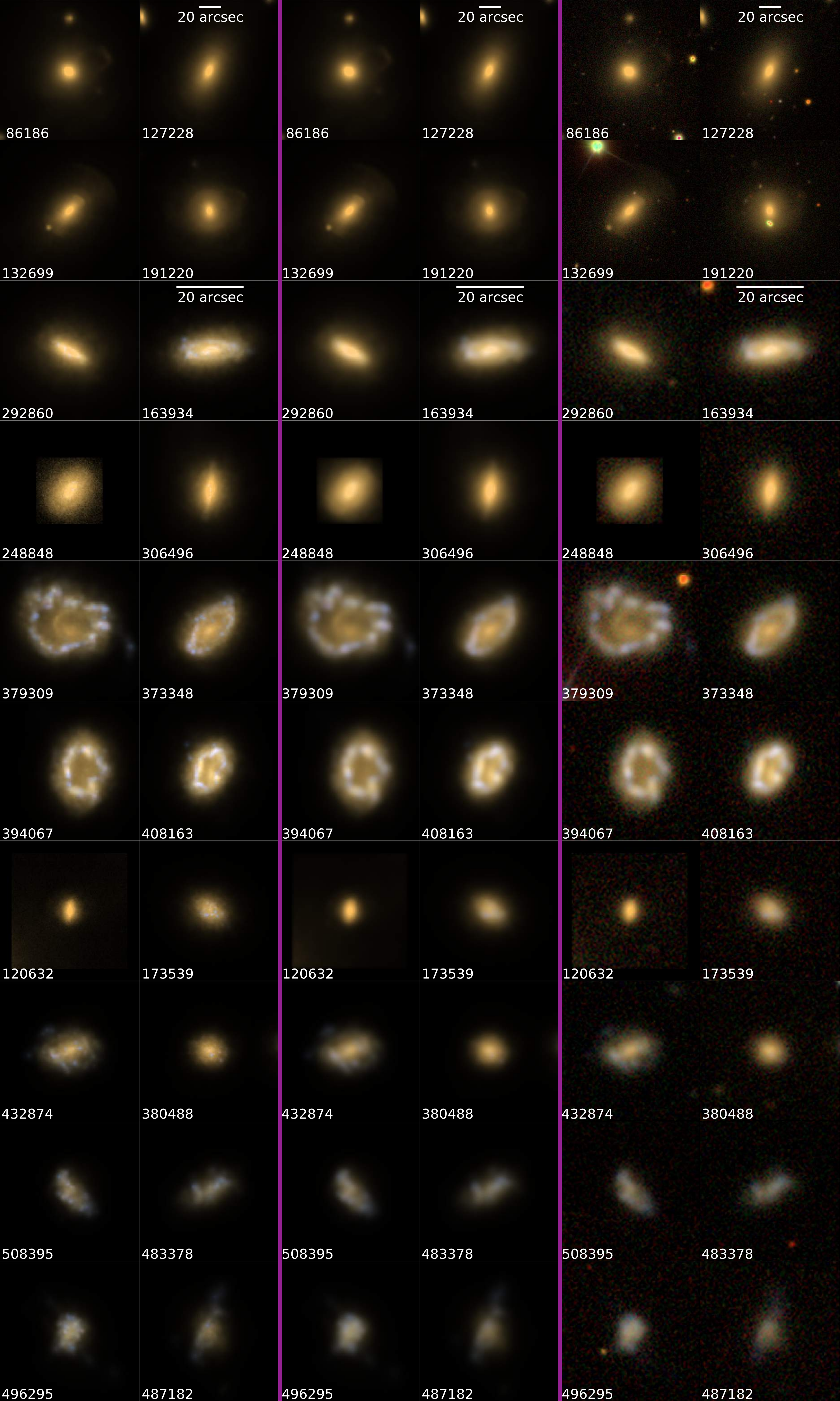}
\caption{Example $g$-$r$-$i$ images of $z=0$ Illustris galaxies at each of three stages of our image realism procedure, arranged by increasing stellar mass from bottom ($M_* \sim 10^{10} M_{\odot}$) to top ($M_* \sim 10^{12} M_{\odot}$) and selected based on their location in the \gmtwenty\ plane (as in Section~\ref{ss:mass}). For displaying, we pretend the galaxy was observed in SDSS at $z=0.05$ with a pixel size of $0.5$ arcsec. Left: Ideal output image directly from \sunrise.  Middle: We convolved the ideal image for each filter by a Gaussian PSF with FWHM$= 2.0$ arcsec.  Right: To the PSF-convolved images, we add random cutouts from SDSS DR10 \citep{Ahn2014}, downloaded from data.sdss3.org/mosaics. The remainder of this paper analyzes measurements of synthetic images that are most like those in the middle column, as described in Section~\ref{ss:realism}.  \label{fig:realism} }
\end{center}
\end{figure*}

\subsection{Realistic Images} \label{ss:realism}

We convert the noise-free, ideal galaxy images (T15) into realistic
synthetic images using the following procedure.  First, we convolve each
high-resolution image with a Gaussian point-spread function (PSF) with
full-width at half-maximum (FWHM) of $1.0$ kpc.  Then, we re-bin the images to
a constant pixel scale of $0.24$ kpc, which is approximately $1/3$ of the
\illustris\ stellar gravitational softening. These parameters correspond roughly to
$1\rm\ arcsec$ seeing for observations of a source located at $z = 0.05$,
where the physical scale is roughly $1\rm\ kpc/arcsec$ (angular-size distance $D_A \approx 200$ Mpc, luminosity distance $D_L
\approx 220$ Mpc).  Thus our realistic images can be thought of as roughly
appropriate comparisons to many sources in the Sloan Digital Sky Survey main
galaxy sample \citep{Strauss2002}.    They also correspond roughly to an
\hst\ WFC3 survey of sources at $z \sim 0.5$, where the scale $\sim
6\rm\ kpc/arcsec$ implies that our PSF FWHM corresponds to $\sim 0.17
\arcsec$.

We save FITS\footnote{Flexible Image Transport System, fits.gsfc.nasa.gov} images, either in the \sunrise\ output units ($W/m/m^2/Sr$), which is a surface brightness independent of distance, or in nanomaggies ($1\rm\ nmy \approx 3.631\times 10^{-6} Jy$) assuming our fiducial setting $z = 0.050$.  To simplify analysis, we also compute AB absolute magnitudes, AB absolute zero-points, original camera distances and pixel scales, and the new implied apparent magnitude at the camera distance.  We store these important metadata in the image headers.  Thus the conversion from the \sunrise\ output can be uniquely specified, and synthetic image fluxes can be recomputed for any assumed distance.  From these FITS files, we then create files containing colour-composite images.  

Finally, we add sky shot noise such that the average signal-to-noise ratio of each galaxy pixel is 25. We assume this sky shot noise is a Gaussian random process independently applied to each pixel. Thus we are assuming that each model galaxy is strongly detected, eliminating biases from potentially noisy morphology measurements.  

For future visual classification projects, we also prepare images for classification by the Galaxy Zoo
project \citep[GZ, e.g.,][]{Lintott2008} in SDSS $g$, $r$, and $i$ filters.  Using our initial radius ($r_P$)
measurements as defined in Section~\ref{ss:morphology}, we re-bin our
SDSS-like FITS images to a new pixel scale ($0.008 \times r_p$) and create images with a fixed pixel count ($424\times 424$).  These choices are such that the galaxy extent defined by $2r_p$ always subtends $\sim 2/3$ of the linear image size, enabling fair visual classifications as a complement to our fixed-scale non-parametric measurements below.  For such visual classification projects, we also add real SDSS background images to create fully synthetic $ugriz$ galaxy images.  To accomplish this, we first downloaded mosaics from the SDSS DR10  \citep{Ahn2014} Science Archive Server with the mosaic web tool ({data.sdss3.org/mosaics}).  From these, we randomly select a region of an appropriate size for each synthetic image, assuming the galaxies are at $z = 0.05$, and add it to the simulated galaxy image.  We demonstrate these steps in Figure~\ref{fig:realism}. This is a simplification from complete image simulations of self-consistent lightcones drawn from the simulation volume \citep[e.g.,][]{Overzier2012,Henriques2012}.  We have created several examples of these simulated fields from \illustris, and such techniques \citep{Kitzbichler2007} will become very useful as the volumes of such simulations grow.  

\begin{figure*}
\begin{center}
\includegraphics[width=6.5in]{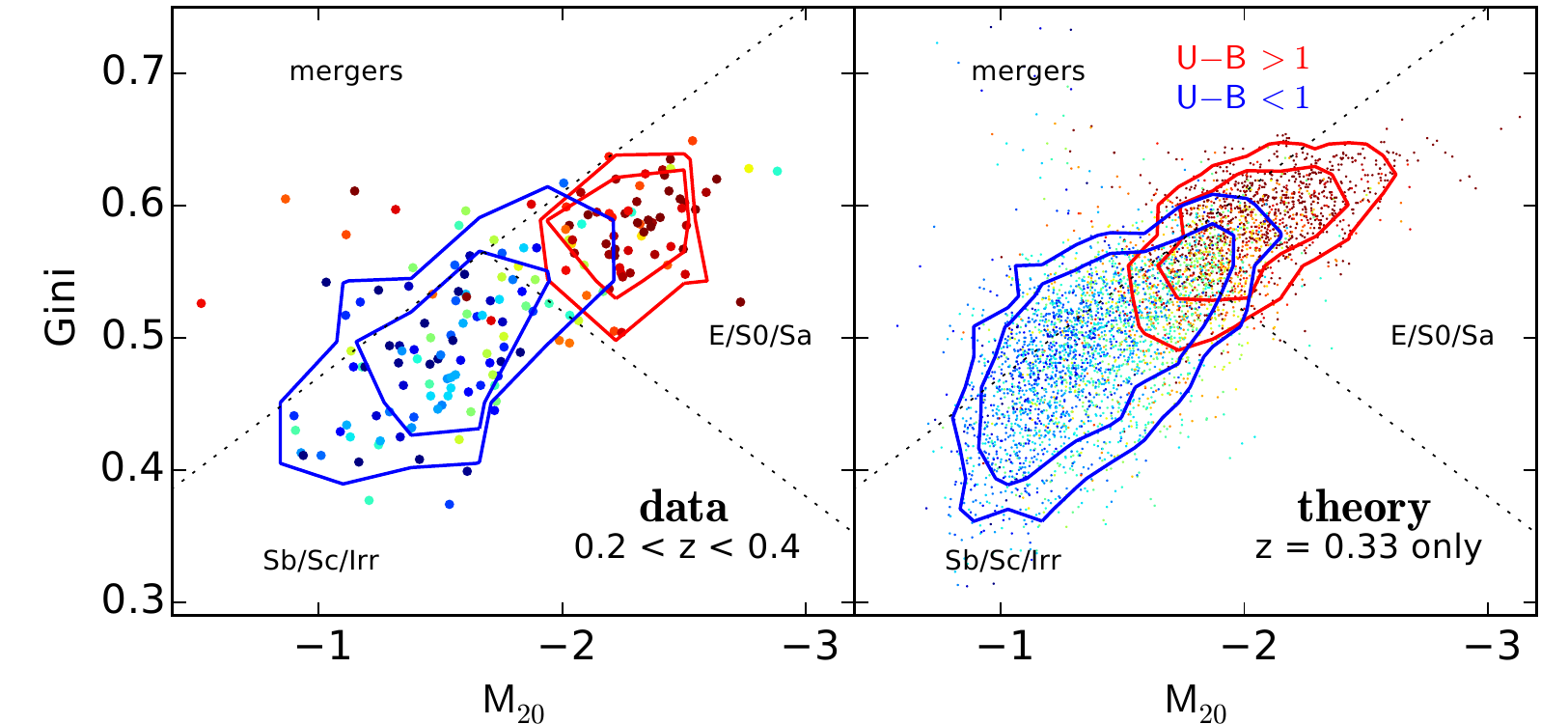}
\caption{Gini--$M_{20}$ as a function of star formation as reflected by $U-B$ colour for galaxies at $z\sim 0.3$.  Left: data from the Extended Groth Strip (EGS) survey compiled by \citet{lotz08_hst}: roughly rest-frame $B$-band.  Right: mock data from T15 at $z=0.33$: rest-frame $r$-band.  Here we show simulated galaxies at $z=0.33$ to roughly match the volume-limited EGS sample, but all other figures show simulated galaxies at $z=0$ only.  Red (blue) contours encircle 68\% and 95\% of the galaxies with $U-B > 1$ ($U-B < 1$).  We find that simulated galaxies of a given colour have roughly the right optical shape, and \illustris\ produces a large population of early-type objects.    \label{fig:gm20_color} }
\end{center}
\end{figure*}

\subsection{Structural Measurements} \label{ss:morphology}

We measure non-parametric morphologies by using code originally developed for
idealized merger simulations and also applied
to galaxy surveys \citep[e.g.,]{lotz08,lotz08_hst,Lotz2011}.  From each image, we
characterize the light profile with a Petrosian radius $r_P$, half-light
radius $R_{1/2}$, Concentration  ($C$), Gini ($G$), and $M_{20}$ \citep{Conselice2003,Lotz2004}, defined below.  Our code also measures merger and disturbance indicators, Asymmetry and the newly proposed MID merger statistics \citep{Freeman2013}, but in this paper we focus on the above simple estimates of galaxy structure. We will return to diagnostics of mergers and disturbances in a future paper.  

We define the Petrosian radius or semi-major axis $r_p$ such that the mean surface brightness in an elliptical annulus with semi-major axis $r_p$ equals 0.2 times the mean surface brightness within this ellipse, following \citet{Lotz2004}.  Here we also compute an elliptical half-light radius $R_{1/2}$ to characterize galaxy sizes, assuming that all of the galaxy's light is contained within an ellipse with semi-major axis $1.5\times r_p$.

We compute the concentration parameter $C$ \citep{Bershady2000}:
\begin{equation}
C = 5\log_{10} \frac{r_{80}}{r_{20}},
\end{equation}
where $r_{80}$ and $r_{20}$ are circular apertures containing $80\%$ and $20\%$ of the total flux within the ellipse with semi-major axis $1.5 r_p$ \citep{Conselice2003} of the galaxy center defined by minimizing the Asymmetry parameter \citep{Abraham1996}.

Gini's coefficient, $G$, measures the inequality in flux value among a galaxy's pixels, varying from $0$ (all pixels equal flux) to $1$ (one pixel contains all flux).  First used by \citet{Abraham2003} to characterize galaxy light profiles, $G$ correlates with $C$ but does not depend on the location of the brightest pixels.  Hence it is sensitive not only to concentrated spheroids but also to galaxies with multiple bright regions.  For a discrete population, \citet{Glasser1962} showed that $G$ can be computed as:
\begin{equation}
G = \frac{1}{ \bar{\left | I_i\right |} n\left (n-1\right )} \sum_i^n{\left (2i - n - 1\right ) \left | I_i\right | },
\end{equation}
where we have $n$ pixels with rank-ordered absolute flux values $\left | I_i\right |$, and $\bar{\left | I_i\right |} = \sum_i{\left | I_i\right |/n}$, the mean absolute flux value.  We follow \citet{Lotz2004} in correcting $G$ using absolute values to mitigate the effect of noise-induced negative fluxes.  This procedure recovers the true $G$ when $S/N \gtrsim 3$ per galaxy pixel, which is true by construction for all of the galaxy images we prepared in Section~\ref{ss:realism}.  

$M_{20}$ measures the second-order spatial moment of a galaxy's bright pixels contributing $20\%$ of the total light, relative to its total moment \citep{Lotz2004}:
\begin{equation}
M_{20} \equiv \log_{10} \frac{\sum_i{M_i}}{M_{\rm tot}},\ \mathrm{for}\ \sum_i{I_i} < 0.2 I_{\rm tot},
\end{equation}
where
\begin{equation}
M_{\rm tot} = \sum_i^n{M_i} = \sum_i^n{I_i \left [  \left (x_i - x_c\right )^2 + \left (y_i - y_c\right )^2 \right ]},
\end{equation}
and $x_c$, $y_c$ are the 2-D spatial coordinates of the galaxy center, defined to minimize $M_{\rm tot}$.  For computing $G$ and $M_{20}$, we define a galaxy's pixels following the segmentation procedure by \citet{Lotz2004}.  

In Figures~\ref{fig:gm20_color} and \ref{fig:gm20_all}, we divide the \gmtwenty\ plane into three regions corresponding to early types, late types, and mergers, based on comparisons with low-redshift visual classifications \citep{Lotz2004}.  These classifications are meant to be loose guidelines and are not to be strictly interpreted.  For future reference, we define a \gmtwenty\ ``bulge statistic'', which depends on an object's location in this diagram and correlates with optical bulge strength.  This will serve as a rough automated assessment of morphological type.  Specifically, we define $F$ as five times the point-line distance from a galaxy's morphology point to the line passing through $(G_0,M_{20,0}) = (0.533,-1.75)$ and parallel to the \citet{Lotz2004} early-type/late-type separation line, which has a slope $m=0.14$ in the space of $(G,M_{20})$.  We chose the scaling factor $5$ so that the resulting values occupy a convenient range ($-2 \lesssim F \lesssim 1$). Starting from the point-line distance formula in two dimensions:
\begin{equation}
d = \frac{\left | a M_{20} + b G + c \right |}{\left ( a^2 + b^2 \right )^{1/2}},
\end{equation}
where $d$ is the distance from a point to the line $G = -(a/b)M_{20}-(c/b)$.  We let $b=1$ and set $a = -m = -0.14$, allowing us to solve for $c = -b(G_0 + a M_{20,0}) = -0.778$, which defines the desired line. We set the sign of $F$ so that positive (negative) values indicate bulge-dominated (disc-dominated) galaxies.  Thus, $\left | F \right | = \left | -0.693 M_{20}  + 4.95 G - 3.85 \right |$, and 
\begin{equation} \label{eq:fgm20}
F(G,M_{20}) =
\begin{cases}
\left | F\right | & G \ge 0.14 M_{20} + 0.778\\
-\left | F\right | & G < 0.14 M_{20} + 0.778
\end{cases} 
\end{equation}
corresponding to the ``\gmtwenty\ bulge statistic'' annotation to Figure~\ref{fig:gm20_all} and shown in panel (d) of Figure~\ref{fig:sfrmass}. For most galaxies, this diagnostic adds little new information beyond $M_{20}$ (or $C$ or \sersic\ index).  However, $F$ is less sensitive to dust, mergers, and other disturbances that move galaxies in a roughly perpendicular direction away from the main \gmtwenty\ locus.  $F$ traces quenched galaxies similarly well, if not a little better, than other structural parameters, and $F$ is closely related to the dominant principal component of a suite of ten non-parametric structural diagnostics (M.\ Peth et al.\ in prep.).  

With these, we can automatically characterize the \illustris\ galaxy
morphologies and compare them directly to measurements from real galaxies.  In
Appendix~\ref{a:morphs} we present the morphology catalog for \illustris\ at
$z=0$.

\section{Simulated Morphologies}  \label{s:morphologies}

Hydrodynamical models have only recently begun to reproduce basic statistics of the galaxy population, such as the galaxy stellar mass function and global star formation history. Without these ingredients, it has been impractical to compare morphology surveys to simulations in interesting numbers. However, given recent achievements, it is worthwhile to evaluate statistically the predicted morphologies and how they depend on mass and star formation. The extent to which they match may provide clues to problems with the assumed galaxy formation physics, or may reveal the relative sensitivity of observed galaxy populations to the assumed physics.  

In this section, we examine our morphology measurements from \illustris\ at $z \sim 0$. This approach will also support studies of how these morphologies change with time or would change if the galaxies were assumed to be at a greater distance, with shallower images, or with various observatories.  

For brevity, we have neglected dust.  Information about the metal and gas content of the simulated galaxies can be used to predict the effects of dust, for example using the 2-dimensional projections of metal density described in Section~\ref{ss:imagepipeline}.  We tested this technique, and found that the most important effect of dust is to lower the $G$ value, and raise the $M_{20}$ value of simulated galaxies, especially where $G$ is very large (for example, in the ``mergers'' region).  In fact, the observed distribution (Figure~\ref{fig:gm20_all}) implies that some kind of attenuation is \emph{required} to bring some high-$G$, high-$M_{20}$ simulated galaxies into better agreement with observed ones.  We have found that these general trends are indeed reproduced by the simple dust treatments we have tested \citep[following, e.g.,][]{DeLucia2007}.  {Specifically, in rest-frame $u$ filter images of galaxies with $G \gtrsim 0.6$, the effect of dust is to reduce $G$ by $\sim 0.05$, on average.  In other words, it tends to equalize the distribution of flux among the galaxies' pixels by obscuring the brightest star-forming regions.  In galaxies with $G \sim 0.5$, $M_{20} \sim -2$ (typically relatively massive composite disc+bulge systems), dust serves a similar role by increasing $M_{20}$ by $\sim 0.2$ on average, i.e. attenuating the bright cores and causing the light profile to appear slightly more extended. However, these effects largely do not change the coarse structural type that we would assign to the simulated images.  Moreover, along the main locus of points, the average effect of dust imparts no systematic bias on the morphology measurements. } These results are consistent with simulated image analyses by \citet{lotz08} and \citet{Snyder2015a}.  We will describe this dust modeling and related issues in a separate paper.

\begin{figure}
\begin{center}
\includegraphics[width=3.3in]{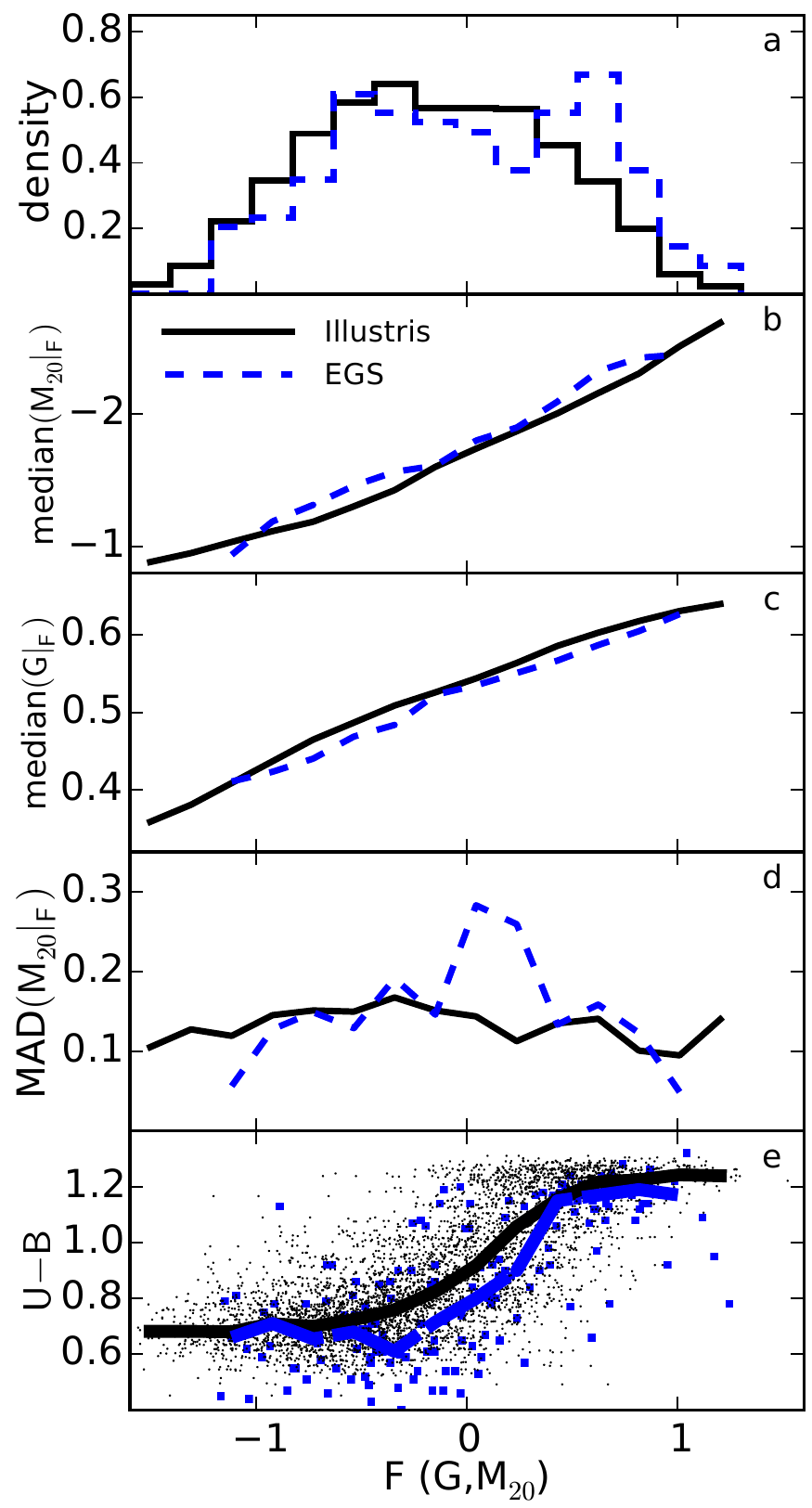}
\caption{{ Statistical comparison of morphology distributions between \illustris\ (black points and curves) at $z = 0.33$ and \hst\ data from EGS at $0.2 < z < 0.4$ (blue points and curves).  Panel a: histogram of \fgmtwenty.  Panels b and c: Position in \gmtwenty\ space versus $F$; where the locii differ in Figure~\ref{fig:gm20_color}, these curves diverge.  Panel d: median absolute deviation of $M_{20}$ values at a given $F$, showing how the widths of the main locii vary.  Panel e: color-morphology plot, showing all original points as well as median trendlines versus $F$. } \label{fig:statmorph} }
\end{center}
\end{figure}

\begin{figure*}
\begin{center}
\includegraphics[width=6.5in]{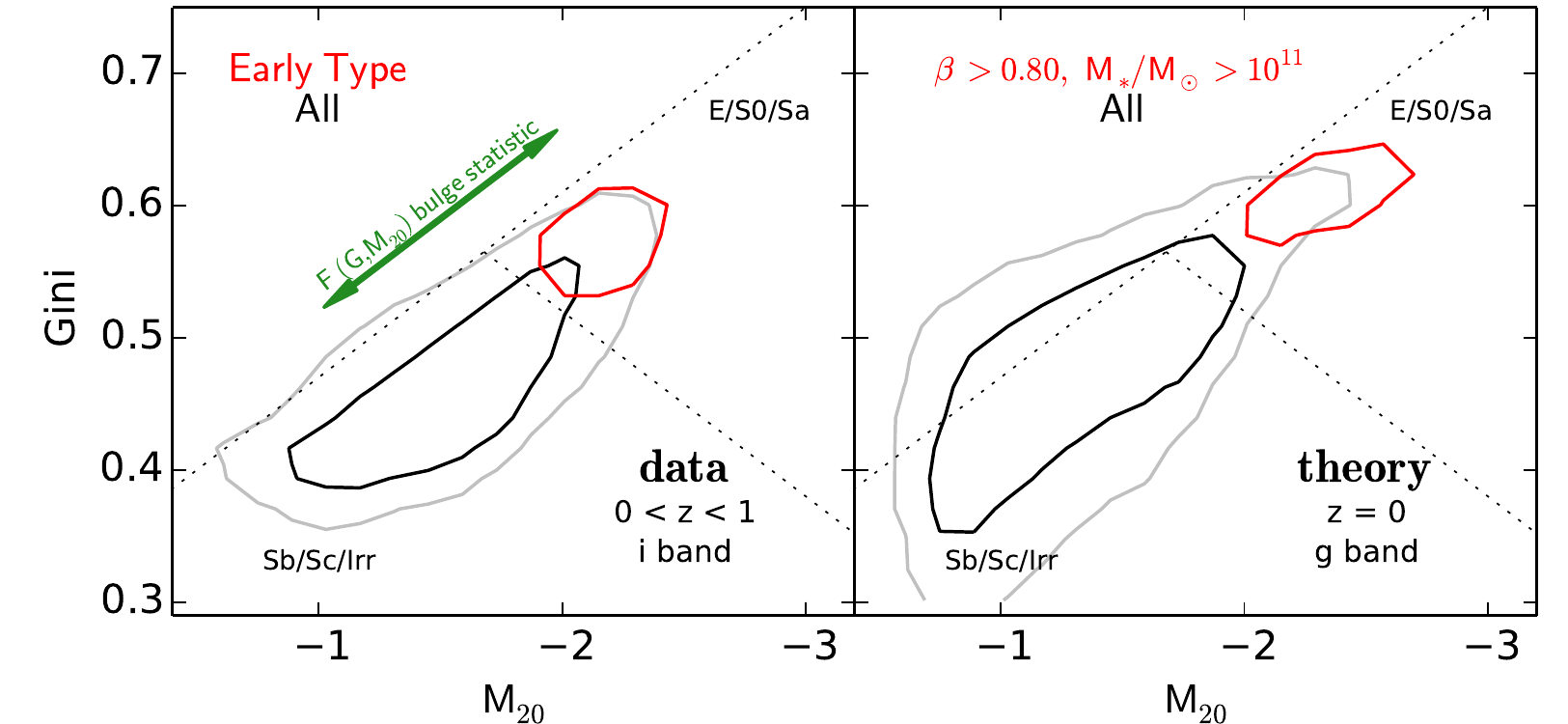}
\caption{Gini--$M_{20}$ as a bulge diagnostic.  Left: data from the Zurich Estimator of Structural Types \citep[ZEST;][]{Scarlata2007} from \hst\ ACS $I$-band images from the COSMOS survey \citep{Scoville2007}.  These data span roughly $0 < z < 1$ and reflect rest-frame $g$, $r$, and $i$-band light.  ZEST defined early type galaxies using a principal component decomposition of observed structures, which roughly matches the results of visual classification.  Right: $\sim 40000$ mock data points from the \illustris\ Simulation in the in rest-frame $g$-band, for all galaxies with $M_* > 10^{9.7} M_{\odot}$.  Solid red and black lines encircle $68\%$ of the data points, and solid gray lines $95\%$. $\beta$ is an estimate of the bulge fraction derived from the orbits of star particles (Section~\ref{ss:kine}).  This shows how the overall population of \illustris\ galaxies approximately matches the observed morphology distribution. However, the main locus is somewhat wider in \illustris\ than in the observed sample: there are more simulated galaxies with excess $G$ and $M_{20}$ than observed.\label{fig:gm20_all} \label{fig:bulgestatistic}}
\end{center}
\end{figure*}

\subsection{Comparison with data} \label{ss:data}

In Figure~\ref{fig:gm20_color}, we compare the 1200 most massive \illustris\ galaxies at $z=0.33$ with a volume-limited sample of galaxies in \hst\ imaging of the Extended Groth Strip (EGS) at $0.2 < z < 0.4$  \citep{lotz08_hst}.  These samples have a similar $B$-band absolute magnitude upper limit ($B \lesssim -19.5$).  For a direct comparison in Figure~\ref{fig:gm20_color}, we used simulated galaxies at $z=0.33$, whereas the rest of this paper discusses $z=0$ only. We added realism to the simulated images at $z=0.33$ so that their resolution and noise levels are similar to the $z=0$ sample described in Section~\ref{ss:realism}.  We also characterize the observed and simulated samples by star formation, in the form of optical colour. We assign to each point in Figure~\ref{fig:gm20_color} a symbol colour based on the source's $U-B$ colour, where the scaling is identical in the theory and data panels.  We split the sample at $U-B = 1.0$ and draw separate contours for the blue and red samples.  We used slightly mismatched filters in Figure~\ref{fig:gm20_color} (rest-frame $B$ or $g$ for \hst\ data but rest-frame $r$ for \illustris), because we created and analyzed here only the simulated $r$-band images at $z=0.33$. In these galaxies, the morphological k-correction is small: on average, switching to the $B$ band will increase $M_{20}$ by $\sim 0.1-0.2$ at most (assuming dust-free), and will change $G$ by a negligible amount ($< 0.01$), compared to the $r$ band images.  With dust, these differences would be even smaller.  

{ In Figure~\ref{fig:statmorph}, we examine summary statistics of the EGS and \illustris\ samples.  In panel a, the histogram of \fgmtwenty\ shows that the two samples both increase in density as $F$ increases above $-1$, have the same average density to within $\sim 10\%$ at $-0.5 \lesssim F \lesssim 0.5$, and then decrease in density to nearly zero at $F \sim 1$.  Real galaxies have a more bimodal morphology distribution than \illustris\ galaxies, since the observed number density drops to $\sim 2/3$ of the peak value at $F \sim 0.1$ where the \illustris\ distribution is flat.  Panels b and c show the position of the \gmtwenty\ morphology locus as a function of $F$: simulated galaxy images have slightly greater $G$ and $M_{20}$ values than observed ones, as hinted by the \illustris\ \gmtwenty\ locus in Figure~\ref{fig:gm20_color} being shifted up and to the left.  Panel d plots the width of the morphology distribution measured by the median absolute deviation from the median $M_{20}$ value at a given $F$.  Outside a narrow range at $0 \lesssim F \lesssim 0.2$, the widths of the distributions are almost the same with MAD$(M_{20|F}) \sim 0.10- 0.15$. The MAD of the EGS sample is large near $F \sim 0.1$ owing to the lack of a clearly defined locus of galaxies through this region in Figure~\ref{fig:gm20_color}; as we noted above, the EGS sample has a much lower number density here than in \illustris.  Panel e shows $U-B$ color versus $F$ for all galaxies in our comparison, with median trendlines overplotted.  \illustris\ galaxies are redder by $\sim 0.1$ mag at a given $F \sim 0$ than EGS galaxies, but otherwise the extrema are well matched with $U-B (F\sim -1) \approx 0.7$ and $U-B (F\sim 1) \approx 1.2$.  }

In Figure~\ref{fig:gm20_all}, we compare all of our imaged \illustris\ galaxies
at $z = 0$ with $M_* > 10^{9.7} M_{\odot}$ to the COSMOS survey
\citep{Scoville2007} at $0 < z < 1$.  Here we show results for all galaxies, including both centrals and satellites.  The $\sim 10000$ model galaxies
translates into $\sim 40,000$ points in this plot, with four viewing
directions per galaxy.  This comparison is not perfectly fair, because we are
neglecting the wavelength dependence of morphology, evolution of the galaxy population from $z=1$ to $z=0$, and
sample selection effects.  However, since the overall locus in \gmtwenty\ is
similar to the fairer volume-limited sample in Figure~\ref{fig:gm20_color}, we
conclude that this limitation does not strongly affect the general agreement. We show data from \citet{Scarlata2007}, who computed morphologies with the Zurich Estimator of Structural Types (ZEST), and show the full samples in the black contours which represent $68\%$ and $95\%$ of the populations.  Red contours represent $68\%$ of the ``early-types'', defined statistically from automated morphologies by ZEST and by orbital distribution of stellar mass in \illustris.  Thus both the real universe and \illustris\ have similar locations and number densities of their late-type and early-type galaxies.  

The main conclusions from Figures~\ref{fig:gm20_color}, \ref{fig:statmorph}, and \ref{fig:gm20_all}
are: \begin{enumerate}
\item{The overall locus of simulated galaxy points has nearly the same size, shape, and location in \gmtwenty\ space as the observed one (Figure~\ref{fig:statmorph}).  }
\item{There exist quenched, early-type galaxies in \illustris\ with a number density relative to star-forming late-type galaxies which is within a factor of two from the observed one (Figure~\ref{fig:statmorph}, panel a).  }
\item{{ \illustris\ achieves a qualitative match to the color-morphology relation (Figures~\ref{fig:gm20_color}, \ref{fig:statmorph}), with the extremes of the morphology population ($|F| \sim 1$) having distinct $U-B$ colors within $0.05$ magnitudes of observed galaxies, with a smooth transition between.}}

\end{enumerate}

This is shown, first, by the existence of a large number of simulated points at $M_{20} < -2$ and $G > 0.55$, in the region occupied by nearby true ellipticals, with very red colours (and low SFRs, see Figure~\ref{fig:gm20_color} and Section~\ref{ss:sfr}).  Second, we show contours relative to the same number density, and so the fact that the outer blue contours have roughly the same position in both panels implies that the relative number density of bulge-dominated and disc-dominated galaxies is reasonably well matched.  {Finally, Figure~\ref{fig:statmorph} confirms that the number densities as a function of morphology and color are broadly matched.}

These successes are essential first steps that any realistic simulation of galaxy formation should satisfy, but which has been difficult to attain until recently owing to limitations in modeling techniques and in computational resources. This agreement implies that \illustris\ is both large enough to simulate a fair sample of galaxies, and also realistic enough that the number density of early and late types are roughly matched to observations.  

We find several basic differences.  The theory contours in Figure~\ref{fig:gm20_all} and median statistics of Figure~\ref{fig:statmorph} both show a tendency to have slightly higher $G$ (by $\lesssim 0.5$) at a given $F$, and greater $M_{20}$ at a given $F$ (by $\sim 0.1$), when compared with data.  { As shown by Figure~\ref{fig:statmorph}, \illustris\ lacks a clear drop in number density (``valley'') between the extremes of the morphology and color distributions, causing disagreements of $\sim 10\%$ in number density and $\sim0.1$ magnitudes in color at a given $F \sim 0$.}  This could be a result of our decision to neglect dust \citep[see, e.g.,][]{Snyder2015a}.   For the simulated galaxies, there are relatively more points in the blue contours (late types), reflected by the wider inner blue contours. This can also be seen in the simulated mass dependence of quenching \citep{Vogelsberger2014b}. \illustris\ has more ``red spirals'' of intermediate shape, as reflected by the extension of the red contours to the lower left of the E/S0/Sa region.  

Despite these differences, Figure~\ref{fig:statmorph} demonstrates that
a general correlation between optical morphology and colour is in place.  Red
galaxies tend to be compact spheroids and blue galaxies tend to be extended
discs.

\begin{figure*}
\begin{center}
\begin{tabular}{cc}
\includegraphics[width=3.4in]{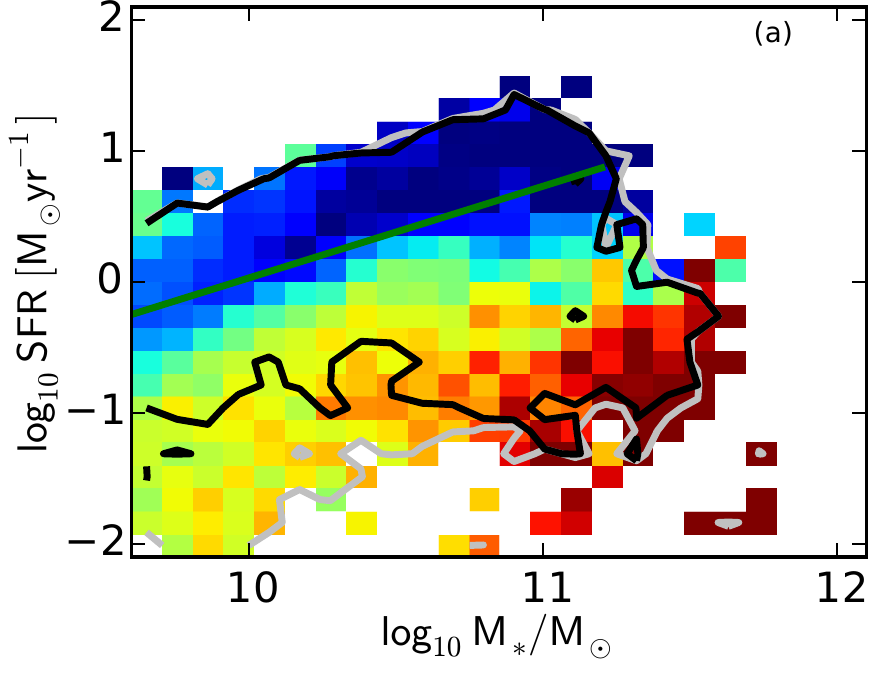} & \includegraphics[width=3.4in]{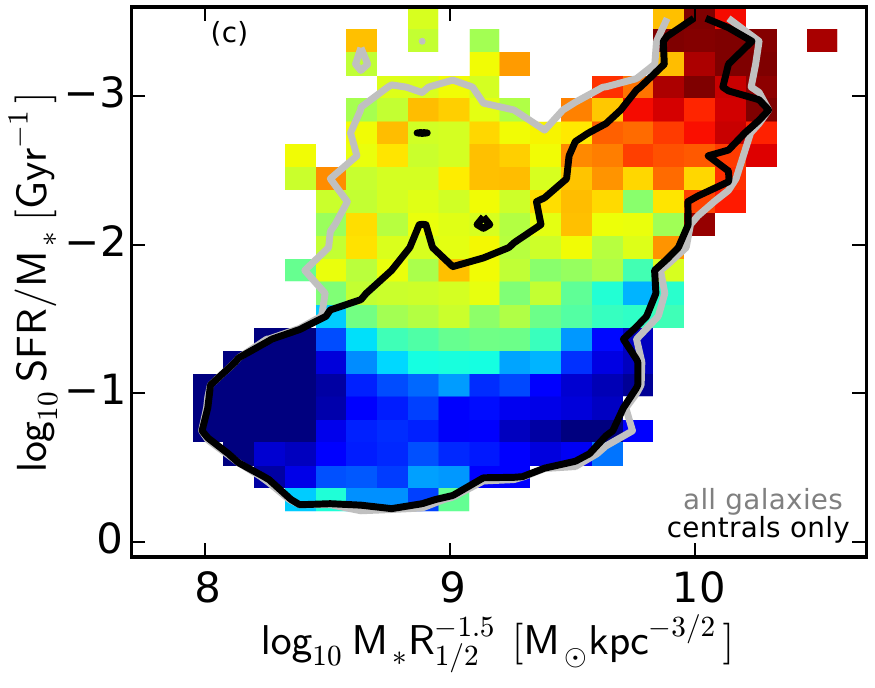} \\
\includegraphics[width=3.4in]{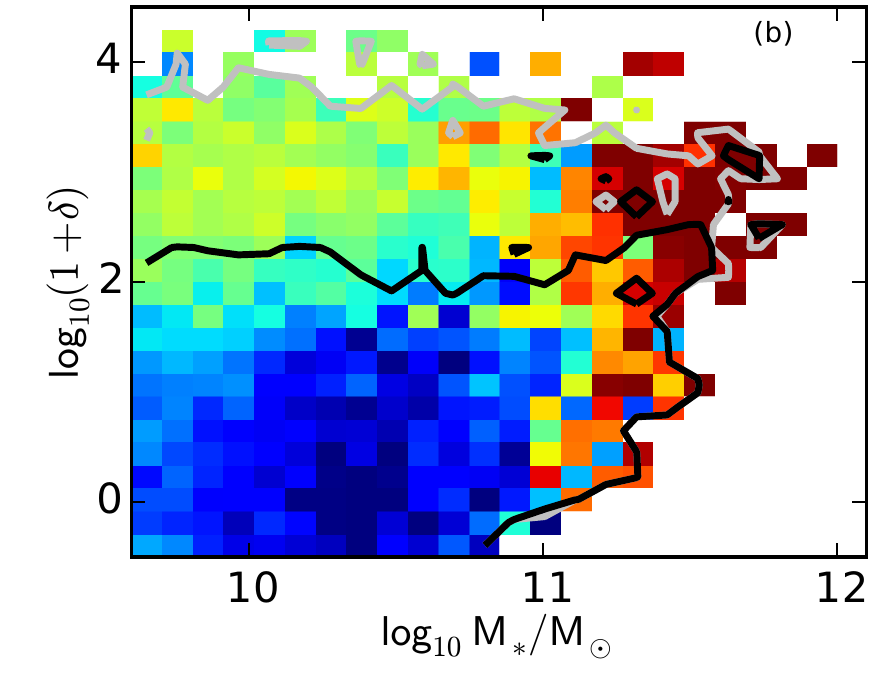} & \includegraphics[width=3.4in]{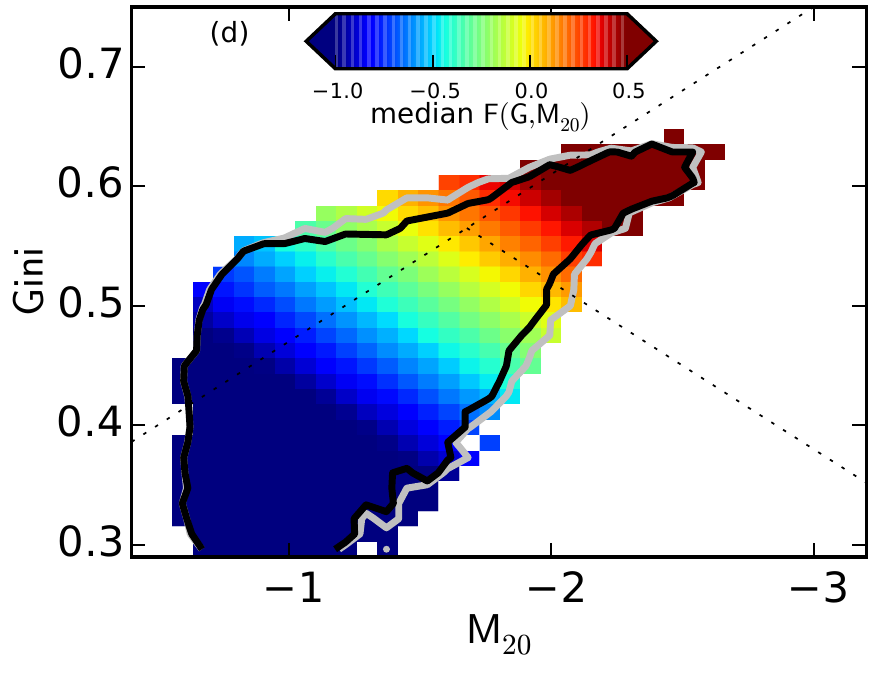} \\ 
\end{tabular}
\caption{SFR versus mass (panel a, top left), overdensity versus mass (panel b, bottom left), and SFR/$M_*$ versus compactness (panel c, top right), with colours proportional to galaxy structure (Equation~\ref{eq:fgm20}). In each bin, we measure the median $F(G,M_{20})$ value from rest-frame g-band images and assign it a colour following panel d.  Bulge-dominated galaxies are colour-coded red and disc-dominated ones are blue. The gray (black) contour outlines a region in which each bin contains 10 or more (central) galaxies. The green solid line in panel (a) is the star-forming main sequence parameterized by \citet{Whitaker2012}. Compared to surveys, \illustris\ recovers roughly the same dependence of structural type on $M_*$ and SFR \citep[e.g.,][]{Kauffmann2003,Wuyts2011}, on $\delta$ and $M_*$ \citep{Peng2010}, and on SFR/$M_*$ and $M_* R_{1/2}^{-1.5}$ \citep[e.g.,][]{Franx2008,Barro2013,Omand2014}. These trends appear to be a natural consequence of galaxy formation processes simulated with physics models crafted to match global star formation histories and stellar mass functions. \label{fig:sfrmass} }
\end{center}
\end{figure*}

\subsection{Dependence on star formation}  \label{ss:sfr}

Observationally, a galaxy's mass and morphology correlate with its star formation rate.  Nearby galaxies occur in a bi-modal distribution of mass and star formation (or luminosity and colour), in that the luminous/red galaxies appear to be a statistically distinct population from star-forming galaxies \citep[e.g.,][]{strateva01_gv}.  Surveys have found that quiescence is strongly correlated with measures of optical structure, such as \sersic\ index and compactness, albeit with large scatter \citep[e.g.,][]{Wuyts2011,Bell2012,Barro2013}.

Section~\ref{ss:data} hinted at the overall reasonableness of the simulated galaxy shapes and that early-type galaxies are redder than late-type ones. In Figure~\ref{fig:sfrmass}, we plot directly our automated non-parametric morphologies from \illustris\ as a function of $M_*$, SFR, and other properties.  For the simulated galaxies in each bin, we compute the median $F(G,M_{20})$ using the definition in Section~\ref{ss:morphology} and Figure~\ref{fig:gm20_all}, measuring from the unattenuated rest-frame $g$-band images.  To each of these median values of \fgmtwenty, we assign a colour from blue (disc-dominated) to red (bulge-dominated), as shown by the legend in Figure~\ref{fig:sfrmass}.  We also plot contours encircling bins that contain more than 10 measurements, black (gray) lines indicating central (all) galaxies.  We measure $M_*$ directly from the simulation outputs as the mass of the stellar particles contained within a subhalo-centered sphere with radius twice the 3D stellar half-mass radius.  We have checked that this definition does not have a noticeable impact on any of our results using $M_*$ (especially Section~\ref{s:otherscience}).

In panel (a) of Figure~\ref{fig:sfrmass}, we plot SFR versus $M_*$ and recover the ``main sequence'' of star-forming galaxies \citep[e.g.,][]{Noeske:2007a,Whitaker2012} in the blue squares. \citet{Torrey2014} showed that the \illustris\ model approximately reproduces this trend at $z=0$, and we find again that its location and slope in this plot is the same as that found by \citet{Whitaker2012}.  \citet{Genel2014} and \citet{Sparre2015} presented the \illustris\ galaxy main sequence as a function of time and show that it roughly matches observations at $z=0$ and $z=4$, but its normalization is too high compared with observations at $z\sim 2$.  Away from this main sequence in Figure~\ref{fig:sfrmass}, at fixed $M_*$, we find that lower-SFR galaxies have, on average, earlier structural types as measured by \gmtwenty. There are hints that compact or spheroidal starbursts populate the highly star-forming sides of this distribution (see also \citealt{Sparre2015}).

We find a clear separation between bulge-dominated and disc-dominated galaxies.  While here we are using a different morphology diagnostic, the transition from disk-dominated to bulge-dominated occurs at a location very similar to the one found by \citet{Wuyts2011}: at SFR $\sim 0.1\ M_{\odot}\rm yr^{-1}$ for $M_* \sim 10^{10} M_{\odot}$ and SFR $\sim 1\ M_{\odot}\rm yr^{-1}$ for $M_* \sim 10^{11} M_{\odot}$. Compared to observations, \illustris\ appears to have too few low-SFR, bulge-dominated central galaxies at $M_* \sim 10^{10.5} M_{\odot}$ in panel (a).  

Panel (b) of Figure~\ref{fig:sfrmass} presents morphology as a function of overdensity and $M_*$.  We use the same definition of 3-dimensional overdensity $\delta$ as \citet{Vogelsberger2014a}, who showed that star formation is correlated inversely with density at fixed $M_*$, matching observed trends \citep[e.g.,][]{Peng2010}. Since \fgmtwenty\ is correlated tightly with SFR (panel a), we recover a similar dependence of galaxy morphology on mass and environment. Thus, \illustris\ reproduces basic features of ``mass quenching'', ``environment quenching'', and the morphology-density relation.

In panel (c) of Figure~\ref{fig:sfrmass}, we plot SFR/$M_*$ versus $M_* R_{1/2}^{-1.5}$, a measure of compactness \citep[e.g., $\Sigma_{1.5}$ from][]{Barro2013}. This quantity is closely related to a surface mass density, which has been shown to correlate with SFR/$M_*$ out to high redshifts \citep[e.g.,][]{Franx2008,Omand2014}.  As also seen in individual galaxy tracks by \citet{Genel2014}, we see that compactness correlates with SFR/$M_*$ in \illustris\ central galaxies: the black solid contour in panel (c) encircles bins containing more than 10 central galaxies, indicating a tight relationship between $M_* R_{1/2}^{-1.5}$ and SFR/$M_*$. Here, we find that both quantities also correlate tightly with bulge strength, as indicated by the change from blue to yellow to red bins as one moves along the relation to the upper right corner.  Compared with results by \citet{Barro2013} at $z \gtrsim 0.5$, \illustris\ galaxies have somewhat smaller SFR/$M_*$ and $M_* R_{1/2}^{-1.5}$.  This difference owes to the decline of SFR/$M_*$ over time \citep[e.g.,][]{Noeske:2007a,Genel2014} and larger-than-observed \illustris\ galaxies (Section~\ref{ss:size}). 

Another feature of panel (c) in Figure~\ref{fig:sfrmass} is the structural difference between central and satellite galaxies, which can be seen by the difference between the gray and black contours. These encircle all and central galaxies, respectively.  There is a single tight correlation between SFR/$M_*$ and $M_* R_{1/2}^{-1.5}$ in central galaxies, but less so among satellite galaxies, which are quenched and early type but have lower $M_* R_{1/2}^{-1.5}$ at a given SFR/$M_*$.  

With the three trends of Figure~\ref{fig:sfrmass}, we conclude that the \illustris\ simulation produces a $z=0$ galaxy population with optical morphologies that successfully match trends emphasized in survey analyses.  These include the SFR-$M_*$-Morphology relation \citep[e.g.,][]{Kauffmann2003,Blanton2003,Wuyts2011}, environment versus mass quenching \citep[e.g.,][]{Peng2010}, and the tight correlation between galaxy compactness, star formation, and morphology \citep[e.g.,][]{Franx2008,Williams2010,Barro2013,Omand2014}. These trends have been thought important for understanding key issues in the formation of massive galaxies, such as bulge formation, quenching, and emergence of the Hubble Sequence. 

The recovery of the median morphology of simulated galaxies as functions of these properties, trends not explicitly matched in the \illustris\ physics model selection, is an intriguing success of such simulations.  The ingredients required to obtain reasonable global stellar mass functions and star formation histories appear to also imply these more detailed relationships between galaxy morphology, mass, and star formation.

\begin{figure*}
\begin{center}
\includegraphics[width=4.5in]{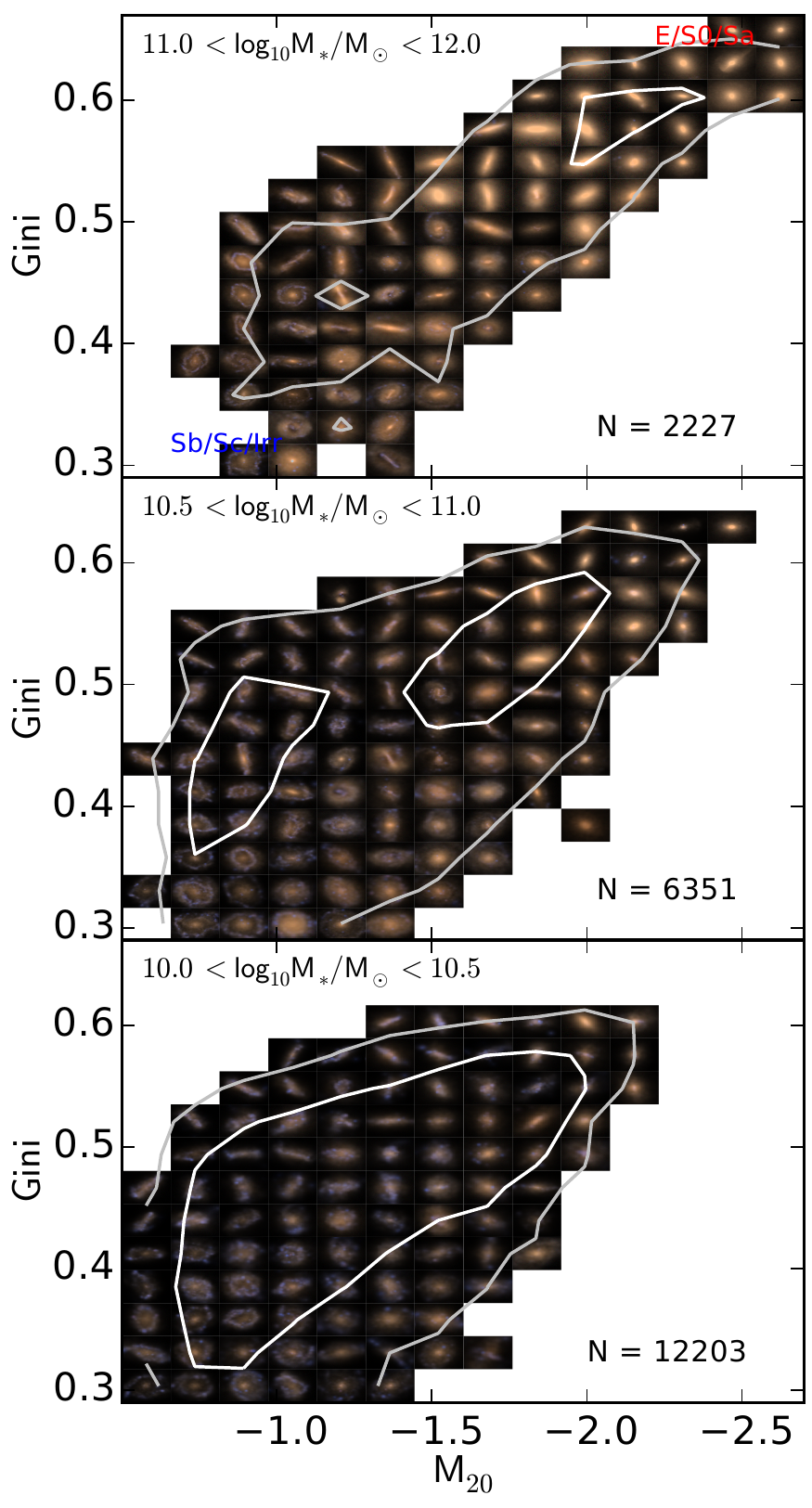}
\caption{ Rest-frame $g$-band morphology distributions and colour-composite
  images as a function of mass. For each bin, we show the galaxy having \fgmtwenty\ nearest the median value.  White (gray) contours encircle bins containing 100 (10) or more galaxies. From these
  we see an expected trend: at lower mass (bottom), most galaxies are still forming
  stars in a disk, and therefore reside in the late-type region ($M_{20} > -2$, $G < 0.6$).  At the
  highest masses (top), galaxies are quenched, and therefore they are concentrated
  mainly in the early-type region.  However, at $\log_{10}
  M_*/M_{\odot} < 11.0$, we identify a peculiar population of
  galaxies: there are two peaks in \gmtwenty\ in the middle panel, one at the expected location
  bridging the early/late type divide as galaxies evolve and quench, and
  another off to the left, at $M_{20} > -1.0$ and $G \sim 0.4$.  See the text
  for more discussion. \label{fig:mass}}
\end{center}
\end{figure*}

\subsection{Dependence on stellar mass} \label{ss:mass}

In this section we show in more detail how the morphologies of \illustris\ simulation galaxies depend on stellar mass.  Figure~\ref{fig:mass} presents the distribution of \gmtwenty\ in \illustris\ as a function of stellar mass at $z=0$, as measured from the rest-frame $g$ band images. We separate the measurements into three bins from $10.0 < \log_{10} M_*/M_{\odot} < 12.0$.  In each panel, we bin galaxies by \gmtwenty\ values and show an example image selected from each bin as having most nearly the median morphology, \fgmtwenty. White (gray) contours encircle bins containing more than 100 (10) galaxies.

The basic trend of Figure~\ref{fig:mass} follows what we would expect given that morphology in \illustris, and most galaxy formation models, is closely correlated with mass.  Simulated galaxies tend to occupy the Sb/Sc/Irr region of the \gmtwenty\ diagram at low mass, and the main locus shifts smoothly to the E/S0/Sa region at $M_* > 10^{11} M_{\odot}$.  This overall shift is visible in the galaxy images: at the top are quenched bulge-dominated galaxies, at the bottom star-forming galaxies.  The star-forming galaxies have a diverse morphology: some are almost entirely blue and disc-dominated, while others are composites, having a compact red bulge.  

At first glance, there are two possible issues with the \illustris\ model in \gmtwenty\ versus mass:
\begin{enumerate}
\item{It appears there are very few, if any, galaxies with $M_* < 10^{10.5} M_{\odot}$ with a bulge-dominated morphology ($M_{20} < -2$, $G > 0.55$); compare also our Figure~\ref{fig:sfrmass} panel a to \citet{Wuyts2011}.  
We believe that this relates to the difficulty of properly quenching star
formation in some galaxies with $M_* \lesssim 10^{11} M_{\odot}$.  
\label{it:nobulge}}
\item{At $10^{10.5} \lesssim M_*/M_{\odot} \lesssim 10^{11}$, the \gmtwenty\ locus appears to separate into two distinct populations.  The main one near the center of the diagram, intermediate between the adjacent panels, and another with very extended morphology: $M_{20} > -1$, $G < 0.45$. \label{it:ring}}
\end{enumerate}

The origin of the second locus of galaxies with masses $M_* \lesssim 10^{11} M_{\odot}$ is unclear. The feature appears to be caused by rings of star formation which are visible in the example images in Figure~\ref{fig:mass} (also Figure~\ref{fig:realism}).  We discuss the origin of this phenomenon in Section~\ref{ss:disc_detailed}.

It is easy to see how $M_{20}$ can be sensitive to this feature.  In short, $M_{20}$ represents the spatial extent of the brightest quintile of a galaxy's pixels.  If there exists a ring of star formation that is brighter than in typical spirals, as we hypothesize here, then the brightest (star-forming) pixels may be unusually separated on the sky.  Figure~\ref{fig:mass} shows an abundance of objects of this character in the rest-frame g-band, but this feature is also present in the rest-frame $i$ and $H$-band morphologies and images.  The images highlighted from this region show both blue pixels and yellow/orange pixels, indicating that these ``rings'' can be sustained long enough to create older stellar populations at that radius.

\begin{figure*}
\begin{center}
\includegraphics[width=6.4in]{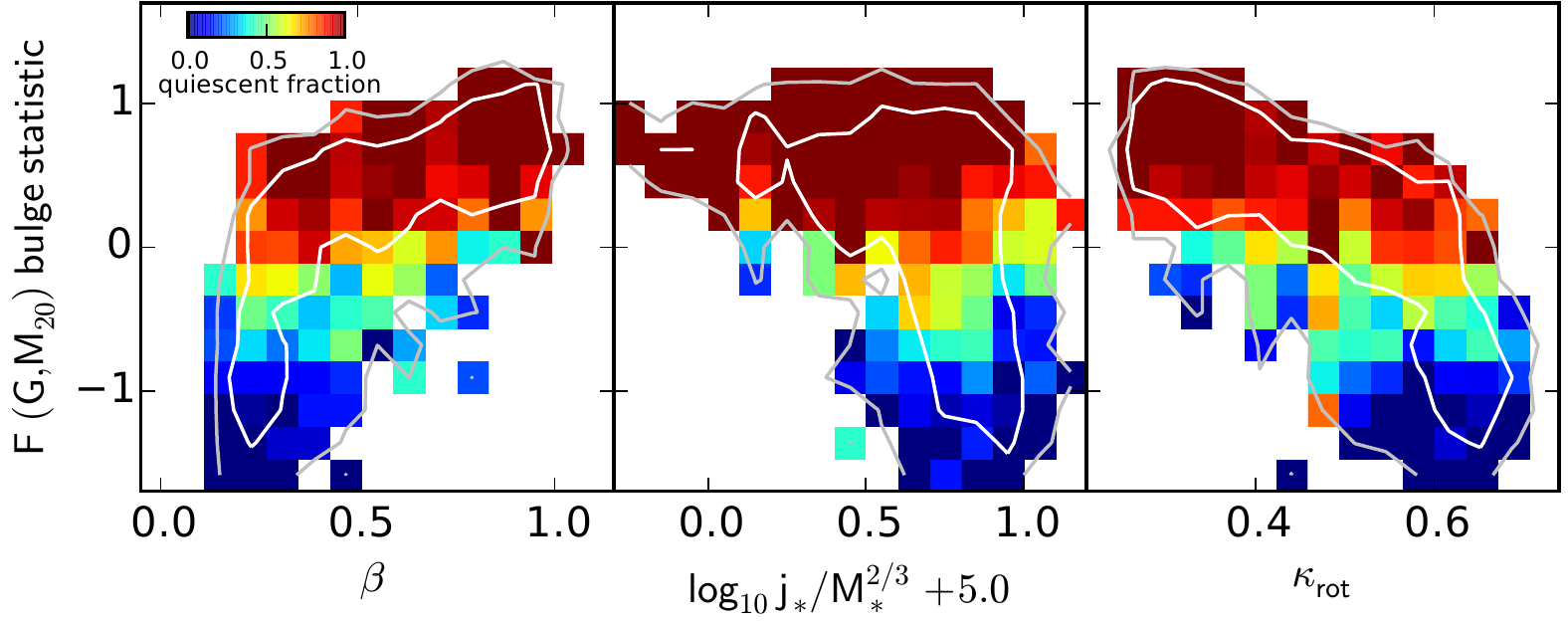}
\caption{Simulated distribution of optical morphology versus kinematic bulge fraction (left), ratio of specific angular momentum to $M_*^{2/3}$ (center), and $\kappa_{\rm rot}$ parameter (right), in galaxies with $M_* > 10^{11} M_{\odot}$.  White (gray) contours enclose regions containing $70\%$ ($95\%$) of the model sources, and colours of the rectangles indicate the fraction of quiescent galaxies in each bin.  We define quiescent fraction as the fraction of galaxies with $\log_{10}$ SSFR$/(\rm yr^{-1})$ $< -0.24 (\log_{10} M_*/M_{\odot}) - 8.50$ \citep{Omand2014}.   We define $\beta$ as twice the fraction of star particles with negative circularity parameter; see text for full definition. Section~\ref{ss:kine} defines $j_*/M_*^{2/3}$ and $\kappa_{\rm rot}$. Light-based morphology traces kinematic-based ones for $M_* > 10^{11} M_{\odot}$ (right), especially at $\beta > 0.4$.  Since the distribution of colours is largely horizontal, the optical morphology traced by \gmtwenty\ is tightly correlated with quiescence.  \label{fig:circbeta} \label{fig:rotation}}
\end{center}
\end{figure*}

\section{Optical Structure and Kinematics}  \label{s:rotation}

The optical morphology of an image is a projection of the SEDs of stellar populations orbiting in the galaxy. Since simulations predict these populations, we can relate structural parameters of simulated galaxy images to the motions of stars. In principle, the kinematic information provided by large-scale simulations could be used not only to interpret observations \citep[e.g.,][]{Kassin2014} but also to constrain galaxy physics models from integral field unit surveys.  Moreover, optical and kinematic morphologies are sensitive in different ways to the stellar mass and SFR distributions, and understanding this connection is important for large imaging surveys.

\subsection{Morphology versus Rotation}  \label{ss:kine}

We first compute several measures of galaxy rotation based on stellar motions. We follow previous studies \citep[e.g.,][]{scannapieco10, Sales2012, Marinacci2013} to parameterize the orbits of star particles, and contrast these quantities with photometric morphologies.  

For each galaxy subhalo, we compute the total specific angular momentum $\vec{j_*}=\sum{\left ( m\ \vec{v} \times \vec{r} \right )}/\sum{m}$ of the star particles within ten times the galaxy's stellar half-mass radius.  {This radius is chosen to correspond with observational studies \citep[e.g.,][]{Fall2013}, which find that a significant fraction of a galaxy's stellar angular momentum lies beyond the half-mass radius \citep{Romanowsky2012}. \citet{Genel2015} found that using all star particles in the subhalo instead of restricting to $10 R_{1/2}$ (or $5 R_{1/2}$) does not change the resulting scaling relations in \illustris\ galaxies.}  We define the direction of $\vec{j_*}$ to be the $z$ unit vector and compute the specific angular momentum of each star particle in that direction, $j_z = (\vec{v} \times \vec{r})_z$. We then define a circularity parameter $\epsilon = j_z/j_z(E)$, where $E$ is the particle's binding energy, and $j_z(E)$ is the maximum specific angular momentum among the 100 particles with a binding energy closest to $E$. Finally, for each galaxy subhalo we compute $j_* = |\vec{j_*}|$ and several summary statistics of the distribution of $\epsilon$. These quantities correlate with rotational support or bulge mass fraction.

In Figure~\ref{fig:rotation}, we plot \fgmtwenty\ of massive galaxies ($M_* > 10^{11} M_{\odot}$) against three such parameters:
\begin{enumerate}
\item{$\beta =$ twice the fraction of star particles with $\epsilon < 0$ \citep[e.g.,][]{Abadi2003}. Perfectly cold discs should have $\beta \sim 0$, and bulge-dominated systems have $\beta \sim 1$.}
\item{$\log_{10} j_*/M_*^{2/3}$.  Since specific angular momentum scales with $M^{2/3}$ \citep{Romanowsky2012},  $j_*/M_*^{2/3}$ is a measure of relative rotational support. This quantity is similar to projected, observed parameters such as $\lambda_R$ used in kinematic surveys \citep[e.g., \atlas][]{Cappellari2011}, and has units of $\rm km\ s^{-1}\ kpc\ M_{\odot}^{-2/3}$. }
\item{$\kappa_{\rm rot} = \frac{1}{K} \sum{\frac{m}{2} \left ( \frac{j_z}{R} \right )^2}$, as defined by \citet{Sales2012}, where $K$ is the total kinetic energy of star particles and $R$ is the distance between the star particle and galaxy center. $\kappa_{\rm rot}$ is the fraction of kinetic energy occupied by ordered rotational motion. Bulge-dominated systems have $\kappa_{\rm rot} \ll 1$, while perfectly cold discs have $\kappa_{\rm rot} \sim 1$. }
\end{enumerate}

In each case, \fgmtwenty\ has some dependence on the kinematic measure. Simulated galaxies with $\beta > 0.5$ are optically early type (\fgmtwenty\ $> 0.0$), and there is a locus of objects with $\sim 0.2$ dex scatter extending from ($\beta \sim 0.3$, \fgmtwenty\ $\sim 0$) to ($\beta \sim 1$, \fgmtwenty\ $\sim 0.7$), so that \fgmtwenty\ correlates positively with $\beta$.  There is a large diversity of optical morphologies when $\beta < 0.5$, with \fgmtwenty\ spanning from $-1$ to $0.5$. 

Similarly, the rotation measures $j_*/M_*^{2/3}$ and $\kappa_{\rm rot}$ are inversely proportional to optical diagnostics of bulge strength.  \fgmtwenty\ versus $\kappa_{\rm rot}$ roughly follows the quantity $1-\beta$, with a wide scatter in disc-dominated galaxies, albeit slightly less scatter than for $\beta$.     This scatter appears to increase in galaxies with lower mass ($M_ \lesssim 10^{10.5} M_{\odot}$), and could reflect more complex structures in these galaxies compared with very massive ones. For example, the $\beta$ indicator assumes that bulges are rotation-free, but we know that many early-type galaxies are significant rotators.  Additionally, structures like bars and counter-rotating disks can confound the kinematic indicators used here.

\subsection{Quiescence versus Morphology and Rotation}  \label{ss:rot_vs_sfr}

\begin{figure*}
\begin{center}
\includegraphics[width=6.8in]{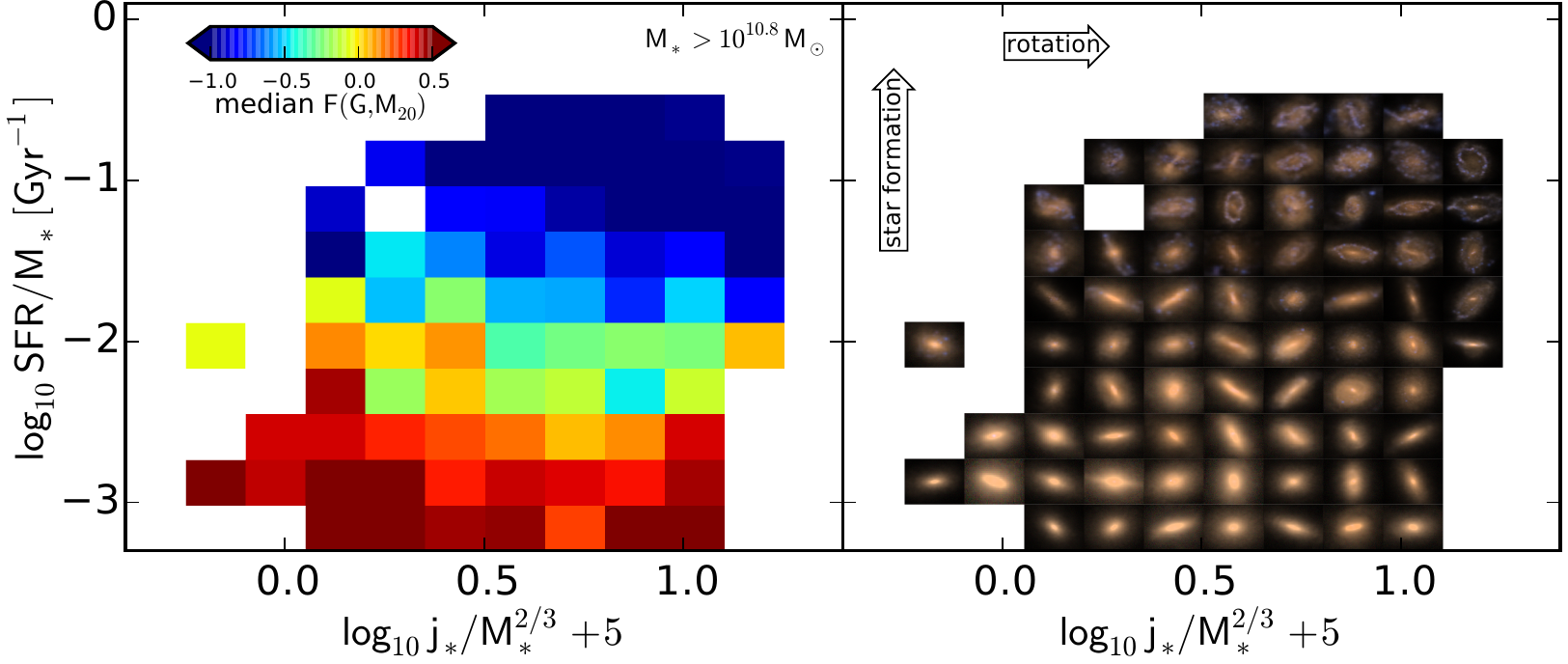}\\
\caption{SFR$/M_*$ versus rotation in \illustris.  Left: We colour each bin by the median \fgmtwenty\ of the galaxies contained therein. Right: We fill each bin with the colour-composite image having most nearly the bin's median \fgmtwenty\ value. Therefore, we obtain a kinematically defined galaxy sequence qualitatively matching observed ones \citep[e.g.,][]{Cappellari2011}. At the bottom left are the slowly rotating early-type galaxies. Moving to the right, the automated or visual classifications remain early-type, while the rotational measure increases and several edge-on examples appear disc-dominated. Having high rotation and high SFR$/M_*$, the top right of the figure contains obviously late-type star-forming galaxies as well as the ring galaxies of Section~\ref{ss:mass}. \label{fig:rotationimages}}
\end{center}
\end{figure*}

In Figure~\ref{fig:rotation}, we colour bins by quiescent fraction, defined as the fraction of galaxies with $\log_{10}$ SSFR$/(\rm yr^{-1}) $ $< -0.24 (\log_{10} M_*/M_{\odot}) - 8.50$ \citep[e.g.,][]{Omand2014}. This roughly matches the definition of quiescent or passive galaxies used in several other studies \citep[e.g.,][]{brammer09, Woo2015}.  As we showed in Figure~\ref{fig:sfrmass}, optical morphology traced by \fgmtwenty\ correlates tightly with quiescence at almost any mass $M_* \gtrsim 10^{10} M_{\odot}$.  

Intriguingly, quenching appears to be correlated less well with kinematic morphology: the bin colours are distributed \emph{horizontally} in the \fgmtwenty\ versus $\beta$ and $\kappa_{\rm rot}$ panels of Figure~\ref{fig:rotation}. By contrast, the colours trace a more diagonal trend in $F$ versus $j_*/M_*^{2/3}$: this x-axis quantity helps to predict quenching at fixed $F$.  However, this is true primarily in rotation-dominated systems. 

This implies that optical morphology closely traces star formation activity in \illustris, even when the simulated kinematic morphology does not.  This could result from several factors, including that the kinematic tracers are mass-based while the image tracers are light-based.  Therefore the optical profile follows the mass profile only if star formation does.  

Figure~\ref{fig:rotationimages} summarizes how \fgmtwenty, SFR/$M_*$, and rotation correlate with each other in \illustris.  In the left panel, we see that at fixed relative rotation, $F$ is inversely proportional to SFR/$M_*$.  And at fixed SFR/$M_*$, $F$ is inversely proportional to $j_*/M_*^{2/3}$.  In the right panel, we over-plot images chosen to have most nearly the median $F$ value of galaxies in the associated bin.  

In Figure~\ref{fig:rotationimages}, the low-redshift \illustris\ galaxies occupy a sequence very similar to the one put forth by observational studies \citep[e.g.,][]{Emsellem2007,Cappellari2007,Cappellari2011}: quenched bulge-dominated galaxies with very little rotation are rare and occupy the bottom left tip of the locus of galaxies.  Moving right from these slow rotators we see the ``fast rotators'', which can be visually very similar to the slow rotators when viewed face-on.  However, there are obviously some disc-dominated galaxies viewed edge-on in the lower right quadrant of Figure~\ref{fig:rotationimages}, as in real galaxies.  In rotating galaxies, there is a continuum of star formation activity which is correlated with the appearance of spiral arms and with having a more disc-dominated distribution of optical light.  

The trends in this figure follow those which kinematic surveys of nearby galaxies have identified recently.  For example, galaxies with bulge-dominated $F$ values occupy a huge range of relative levels of rotation.  Thus, in red galaxies, it is very difficult to tell from optical morphology alone, either visually or automatically, whether a galaxy has significant rotation.  Moreover, at fixed nonzero rotation level, simulated galaxies have a wide range of SFR activity.  At minimal rotation ($\log_{10} j_*/M_*^{2/3} + 5 \lesssim 0$), most simulated galaxies are quenched.  This is reminiscent of recent work emphasizing a kinematic perspective on the Hubble sequence \citep[e.g.,][]{Emsellem2007,Krajnovic2008,Cappellari2011}, whereby early type galaxies with high levels of rotation are more closely related with ``anemic spirals'' \citep[e.g.,][]{VanDenBergh1976} than to true slowly-rotating elliptical galaxies.

\subsection{Morphology-density}  \label{ss:env}

\begin{figure}
\begin{center}
\includegraphics[width=3.2in]{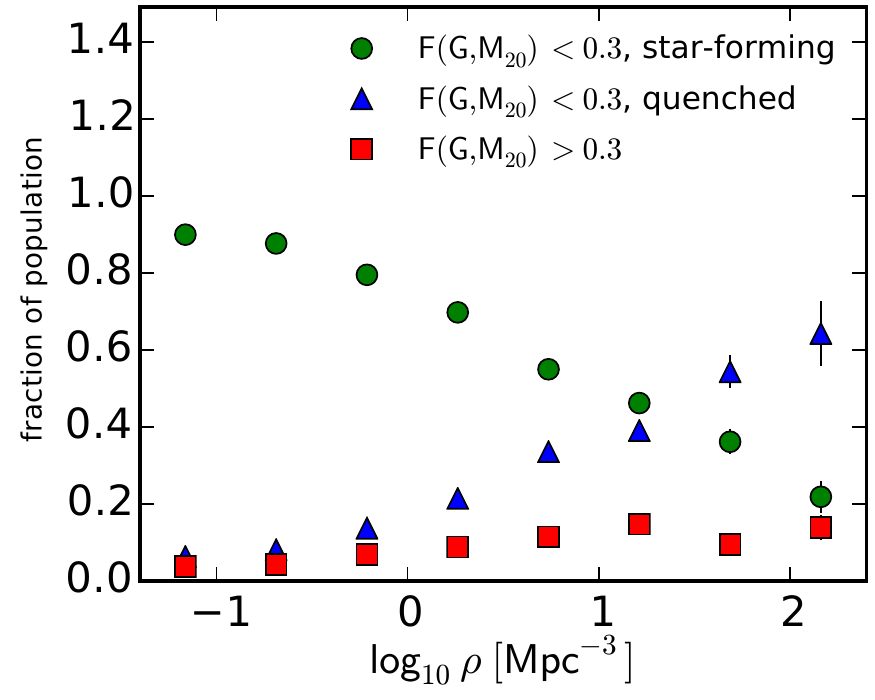}
\includegraphics[width=3.2in]{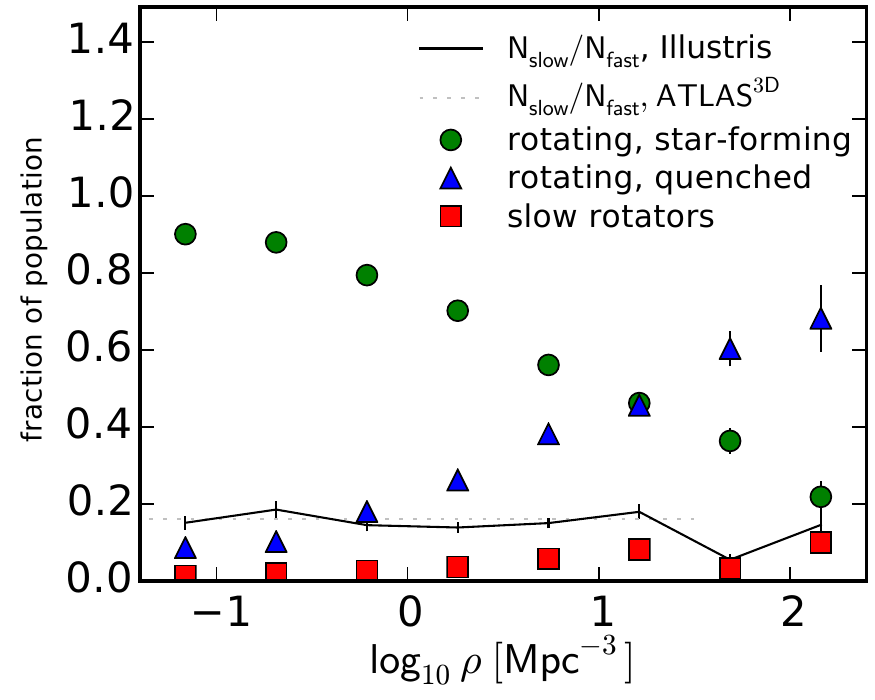}
\caption{Simulated morphology-density relations.  Top: Type fractions, classified by SFR/$M_*$ and \fgmtwenty, as a function of local galaxy volume number density $\rho$ for $M_* > 10^{9.7} M_{\odot}$.  Bottom: Type fractions, classified using SFR/$M_*$ and rotation.  We follow integral field unit surveys, such as \atlas\ \citep{Cappellari2011a}, to classify simulated galaxies as slow rotators, rotators with star formation (a.k.a. late-type galaxies), and quenched rotators.  The share of late-type discs falls smoothly as density increases, falling below the share of early-type discs at $\rho \sim 10-20\rm\ Mpc^{-3}$.  We plot the ratio between the numbers of slow and fast rotators among quenched galaxies (slow rotators are almost exclusively quenched, i.e. Figure~\ref{fig:rotation}), from \illustris\ (black line with error bars indicating the $1\sigma$ Poisson noise) and the global average from \citet[][dotted line]{Cappellari2011}.  At much higher densities (i.e., the richest clusters), there are insufficient statistics in the $z=0$ \illustris\ snapshot to continue following this relation.  
\label{fig:env}}
\end{center}
\end{figure}

\citet{Vogelsberger2014b} showed how quiescence traces environment in \illustris\ galaxies, and we showed in Figure~\ref{fig:sfrmass} that this translates to the expected dependence of galaxy structure on density. In Figure~\ref{fig:env}, we explore in greater detail such a ``morphology-density'' relation, and how it appears in \illustris\ when using the slightly finer definitions of morphology discussed in Section~\ref{ss:rot_vs_sfr}.  We use $\rho$ as the volume number density of galaxies following the conventions of observational studies \citep[e.g.,][]{dressler80,Cappellari2011}. 

First, in the top panel of Figure~\ref{fig:env}, we classify simulated galaxies as late-type discs, early-type discs, or ellipticals using galaxies' SFR/$M_*$ and \fgmtwenty.  At low densities, $\rho \lesssim 1\rm\ Mpc^{-3}$, the population is $> 80\%$ star-forming disc-like galaxies ($F < 0.3$, SFR/$M_* > 0.01\rm\ Gyr^{-1}$). As the density increases, the share of late-type discs shrinks smoothly, falling below $0.5$ when $\rho \gtrsim 10\rm\ Mpc^{-3}$, while the share of early-type discs smoothly increases.  The share of near or pure elliptical galaxies ($F > 0.3$) is small but increases smoothly to $\sim 10\%$ over this range.  

To choose these classification boundaries, we followed an approach similar to those taken by kinematic studies, which have highlighted the difference between slowly rotating ellipticals and red, rotating disc galaxies \citep[e.g.,][]{Cappellari2011}, both of which could be classified visually or automatically in projected light as early type. Galaxies with $F > 0.3$ are almost entirely classical bulge-dominated ellipticals, while those with $F < 0.3$ are a mix of bulge- and disc- dominated galaxies, often with significant rotation.  

In the bottom panel of Figure~\ref{fig:env}, we classify galaxies directly using their stellar mass kinematics and SFR/$M_*$.  We follow an approach similar to above, selecting slow rotators as those with extremely low specific angular momentum relative to their mass: $\log_{10} j_*/M^{2/3}+5 < 0.0$.  These are almost exclusively quenched, massive, and optically early type ($F>0.0$: Figure~\ref{fig:circbeta}). By contrast, as we saw in Section~\ref{ss:kine}, simulated rotating galaxies have a broad range of properties. Therefore, we separate the rotating galaxies at SFR/$M_*=0.01\rm\ Gyr^{-1}$ into quenched and star-forming classes.

We find that the kinematic morphology-density relation is very similar to the optically defined one: At $\rho \lesssim 1\rm\ Mpc^{-3}$, the population is $> 80\%$ rotating star-forming galaxies. As the density increases, the share of rotating quenched galaxies smoothly increases, reaching $50\%$ at $\rho \sim 10\rm\ Mpc^{-3}$. The share of slowly rotating galaxies increases smoothly from $\sim 1\%$ to $\sim 10\%$ over this range.  At the lowest densities, there is a hint that the simulated galaxy population is saturating as star-forming discs, in contrast to observations \citep[e.g.,][]{Cappellari2011}.  However, this is an expected consequence of the fact that the \illustris\ model yields too little quenching in the low-mass central galaxies dominating these fractions.  

In both panels of Figure~\ref{fig:env}, the share of early-type discs equals the share of late-type discs at a volume density $\rho \sim 10-20\rm\ Mpc^{-3}$, closely matching the results of \atlas\citep{Cappellari2011}, which has a very similar selection limit ($M_* \gtrsim 6\times10^{9} M_{\odot}$) to the \illustris\ sample here ($M_* \gtrsim 4.5\times 10^{9} M_{\odot}$). The transition from star-forming (low densities) to quenched (high densities) is extremely smooth, with no evidence for a break or dramatic change at any density threshold, even beyond the \atlas\ sample range at $\log_{10} \rho \gtrsim 1.5$.  Moreover, the fraction of slow rotators, relative to all quenched galaxies, is nearly constant at almost $20\%$, also roughly matching the results of \atlas\ \citep{Cappellari2011}. The normalizations of both simulated and observed ratios is somewhat arbitrary, since they depend on the selection cut between slow and fast rotation. Regardless, we find that the simulated $N_{\rm slow}/N_{\rm fast}$ is nearly constant over three orders of magnitude in density.  

Above a density of $\rho \sim 100\rm\ Mpc^{-3}$, the statistical uncertainty on the population fractions is too large to continue following these trends, {owing to the finite volume of \illustris\ ($106$ Mpc)$^3$.  While this is adequate for comparing to very local volumes (e.g., Virgo cluster samples), we require simulations with volumes $\gtrsim (200\rm\ Mpc)^3$ to probe this regime more with greater statistical significance and contrast with data from forthcoming wide-area, high-resolution imaging surveys from \citep[e.g., {\sc Euclid} and WFIRST,][]{Laureijs2011,Spergel2015}. } 

{ However, to improve the situation at lower galaxy masses (and lower number densities) may require qualitatively improved quenching mechanisms.  While the colours of \illustris\ satellite galaxies are in agreement with data \citep{Sales2014}, it is possible that numerical resolution better than $\sim 1$ kpc is required to implement efficient enough feedback physics, or to allow the ISM to quench through different mechanisms in isolated or central galaxies (See also Section~\ref{ss:quenchingproblem}).  }

\section{Morphology, Quenching, and Galaxy Properties}  \label{s:otherscience}

In Sections~\ref{s:morphologies} and \ref{s:rotation}, we showed how optical galaxy morphologies in the \illustris\ simulation at $z=0$ are a reasonable match to several observed trends broadly related to massive galaxy evolution. This appears to be a consequence of crafting the sub-grid supernova and AGN feedback models to roughly recover the total stellar mass content of halos over cosmic time.  An interesting question becomes: Given such tuning, is the success with galaxy morphology inevitable, or is this particular class of comprehensive models special in some way?

In this section, we explore how \illustris\ galaxy morphologies depend on other aspects of the simulated galaxies, such as their optical sizes (Section~\ref{ss:size}), SMBH masses (Section~\ref{ss:smbh}), and dark matter (DM) halo masses (Section~\ref{ss:halo}).  The idea is to search for predictions that could be unique to the specific galaxy physics assumed in \illustris.  

\begin{figure}
\begin{center}
\includegraphics[width=3.4in]{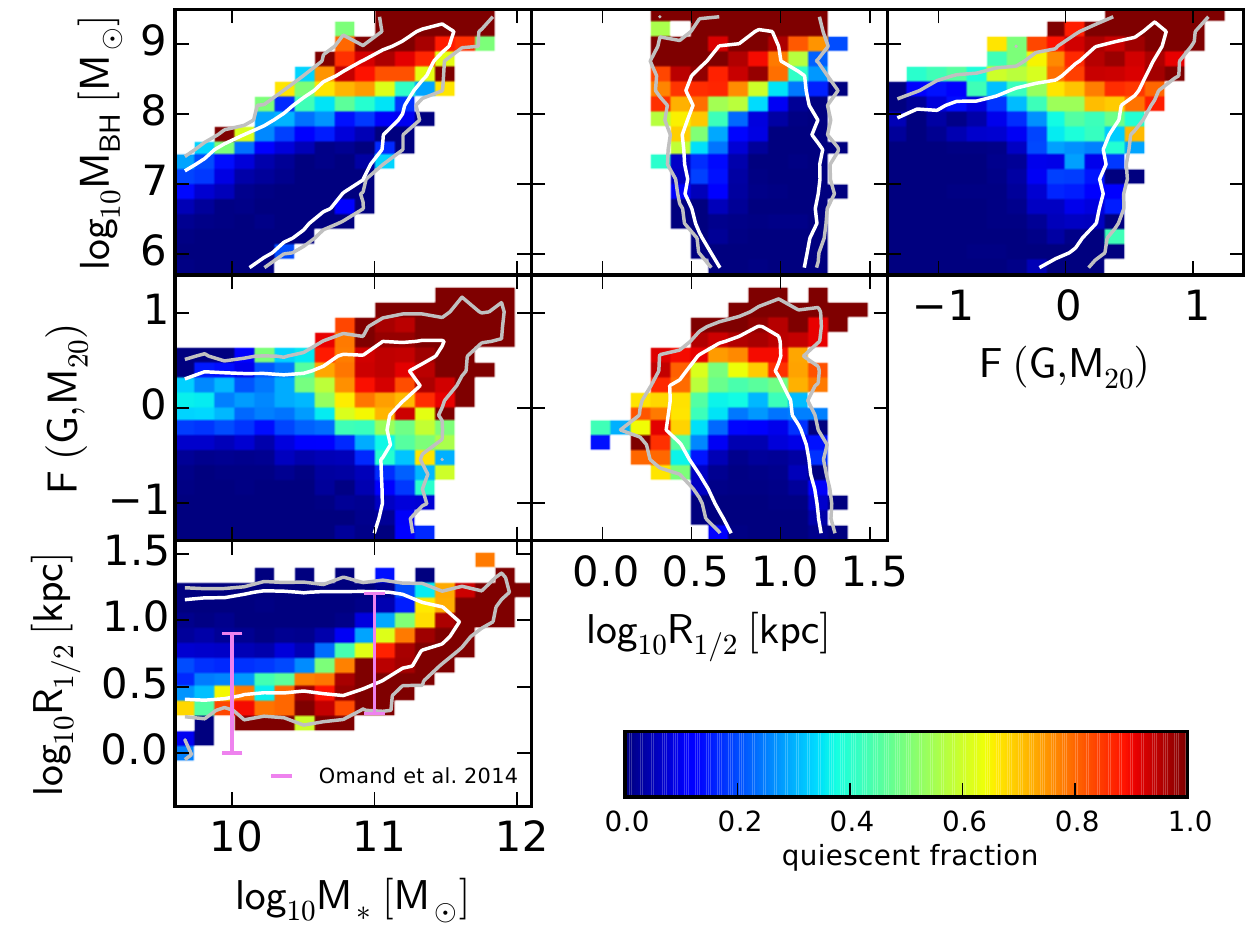}
\caption{Simulated relationships between size, mass, morphology, and star
  formation.  Sizes and morphologies are as measured from the unattenuated rest-frame $g$ band images.  White (gray) contours encircle bins with 50 (10) or more simulated
  points, and the rectangle colours reflect the fraction of quiescent galaxies in
  each bin.  In the lower left panel, we plot half-light radius versus stellar
  mass, showing that the simulated distribution is similar to the observed one
  at $M_* \sim 10^{11} M_{\odot}$ but too large by a factor $\sim 2$ at
  $10^{10} M_{\odot}$.  \illustris\ recovers the observed trend
  \citep[e.g.,][]{Omand2014} that the distribution of quiescence is tilted with respect
  to the $M_*$ axis.  This means at fixed $M_*$, smaller galaxies are more
  likely to be quenched, known to be true for both nearby
  \citep{Kauffmann2003} and distant \citep{Barro2013} galaxies.  The middle
  row reflects that morphology is also an effective predictor of quenching,
  both at fixed $M_*$ (left) and also at fixed $R_{1/2}$ (right).  The top
  panel plots these same variables against SMBH mass. We find that objects residing on or above the
   mean $M_{\rm BH}$-$M_*$ relation are slightly more likely to be quenched (top left), as expected and found by \citet{Sijacki2015}. \label{fig:contours}}
\end{center}
\end{figure}

Figure~\ref{fig:contours} presents several correlations among galaxy properties.  In addition to raw correlations, we over-plot contours indicating the ``quiescent fraction'' of galaxies in each of the 2-dimensional bins, in order to give some idea of the relationship between these quantities and galaxy star formation (i.e., Section \ref{ss:sfr}).  We define quiescent fraction as the fraction of galaxies satisfying:
\begin{equation} \label{eq:passive}
 \log_{10} \mathrm{ SSFR/(yr^{-1}) } < -0.24 (\log_{10} M_*/M_{\odot}) - 8.50,
 \end{equation}
 as used also in Section~\ref{ss:rot_vs_sfr} and following \citet{Omand2014}.  This corresponds to $\rm\ SSFR\sim 10^{-11} yr^{-1}$ in galaxies with $M_* \sim 10^{10} M_{\odot}$.  All sizes and morphologies presented in this section are measured in the rest-frame SDSS-$g$ band images.

\subsection{Size-mass-morphology} \label{ss:size}

We show how size, mass, and quenching are related in the bottom left panel of Figure~\ref{fig:contours}.  Here we make a rough comparison to the distribution of SDSS galaxies presented by \citet{Omand2014}.  Genel et al. (in prep.) will study the size distributions in more detail.  We use sizes measured from the rest-frame $g$ band images, as described in Section~\ref{ss:morphology}.  There are thus four sources with morphology and size measurements for each simulated galaxy, one each for the four default random viewing angles.  

The \illustris\ size-mass correlation is shallower than the observed one for $M_*/M_{\odot} > 10^{9.5}$.  At $M_*/M_{\odot} \sim 10^{11}$, the observed and simulated distributions are very similar.  However, observed galaxies shrink more quickly as $M_*$ decreases than do \illustris\ simulation galaxies.  The lower end of the size distribution at $R_{1/2} \sim 1$ kpc may be affected by the relatively coarse spatial resolution ($\gtrsim 0.7$ kpc) of the simulated galaxies.  Indeed it may be impossible for \illustris\ galaxies to reflect a size smaller than a few times this resolution scale.  However, this effect alone would not necessarily make the largest simulated galaxies significantly larger.  Thus we conclude that the upper end of the simulated galaxy $R_{1/2}$ values are a factor of $\sim 2$ larger than observed at $M_*/M_{\odot} \sim 10^{10}$.  

Although \illustris\ has fewer quenched/compact galaxies (or larger galaxies on average) at $M_* \lesssim 10^{10.5} M_{\odot}$, the diagonal shape of the quenching contours in the $M_*$--$R_{1/2}$ plane is nearly a perfect match to the results of \citet{Omand2014}, where quenching is most tightly correlated with $M_*/R_{1/2}^{\alpha}$ with $\alpha \sim 1$-$2$.  As we showed also in Section~\ref{ss:sfr}, this implies that quenching tends to trace the compactness of stellar density profiles in roughly the correct manner in the \illustris\ simulation.  

The left middle panel of Figure~\ref{fig:contours} shows \gmtwenty\ versus mass, also discussed in Section~\ref{ss:mass}.  The \illustris\ simulated galaxies reflect the well known rough correlation of automated morphology with stellar mass or luminosity \citep[e.g.,][]{Kauffmann2003,strateva01_gv}, with morphologically early type ``red sequence'' at $M_*/M_{\odot} \gtrsim 10^{11}$.  As in the $R_{1/2}-M_*$ panel, horizontal or diagonal contours in the \gmtwenty$-M_*$ panel imply that morphology is an informative predictor of star formation at fixed $M_*$.  In fact, the central panel of Figure~\ref{fig:contours} implies that a galaxy's morphology is also a better predictor of quenching in \illustris\ than its radius alone.  At fixed $R_{1/2}$, average star formation activity is inversely proportional to \fgmtwenty.  

\subsection{SMBH-morphology}  \label{ss:smbh}

Since the AGN feedback model suppresses the low-redshift SFR of galaxies, we might expect quantities correlated with the SMBH output energy (for example, their mass) to be also correlated with star formation activity or quenching.  Indeed, \citet{Sijacki2015} demonstrated that at a given $M_*$, galaxy colour (SFR) is directly (inversely) proportional to $M_{\rm BH}$. In the top row of Figure~\ref{fig:contours}, we show the predicted relationship between  SMBH mass ($M_{\rm BH}$) and optical morphology. 

In the top left panel, we recover the same result as \citet{Sijacki2015}: at fixed $M_*$, galaxies with higher $M_{\rm BH}$ are more likely to be quiescent.  There is also a hint that at fixed $M_{\rm BH}$, lower $M_*$ implies more quiescence, a feature that we explore in more detail in Section~\ref{ss:halo}.  Similarly, at fixed $R_{1/2}$, and at fixed \fgmtwenty$ \gtrsim 0$, galaxies with higher $M_{\rm BH}$ are more likely to be quiescent (top center and right).  However, at intermediate morphology ($F \sim 0$), and at any $R_{1/2}$, at fixed $M_{\rm BH}$, higher $F$ or lower $R_{1/2}$ implies more quiescence, on average.

\begin{figure}
\begin{center}
\includegraphics[width=3.0in]{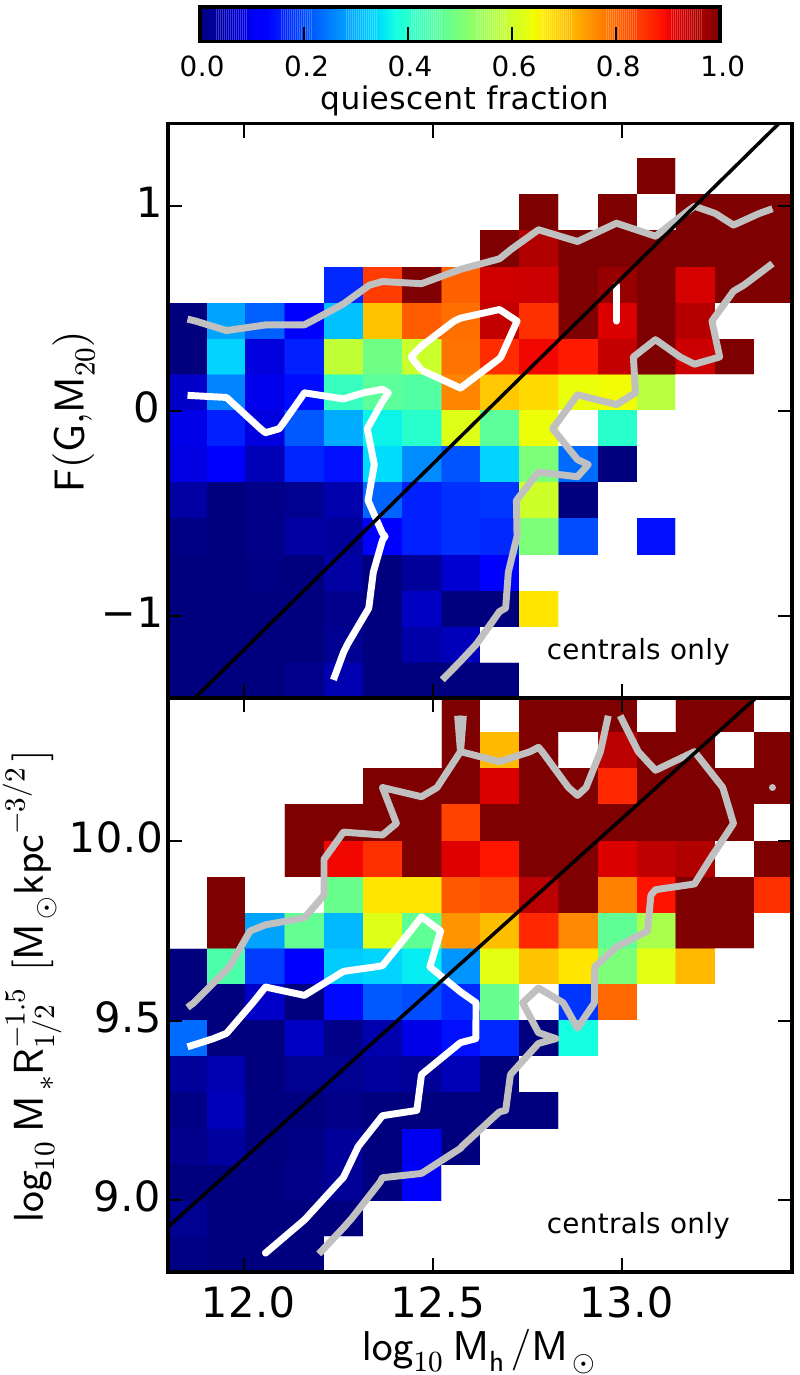}
\caption{Quiescent fraction and optical morphology versus total mass $M_h$ for massive galaxies in \illustris, with colours representing the fraction of quiescent galaxies in each bin, as defined in equation (\ref{eq:passive}).  {At a given $M_h$, the quiescent fraction increases (bins become red) as $F$ or $M_* R_{1/2}^{-1.5}$ increases.  \illustris\ galaxies at fixed $M_h$ are much more likely to be quenched if they have a dominant central bulge component, as inferred in real galaxies \citep[e.g.,][]{Woo2015}. } \label{fig:morphology_mhalo}}
\end{center}
\end{figure}

\subsection{Halo mass-morphology} \label{ss:halo}

From single trends alone, it is difficult to assess directly the impact of astrophysical processes on the resulting galaxy structures. In part, this is because these processes alter the structural quantities themselves, and so the baryons in a given galaxy may have a very different history under different model assumptions.  However, a galaxy's total mass is likely far less sensitive to this issue.  

Therefore, investigating how closely quenching and morphology depend on halo mass in \illustris\ may be a productive avenue for communicating its testable predictions.  For example, if the mechanism most responsible for setting galaxy morphology depends purely on DM halo mass ($M_h$), then we might expect there to be no residual correlations among important galaxy properties at fixed $M_h$. In contrast, some models assume that baryonic processes such as feedback regulate galaxy evolution, while their implementation may also depend on the galaxy's halo properties \citep[e.g.,][]{Vogelsberger2014a}.  Therefore it is uncertain how halo mass might relate to galaxy morphology in a simulation like \illustris.  Even assuming that all comprehensive models matching global stellar mass quantities also reproduce basic morphology trends (Section~\ref{s:morphologies}), galaxy properties at fixed halo mass might be an effective means by which to choose among them.

In Figure~\ref{fig:morphology_mhalo}, we plot \fgmtwenty\ and compactness ($M_* R_{1/2}^{-1.5}$; Section~\ref{ss:sfr}) against total mass $M_h$ and quiescent fraction, as defined in equation (\ref{eq:passive}).  For a given galaxy, we define $M_h$ as the total mass of all particles (all types) and cells bound to the subhalo, not including mass associated with subhalos of this subhalo.  Qualitatively, \illustris\ galaxy morphologies depend on $M_h$ in a very similar way to observed low-redshift galaxies \citep[e.g.,][]{Woo2015}: at fixed $M_h$, optical structure is very strongly correlated with whether a galaxy is quenched or not.  However, it is difficult to immediately assign causality.

\begin{figure}
\begin{center}
\includegraphics[width=3.0in]{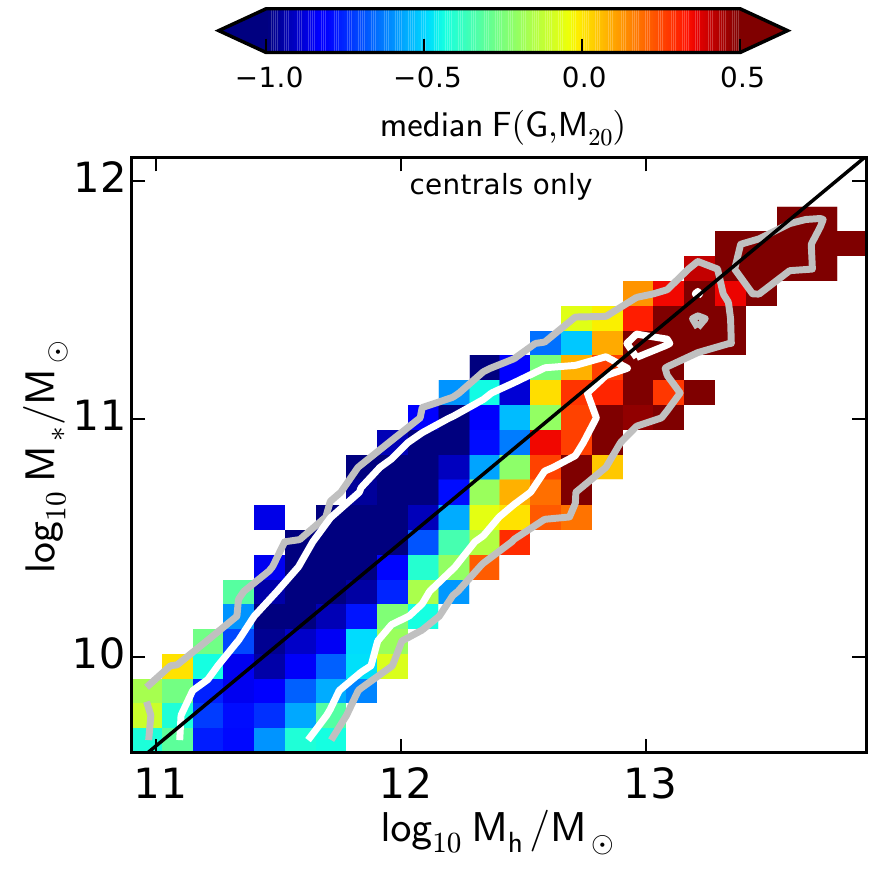}\\
\includegraphics[width=3.0in]{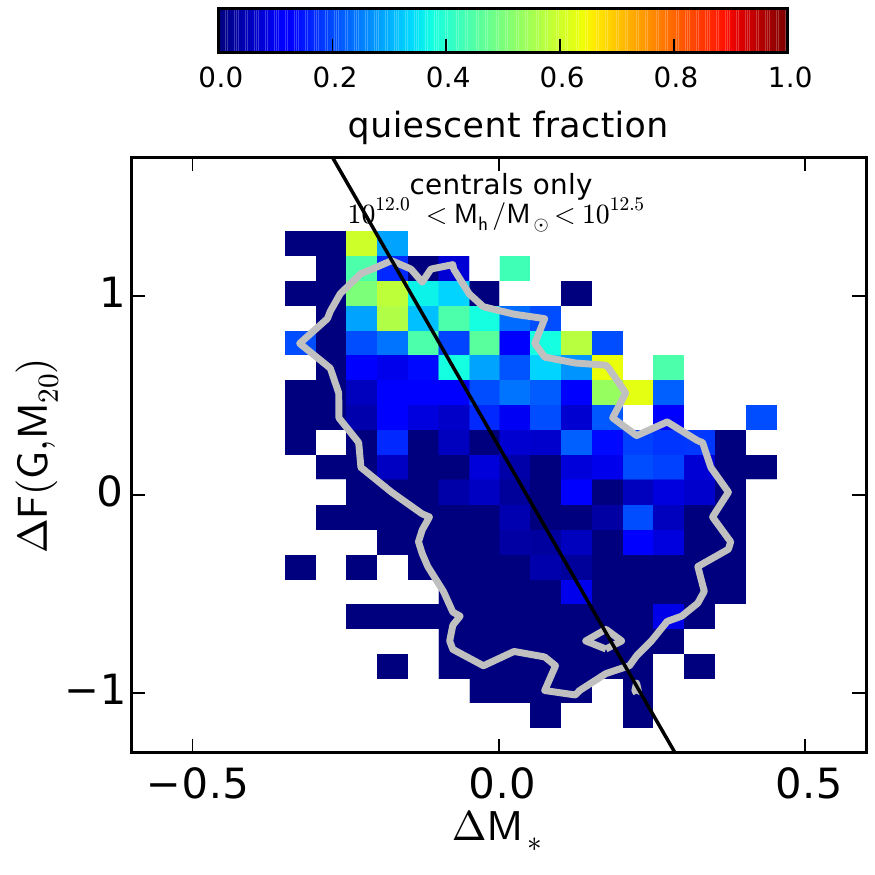}
\caption{Morphology versus stellar mass at fixed halo mass for central galaxies in \illustris. Top: $M_*$-$M_{h}$ relation, with bin colours representing optical morphology, \fgmtwenty.  Bottom: Residual correlation between \fgmtwenty\ and $M_*$, holding $M_h$ fixed; see text for definitions. In this panel we colour the bins by quiescent fraction.  Black lines are orthogonal distance regression \citep{Boggs1990} fits to the raw data.   We find a statistically significant ($\log_{10} p \ll -3$) negative residual correlation between $F$ and $M_*$ at a given $M_h$, implying that galaxies with higher stellar mass have later morphological type, on average. We interpret this as a consequence of the \illustris\ physics model suppressing galaxy star formation: in a given halo, less quenching leads to later type galaxies and more total stars. More quenching leads to earlier types and fewer total stars. \label{fig:quenching_implies_morphology}}
\end{center}
\end{figure}

Over time, the existence of quenching implies that a galaxy will form fewer stars than it would have without the quenching mechanism(s).  This implies that in a given halo, quantities which trace the \emph{integrated} star formation history (SFH) could provide a direct constraint on the key physical processes.  For example, Figure~\ref{fig:quenching_implies_morphology} shows how morphology depends on $M_*$ in simulated massive central galaxies at fixed $M_h$. The bin colours as a function of $M_h$ and $M_*$ (top panel) indicate that at $M_h \sim 10^{12} M_{\odot}$, galaxies with lower $M_*$ have higher \fgmtwenty, on average.  In other words, galaxies with higher $M_*$ at fixed $M_h$ are more disc-dominated.  \citet{Pillepich2014} also discussed this idea, in which these more massive disc-dominated galaxies also have shallower stellar halo mass profiles than their less massive, bulge-dominated counterparts at fixed $M_h$.  

The bottom panel of Figure~\ref{fig:quenching_implies_morphology} shows explicitly the residual correlation between $F$ and $M_*$ at fixed $M_h$.  To define this relation, we fit the $F$ versus $\log_{10} M_h$ and $\log_{10} M_*$ versus $\log_{10} M_h$ correlations with orthogonal linear regression \citep{Boggs1990} and subtract the predicted from actual values: $\Delta F$ = $F - F |_{M_h}$, $\Delta M_*$ = $\log_{10} M_* - \log_{10} M_* |_{M_h}$. With these, we can understand the internal correlations among galaxy properties and determine which, if any, predict galaxy morphology better than others \citep[e.g.,][]{hopkinsfp_07b}.  For the fitting procedure, we assume uniform measurement uncertainties of $0.1$ dex for masses and $0.2$ for $F$, so that the uncertainties are a similar fraction of the intrinsic scatter in each quantity.  For clarity, we show $\Delta F$ versus $\Delta M_*$ only for galaxies in the halo mass range $12.0 < \log_{10} M_h/M_{\odot} < 12.5$, which is a range during which the simulated galaxy population transitions from mostly star-forming to largely quenched, and therefore we might expect the most obvious consequences. 

We find a negative residual correlation between $M_*$ and $F$ at fixed $M_h$ in $10^{12} < M_h/M_{\odot} < 10^{12.5}$.  According to t-tests, the null hypothesis (no correlation) is overwhelmingly unlikely to yield these data.  We also split the sample into star-forming and passive subsamples, at $\log_{10}($SFR/$M_*) = -2.0$, and find in each a negative correlation at high significance.

Therefore, in halos of the same total mass, simulated galaxies with above-average stellar mass are more disc-dominated than average, and galaxies with below-average stellar mass are more bulge-dominated than average. The reason could be simple: stars typically form in a disc configuration, and therefore if there is more total star formation in a particular galaxy, there will be both more stars and their morphology will be more disc-like, on average.  

The colours in the bottom panel of Figure~\ref{fig:quenching_implies_morphology} indicate quiescent fraction, which tends to be higher in galaxies that are more bulge-dominated than average (higher $\Delta F$).  In contrast, the average current SFR/$M_*$ or quiescent fraction is roughly constant as a function of $\Delta M_*$.  This was also seen in the red fractions of massive simulated galaxies by \citet{Vogelsberger2014a} and quenched fractions in SDSS by \citet{Woo2012}. Therefore, galaxy morphology may be a clearer diagnostic of the time-integrated effect of feedback in \illustris\ than current SFR.

A candidate for the \emph{cause} of this effect is the radio-mode AGN feedback, which is the mechanism most responsible for reducing star formation at late times in massive \illustris\ galaxies \citep{Vogelsberger2013}.  If so, we might expect galaxy properties to have residual correlations with the simulated SMBH mass at fixed $M_h$.  Indeed, \citet{Sijacki2015} found that galaxies with over-massive SMBHs are redder than average, which we showed also in Figure~\ref{fig:contours} with respect to $M_*$.  

\begin{figure}
\begin{center}
\includegraphics[width=3.0in]{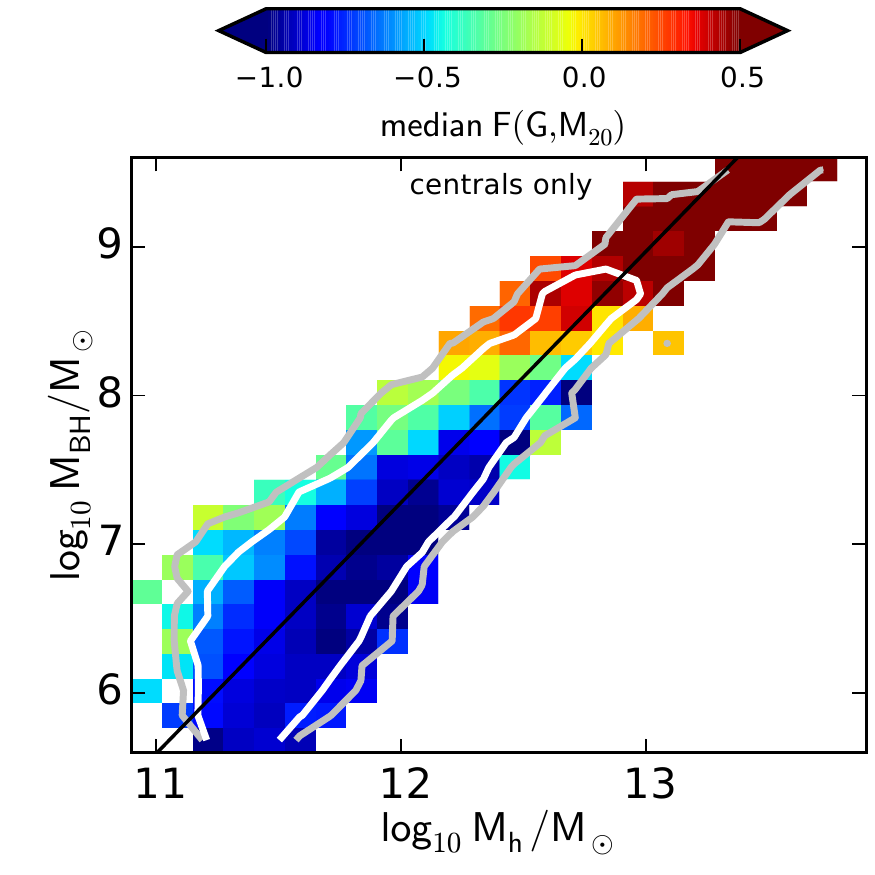}\\
\includegraphics[width=3.0in]{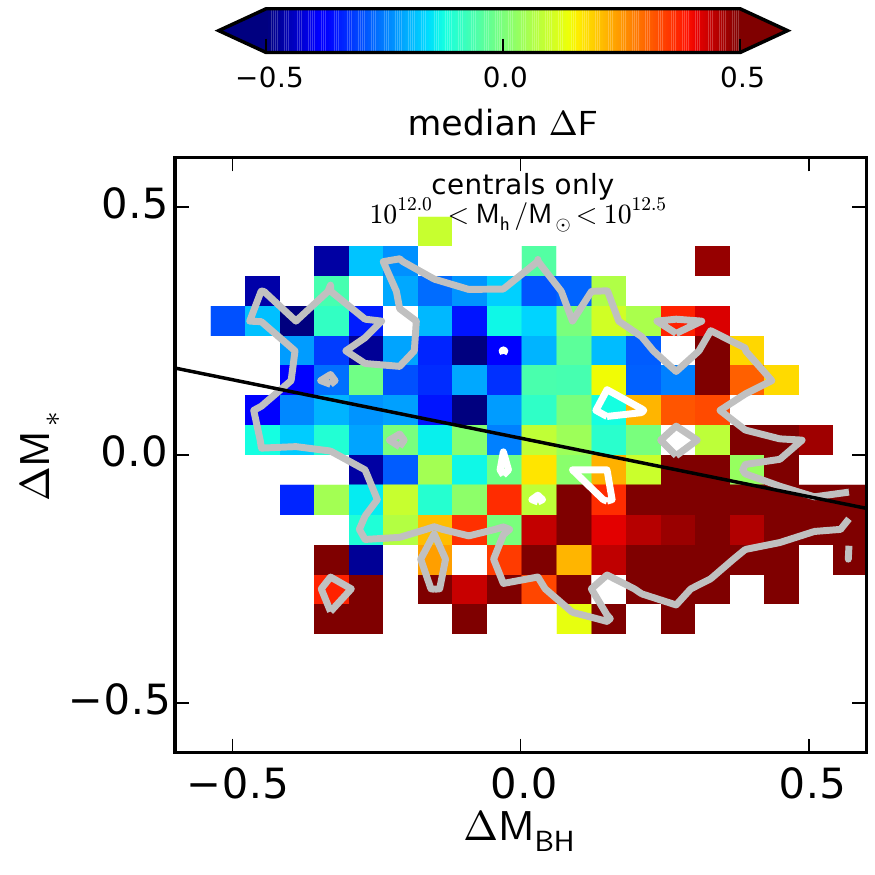}
\caption{Stellar mass versus black hole mass at fixed halo mass in \illustris, following same procedure as Figure~\ref{fig:quenching_implies_morphology}. Top: $M_{\rm BH}$-$M_{h}$ relation. Bottom: Residuals of $M_*$ and $M_{\rm BH}$ at fixed $M_h$. Bin colours represent optical morphology, \fgmtwenty.  We find a statistically significant ($\log_{10}\rm p \ll -3$) negative correlation between $\Delta M_*$ and $\Delta M_{\rm BH}$, implying galaxies with higher SMBH mass have lower stellar mass.  Again, we can interpret this as a consequence of the \illustris\ physics model: In a given halo, higher SMBH mass implies more total energy deposited by feedback and a greater reduction in the total stellar mass formed.  \label{fig:smbh_implies_quenching}}
\end{center}
\end{figure}

In Figure~\ref{fig:smbh_implies_quenching}, we present the $M_{\rm BH}$-$M_h$-morphology relation in \illustris. The top panel shows that above-average $M_{\rm BH}$ implies higher median $F$, as expected from any reasonable SMBH-bulge correlation.  Since $F$ and $M_*$ are negatively correlated at fixed $M_h$ (Figure~\ref{fig:quenching_implies_morphology}), it is likely that these same galaxies with above-average $M_{\rm BH}$ have below-average $M_*$.  The bottom panel of Figure~\ref{fig:smbh_implies_quenching} confirms this: at fixed $M_h$, simulated galaxies with above-average $M_{\rm BH}$ have below-average $M_*$. In addition, above-average $M_{\rm BH}$ implies earlier-than-average morphological type (redder bins).

If true, then the effect of AGN feedback on galaxy morphologies at fixed $M_h$ is indirect.  It results from the two facts that 1) AGN feedback quenches star formation over long periods, and 2) star formation over an extended period leads to disc-dominated morphologies.  Differences in evolution of baryons in similar-mass halos would lead to a varying integrated feedback efficiency.  This would have an effect in the sense described above for the $M_*$-$M_h$-morphology relation.  Since morphology is so tightly linked with the extended star formation history (SFH), other physical processes act less signficantly on morphology, and the overall SFH potentially masks these effects.  For example, a merger may cause a late-type galaxy to temporarily evolve into an early-type.  However, if quenching isn't sufficient owing to an undermassive SMBH, then this galaxy might continue forming stars and re-grow a disc over a few Gyr \citep[c.f.,][]{Snyder2015a}, leading to a higher $M_*$ and more disc-dominated morphology at late times.

\subsection{$M_*$ and quiescent fraction}

\begin{figure}
\begin{center}
\includegraphics[width=3.0in]{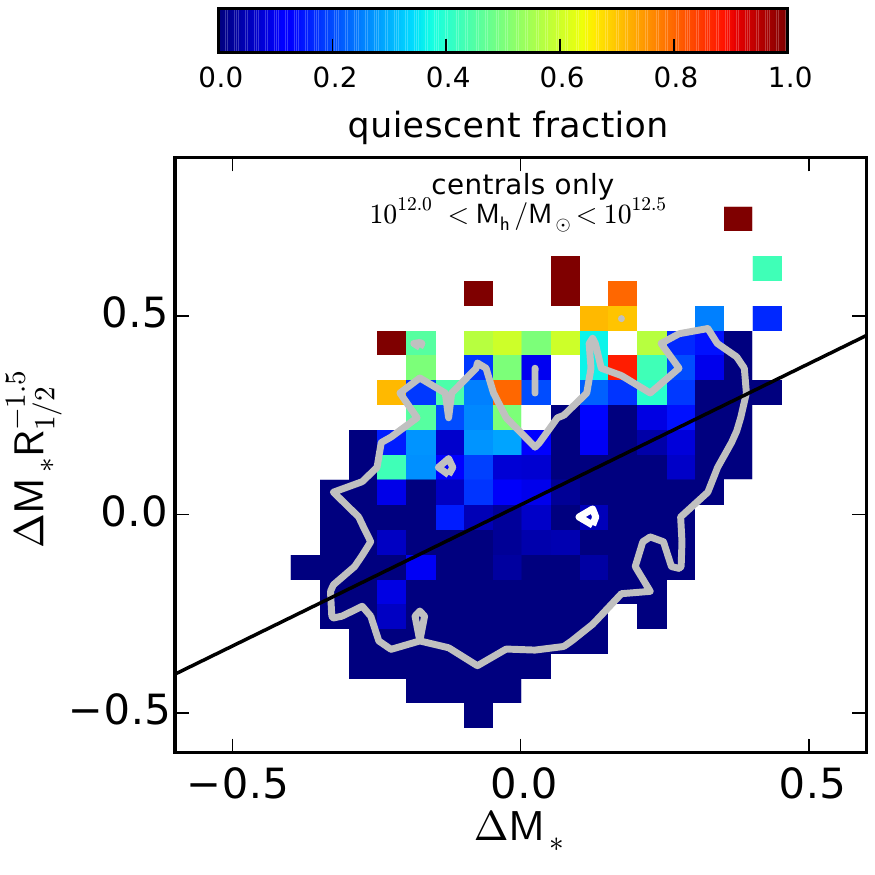}
\caption{ Residuals of compactness and $M_*$ at fixed halo mass, with bin colours representing quiescent fraction.  $\Delta M_* R_{1/2}^{-1.5}$ correlates positively with $\Delta M_*$, in direct contrast to the behavior of $\Delta F$ in Figure~\ref{fig:quenching_implies_morphology}.   \label{fig:compactness_resids}}
\end{center}
\end{figure}

If we assume that quenching determines $M_*$ at fixed $M_h$ in \illustris\ in the sense described in Section~\ref{ss:halo}, one consequence is that we should be cautious with quantities depending directly on $M_*$.  For example, the compactness quantity we used above \emph{correlates positively} with $M_*$ at fixed $M_h$, in contrast with \fgmtwenty\ and $M_{\rm BH}$; see Figure~\ref{fig:compactness_resids}. Thus, it is possible that using quantities directly proportional to $M_*$ can obscure some quenching signatures we hope to observe. 

Both $\Delta M_* R_{1/2}^{-1.5}$ (Figure~\ref{fig:compactness_resids}) and $\Delta F$ (Figure~\ref{fig:quenching_implies_morphology}) correlate with quiescent fraction, while $\Delta M_*$ does not.  Thus, compactness so defined can be thought of as tracing well the average current SFR/$M_*$, but perhaps less well the integrated effects of quenching on $M_*$ over time.  Another way to say this is that a galaxy's $M_*$-independent structure may provide additional insight into the long-term effects of feedback processes in massive galaxies.  

\section{Discussion} \label{s:discussion}

Analyzing synthetic data from simulations alongside galaxy surveys promises to be a valuable approach to constrain the physics of galaxy evolution.  Powerful pipelines have been developed to convert theoretical models of galaxy formation into useful mock surveys, such as the Millennium Run Database and Observatory \citep{Lemson2006,Overzier2012}, the Theoretical Astrophysical Observatory \citep{Bernyk2014}, and the simple semi-empirical approach by \citet{Taghizadeh-Popp2015}.  Moreover, such models now include treatments motivated by high-resolution simulations for the morphological transformations that may be a key aspect of galaxy evolution \citep[e.g.,][]{Porter2014,Brennan2015}.  Now, owing to advances in our ability to evolve hydrodynamics coupled to sub-grid galaxy formation physics, multiple groups have simulated ever-larger volumes of a $\Lambda$CDM universe with high enough spatial resolution to predict the internal structures of galaxies to varying degrees \citep[e.g.,][]{Munshi2013,Ceverino2014,Christensen2014,Dubois2014,Cen2014,Khandai2015,Vogelsberger2014a,Schaye2014}.  

An obvious way forward is to begin to combine synthetic observatory pipelines with cosmological hydrodynamical simulations, because such calculations follow directly the gravitational and physical processes which ultimately control galaxy morphologies.  By constraining our models with observed galaxy populations, we can make progress toward improving not only our understanding of galaxy physics, but also our ability to link these processes to observations.  

This requires two actions: creating realistic synthetic data and connecting observational diagnostics measured from such data to these physical processes. For example, this technique has been used to measure the galaxy merger rate at $z < 1$ by measuring how morphologies evolve in merger simulations \citep{lotz08,lotz10,Lotz2010}.  Recently, studies have used cosmological simulations to determine the impact of powerful feedback on clumpy high-redshift galaxies \citep{Mandelker2014,Moody2014}, and how galaxy structures might evolve \citep[e.g.,][]{scannapieco10,Pedrosa2014,Wellons2015}, including by direct analogy with \hst\ surveys of distant galaxies \citep{Snyder2015a}.

To begin expanding this effort to very large-scale hydrodynamical simulations, T15 converted the $z=0$ output of the \illustris\ Simulation \citep{Vogelsberger2014b,Genel2014} into synthetic optical images and spectra, a public dataset with which it is possible to address a number of open scientific questions.  Moreover, this observationally oriented approach is an effective way to compare multiple large-scale yet moderate-resolution simulation projects, for example the {\sc Eagle} simulation \citep{Schaye2014,Crain2015}, {\sc Horizon-AGN} \citep{Dubois2014}, MassiveBlack-II \citep{Khandai2015}, and others, as well as with future larger and higher-resolution calculations.  

In the present work, we measured a suite of non-parametric morphology statistics of galaxies in the \illustris\ Simulation at low redshift, and began to explore the extent to which these resemble real observations.  We have made several simplifying choices, including to separate galaxies from their neighbors (Section~\ref{ss:imagepipeline}) and to neglect for now the effect of dust on optical morphology.  Of course, the quantitative morphology of individual galaxies does depend on dust \citep{lotz08,Snyder2015a}, but primarily in cases such as starbursts which are rare in the local galaxy population. Moreover, we have tested that the optical morphology \emph{trends} we analyzed are, to first order, unchanged by including a treatment for dust.

\subsection{General optical morphologies}

In Section~\ref{s:morphologies}, we presented the optical morphology distributions of \illustris\ galaxies at $z=0$.  We found that simulated galaxies occupy roughly the observed \gmtwenty\ space, with correct correlations of morphology with stellar mass and SFR.  This is a noteworthy success: hydrodynamical simulations can now produce galaxy populations with reasonable morphology distributions.  

That morphologies occupy the space of SFR, $M_*$, galaxy density, and $R_{1/2}$ (Figures~\ref{fig:sfrmass} and \ref{fig:contours}) each in a realistic fashion was somewhat surprising.  The parameters of the \illustris\ physics models were tuned to match observed distributions in both $M_*$ and SFR (the axes of Figure~\ref{fig:sfrmass} panel a), but not to recover the median morphology as a function of these quantities.  Moreover, \citet{Snyder2015a} showed the dependence of \gmtwenty\ on $M_*$ and SFR to be moderately insensitive to the precise physics of feedback, using very different simulations. 

One explanation is that in a simulation such as \illustris, a galaxy's morphology and star formation rate are not independent: the rate and mode of the feedback-regulated gas supply \citep{Nelson2015} determines the subsequent conversion into stars.  If those two things are modeled reasonably accurately, then realistic galaxy structures are a natural consequence.  

However, another interpretation is to say that the distribution of galaxy morphology -- the coarse breakdown of bulges and discs -- does not add significantly more information to galaxy formation models beyond that provided by constraints on the $M_*$ and SFR distributions. Indeed, it seems that the success of \illustris\ with the broad features of \gmtwenty\ space could be a direct consequence of tuning the feedback models to reproduce the stellar mass function and stellar halo occupation function at $z=0$ and total SFR versus time.  

If true, then the regulation of star formation in a crudely correct manner, combined with jointly modeling baryons and dark matter to accurately follow collapse, cooling, and angular momentum transport, would be sufficient to reproduce the distribution of automated galaxy morphologies.  This hypothesis can be tested by implementing more realistic treatments for galaxy physics, by analyzing more detailed morphological and kinematic tracers, and by studying directly galaxy structures at fixed DM halo mass or other properties (Section~\ref{ss:halo}), where possible.  Even if we reach a point where these do not change drastically with further improvements to galaxy physics, it is clearly desirable to learn how the microphysics of star formation and feedback \citep[e.g.,][]{Hopkins2014} translate into the assumptions used by large-scale models \citep[e.g.,][]{Muratov2015}.

\subsection{Rotation versus optical structure}

In Section~\ref{s:rotation}, we showed how optical morphology statistics relate to the physical kinematics of massive galaxies in \illustris. Broadly, we measure bulge-dominated light profiles in galaxies lacking significant rotation, and a diversity of light profiles in rotating galaxies with $M_* > 10^{11} M_{\odot}$. Although quiescence is correlated tightly with bulge strength measured by \fgmtwenty (Equation~\ref{eq:fgm20}), it correlates less with kinematic measures of bulge strength and lack of rotation. One interpretation of this is that it simply follows recent thinking about galaxy structures triggered by kinematic surveys like \sauron\ and \atlas\ \citep[e.g.,][]{Emsellem2007,Cappellari2011a}: flat, rotating galaxies are closely related, and differ primarily by having differing levels of star formation and therefore appear to have stronger or weaker spiral arms \citep[e.g., as proposed and discussed by][]{VanDenBergh1976, Bender1994, Kormendy1996}.  Slowly rotating galaxies are rare, quenched, and occur at a similar rate relative to the total quenched fraction independent of environment (Section~\ref{ss:env}), as they do in e.g., \atlas.  Also, the relative populations of late-type galaxies and fast-rotating early type galaxies evolve smoothly as a function of environment in \illustris, similar to the results by \citet{Cappellari2011}.

As in real nearby galaxies, \illustris\ galaxies appear to have structures implying that meaningful intrinsic classification is more like a ``comb'' than a ``tuning fork'' \citep[e.g., Figure 1 by][]{Cappellari2011}.  Therefore, large-scale galaxy formation simulations like \illustris\ and {\sc Eagle} are capable of producing not only a roughly correct distribution of bulge- and disc-dominated galaxies, but also a roughly correct phenomenology of galaxy morphology and kinematics at $z=0$.  Moreover, \citet{Genel2015} found that the angular momentum content of \illustris\ galaxies depends on mass in roughly the same manner as observations and predictions of analytic models \citep[e.g.,][]{Romanowsky2012}.  Therefore, observational constraints which have been tested with idealized merger simulations \citep{cox06_kinematics,Moody2014a} or small samples of cosmological simulations \citep{Naab2014} can also now be tested with large-scale hydrodynamical simulations.  This will allow us to exploit not only the rough diversity of galaxy types, but also their detailed statistical distributions, in order to better constrain models of galaxy formation.  

\subsection{Detailed morphologies} \label{ss:disc_detailed}

Although we are finding that simulations can reproduce average kpc-scale morphologies, we are also beginning to find hints of new constraints available in higher-order morphology measurements and residual correlations.  

For instance, a significant number of \illustris\ galaxies appear to have a ring-like morphology.  One possibility is that the ring-like shapes result from the equation of state ISM pressurization in the \citet{springel03} model.  Another possibility is that the Illustris ISM and feedback model is leading to the formation of ring-like star formation patterns driven by the choice of density scale at which to either form stars or recouple the wind material to the galaxy material. In large galaxies this wind mass can recouple at a location internal to the galaxy, enhancing the ISM density and causing additional star formation there (a positive feedback effect similar to, e.g., \citealt{Dugan2014}). 

The fact that we can detect this ring-like feature in a statistical sample of simulated galaxies, both in young and old stars, implies that we may now be able to use it as a constraint for large-scale models of galaxy formation.  The number densities, masses, ages, SFRs, and sizes of ring-like galaxies may rule out or better constrain the numerical treatment of the ISM and feedback models like the one used in \illustris.  This could require treating dust faithfully in these images.

\subsection{Morphologies at intermediate and low mass} \label{ss:quenchingproblem}

There is a lack of quenching in \illustris\ galaxies with $M_* \lesssim 10^{10.5} M_{\odot}$, leading both to stellar mass functions that are too high and also quenched fractions that are too low \citep{Vogelsberger2014b,Genel2014}.  This appears also as an apparent shortfall in the numbers of quenched spheroids in the distribution of morphology versus $M_*$ and SFR at intermediate masses, such as we see at $M_* \sim 10^{10.5} M_{\odot}$ in Figure~\ref{fig:sfrmass} (Sections \ref{ss:sfr} and \ref{ss:mass}).  One obvious interpretation is that feedback is too weak in these systems.  However, the morphologies in low mass galaxies may also be affected by other factors, such as the inability to follow gas flows toward galaxy centers on very small scales, a mechanism which could cause the remnants to contract. Also, the artificial ISM pressurization used by \illustris\ and other simulations of comparable volume could prevent gas from being influenced by turbulence and collapse on very small physical scales $\ll 1$ kpc. This could prevent fragmentation and inward migration of gas during periods of ``violent disc instability'' (VDI), which could assist the formation of compact spheroids \citep[e.g.,][]{Dekel2009,Zolotov2015} in intermediate mass galaxies.  Also, as in \illustris, lower-mass halos ($M_h < 10^{11} M_{\odot}$) in both semi-analytic and large-scale hydrodynamical models overproduce stars at high redshift \citep{White2015}.

\subsection{Residual morphology correlations}

If multiple galaxy formation models can broadly reproduce the diversity of galaxy types, then we should appeal to residual or higher-order correlations in order to make new falsifiable predictions.  The \illustris\ galaxy physics model includes a component based on feedback from SMBHs, which imprints a residual dependence on morphology at fixed $M_*$ and $R_{1/2}$ (Section~\ref{ss:smbh}).  \citet{Sijacki2015} presented the prediction for this residual correlation with galaxy colours.

Similarly, this model imprints residual correlations in galaxy morphologies \emph{at fixed halo mass} \citep[see also][]{Pillepich2014}.  In Section~\ref{ss:halo}, we found that at fixed halo mass (for $12 < \log_{10} M_{h}/M_{\odot} < 12.5$), $M_*$ is inversely proportional to optical bulge strength as measured by \fgmtwenty\ (Equation~\ref{eq:fgm20}). In other words, in galaxies at fixed halo mass, more disc-dominated systems have greater stellar mass than more bulge-dominated ones.  Relatedly, at fixed halo mass, $M_*$ is inversely proportional to $M_{\rm BH}$, so that in galaxies with fixed halo mass, over-massive SMBHs imply less total stellar mass than in galaxies with under-massive SMBHs.  Necessarily then, $M_{\rm BH}$ and \fgmtwenty\ are positively correlated at fixed $M_h$.  

This appears to be a natural consequence of the physics of feedback assumed in \illustris.   In a given halo, more SMBH accretion ultimately deposits more energy into the galaxy's gas reservoir, preventing it from forming as many stars as it otherwise would. Since stars form most often in cold gas that has collapsed into a disc, this quenching signature is also imprinted in galaxy morphology: more quenching leads to fewer young stars and a fainter disc component, leading eventually to a lower $M_*$ and earlier type in galaxies of the same halo mass.  This reduction in $M_*$ at fixed $M_h$ may imply that we should contrast the behavior of $M_*$-based structural tracers (e.g., some measures of compactness) with $M_*$-independent ones like \gmtwenty, S\'{e}rsic index, and $C$, for example.

This hypothesis may be difficult to disentangle from alternative ones.  For example, what causes a SMBH to be over-massive for a given $M_h$ in the first place?  A common answer might invoke processes leading to the co-evolution of bulges and SMBHs, in which case galaxy morphology variations might precede quenching variations at fixed $M_h$.  To find out what \illustris\ predicts for this question, it will be important to map how individual simulated galaxies evolve over time in distributions of these important properties.

\section{Summary and Conclusions} \label{s:conclusions}

We converted $z=0$ galaxies formed in the \illustris\ simulation into synthetic images and measured their optical morphology using automated tools.  From idealized images, we created $42531$ SDSS-like sources for $10808$ simulated galaxies (up to $4$ angles per galaxy) with $10^{9.7} M_{\odot} < M_* < 10^{12.3} M_{\odot} $, in each of rest-frame $u$, $g$, $i$, and $H$ filters.  From these synthetic data, we measured non-parametric diagnostics of galaxy morphology, catalogs of which we present in Appendix~\ref{a:morphs}.  In equation~(\ref{eq:fgm20}), we defined a non-parametric bulge-strength parameter, \fgmtwenty, which is easy to measure and relatively robust against the effects of dust and disturbances.  

Our main conclusions about the morphology of galaxies formed by the \illustris\ Simulation at $z=0$ are:
\begin{enumerate}
\item{ The distribution of synthetically observed morphology agrees well with the observed one at $z \sim 0$.  Crucially, this includes a substantial population of quenched bulge-dominated galaxies having $G \approx 0.6$ and $M_{20} \approx -2.5$, and a locus of disc-dominated galaxies with realistic concentrations and pixel light distributions.  \label{it:1} }
\item{ At $M_* \sim 10^{10} M_{\odot}$, we find that \illustris\ galaxy
  structures are a combination of spirals and composite systems with $G < 0.5$
  and $M_{20} > -2$.  As $M_*$ increases above $10^{11} M_{\odot}$, the
  primary locus shifts smoothly toward the region occupied by ellipticals. }
\item{ At $10^{10.5} \lesssim M_*/M_{\odot} \lesssim 10^{11}$, simulated \gmtwenty\ values are bimodal, with a separate peak at $G \sim 0.4$ and $M_{20} > -1$.  The extreme location of this second peak is a consequence of a large number of galaxies having a ``ring-like'' morphology. A detailed comparison of the population of such galaxies in \illustris\ with observed samples may be a new constraint on models of large-scale galaxy formation.  }
\item{As a function of $M_*$ and SFR, the simulated optical morphologies follow trends observed in galaxy surveys:  at fixed $M_*$, lower-SFR galaxies have earlier structural types, and at fixed SFR, higher-mass galaxies have earlier structural types.}
\item{Optically defined morphologies correlate with kinematically defined ones, especially in simulated galaxies with $M_* > 10^{11} M_{\odot}$.  Some kinematically defined tracers are less correlated with quiescence than is \gmtwenty. }
\item{Simulated optical morphologies trace the same rotation-SFR-environment relation as found by integral field spectroscopy surveys such as \atlas: slowly rotating, quenched ellipticals are rare and occur as a constant share of total quenched galaxies. Rotating galaxies are diverse, some with high SFR and spiral arms while others are ``anemic spirals'': flat, rotating, but quenched. }
\item{At $M_* \sim 10^{11} M_{\odot}$, the $g$-band size distribution of \illustris\ Simulation galaxies roughly matches the observed one.  At $M_* \sim 10^{10} M_{\odot}$, the \illustris\ sizes are a factor of $\sim 2$ larger than observed.}
{\item{At fixed halo mass $M_h$ and at fixed relative rotation, quiescence correlates strongly with optical bulge strength.}}
\item{At fixed intermediate halo mass ($M_h \sim 10^{12} M_{\odot}$), $M_*$ correlates inversely with optical bulge strength \fgmtwenty\ and SMBH mass, which we believe results from the \illustris\ AGN feedback model.  In other words, AGN quenching acts to reduce the total stellar mass below what would have otherwise formed, with a side effect of reducing the extent to which the light profile is disc-dominated.  }

\end{enumerate}

The \illustris\ simulation is a large-scale calculation of galaxy formation with physics model parameters set to approximately reproduce selected global statistics.  These include the galaxy stellar mass function, the stellar mass-halo mass relation, and the evolution of total star formation density.  The field has made significant efforts to overcome numerical challenges that had previously limited the ability of cosmological simulations to form and maintain a diverse population of star-forming disks and quiescent spheroids.  

We have shown that these ingredients lead to a diverse distribution of galaxy structures that includes both spheroids and disks at $z=0$.  The non-parametric morphologies of our simulated galaxies agree well with observed populations.  Certainly the balance of these populations is imperfect in \illustris, since, for example, quenching is not as complete as observed in galaxies with $M_* \sim 10^{10} M_{\odot}$.  This leads in part to the simulated galaxies having larger radii, a less-clear structural bimodality, and more extended disks than observed. However, we find an approximately correct \emph{correlation} between star formation and galaxy morphology, which we believe merits optimism. It implies that the ingredients above, plus physics improvements and/or higher spatial resolution, could be sufficient to reproduce a more fully realistic galaxy population.  In other words, we hypothesize that for a model which properly regulates a galaxy's star formation, and which mitigates possible impacts on second-order structural parameters (see Section~\ref{ss:mass}), structure formation plus galaxy physics leads naturally to the distribution of galaxy morphologies.  {Proving this hypothesis will require comparing the several recent large-volume hydrodynamical simulations, performing a parameter study of large-volume simulations with different feedback implementations, or both.}

We have also presented our methods for modeling synthetic observations of galaxies.  This includes the release of Python modules for the manipulation of idealized synthetic images (in our case, output of the \sunrise\ code) into observational analogues, which can then be analyzed with automated tools and by visual classifiers.  

We expect to use these methods to enable additional science with galaxy surveys.  With the emergence of statistically relevant cosmological simulations that directly include much of the required galaxy physics, we can, for instance, measure the predicted observability time of key galaxy formation phenomena, and improve the robustness of the necessary galaxy classification schemes.  Moreover, T15 demonstrated several projects based on detailed simulated images, including how shells observed in massive spheroids might reflect their formation histories, how SED fitting routines can be tested against simulated SED maps, and generally how the spectra of the galaxy population could be used to further constrain galaxy formation physics.  Such calculations will improve measurements of the galaxy-galaxy merger rate from disturbed morphologies, the formation of disks and bulges, the impact of feedback, and the role of star-forming clumps.  


\section*{Acknowledgements}

For discussions that contributed to this paper, we thank Chris Hayward, Ann Zabludoff, Joel Primack, Michael Peth, S.\ Alireza Mortazavi, Rachel Somerville, Roger Davies, David Spergel, Susan Kassin, and Harry Ferguson.  We thank the anonymous referee for a constructive and productive review.   We thank Patrik Jonsson for writing, updating, and supporting the \sunrise\ code, which we used extensively in this work. LH acknowledges support from NASA grant NNX12AC67G and NSF grant AST-1312095. VS acknowledges support by the European Research Council through ERC-StG grant EXAGAL-308037.  GS and JL appreciate support from \hst\ grant numbers HST-AR-$12856.01$-A and HST-AR-$13887.004$-A. SG acknowledges support provided by NASA through Hubble Fellowship grant HST-HF2-51341.001-A.  Support for the Hubble Fellowship and \hst\ programs \#12856 (PI J.\ Lotz) and \#13887 (PI G.\ Snyder) was provided by NASA through a grant from the Space Telescope Science Institute, which is operated by the Association of Universities for Research in Astronomy, Inc., under NASA contract NAS 5-26555. The computations in this paper were run on the Odyssey cluster supported by the FAS Division of Science, Research Computing Group at Harvard University.  This research has made use of NASA's Astrophysics Data System, the Flexible Image Transport System \citep[FITS;][]{Wells1981,Hanisch2001} standard, the PyFITS Python module, the SciPy library \citep{SCIPY}, and AstroPy \citep{Robitaille2013}.  Figures in this paper were constructed with the Matplotlib Python module \citep{Hunter:2007}.  

We used SDSS data as backgrounds for one form of synthetic images.  Funding for SDSS-III has been provided by the Alfred P. Sloan Foundation, the Participating Institutions, the National Science Foundation, and the U.S. Department of Energy Office of Science. The SDSS-III web site is www.sdss3.org/.  SDSS-III is managed by the Astrophysical Research Consortium for the Participating Institutions of the SDSS-III Collaboration including the University of Arizona, the Brazilian Participation Group, Brookhaven National Laboratory, Carnegie Mellon University, University of Florida, the French Participation Group, the German Participation Group, Harvard University, the Instituto de Astrofisica de Canarias, the Michigan State/Notre Dame/JINA Participation Group, Johns Hopkins University, Lawrence Berkeley National Laboratory, Max Planck Institute for Astrophysics, Max Planck Institute for Extraterrestrial Physics, New Mexico State University, New York University, Ohio State University, Pennsylvania State University, University of Portsmouth, Princeton University, the Spanish Participation Group, University of Tokyo, University of Utah, Vanderbilt University, University of Virginia, University of Washington, and Yale University. 

\appendix

\section{Light Assignment Tests} \label{a:starsizes}

{ Here we show how our method of assigning sizes to star particles affects the morphology measurements studied in this paper.  In \sunrise, star particles emit light with a distribution approximating a simple SPH kernel with smoothing length $r_*$ \citep{jonsson06}.  Our fiducial method assigns $r_*$ equal to the radius which encloses a star particle's 16 nearest star particle neighbors (T15), which we denote $r_*^{16}$.  The most common alternative method \citep[e.g.,][]{scannapieco10,bush10,snyder11a} assigns star particles a constant radius equal to the gravitational softening length for stars; in \illustris\ at $z=0$, $r_*^{\prime} = 0.710$ kpc.  }

\begin{figure*}
\begin{center}
\includegraphics[width=7.0in]{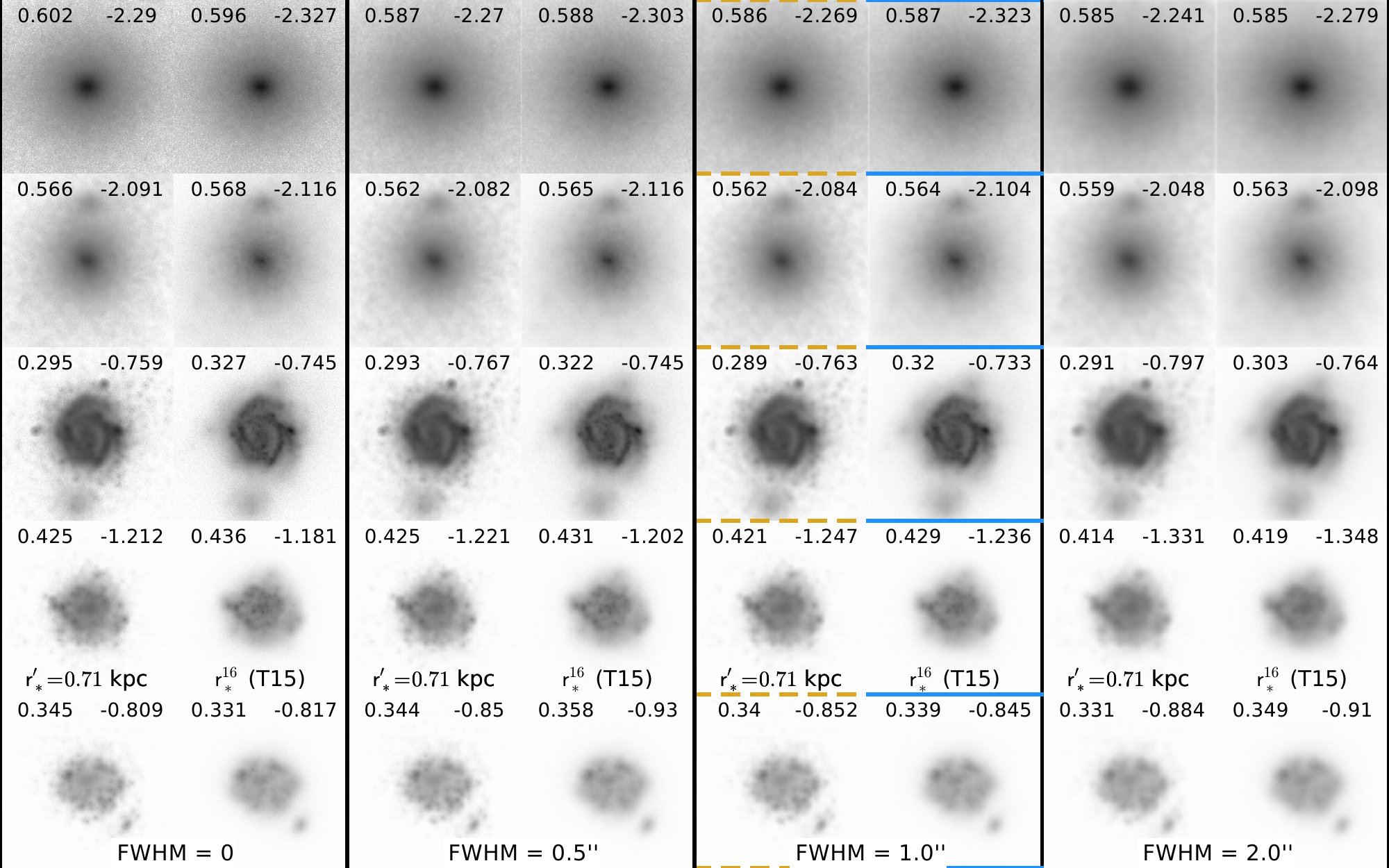}
\caption{ { Face-on example $g$-band images from \illustris\ at $z=0$, as a function of star particle radius assignment method and PSF FWHM, labeled with $G$ and $M_{20}$ values at the top left and top right of each panel, respectively. The blue solid bordered column represents the images used throughout this paper, while the orange dotted border shows the most common alternative (fixed $r_*^{\prime} = 0.71$ kpc). }  \label{fig:starsizes}}
\end{center}
\end{figure*}

{ Figure~\ref{fig:starsizes} compares these two methods as a function of PSF FWHM for five example galaxies at various masses ($9.5 < M_*/M_{\odot} < 11.5$) placed at $z=0.05$ and a pixel size of $0.25$ arcsec.  Since we model observations with PSF FWHM near the \illustris-1 simulation spatial resolution, which sets the star particle sizes in either method, these images are not especially sensitive to PSF size.  We find that using a $2$ arcsec PSF FWHM instead of $1$ arcsec changes \fgmtwenty\ (Equation~\ref{eq:fgm20}, the quantity used throughout this paper) by a median (mean) of $-0.01$ (0.03), with a median absolute difference of $0.06$. }

{ In this paragraph, we fix PSF FWHM$=1.0$ arcsec $\approx 1$ kpc, the value we employed in this paper.  Across all five galaxies with four viewing angles each, the median (mean) difference in $M_{20}$ between the two light assignment methods is $0.009$ ($0.026$), or $0.85\%$ (0$.89\%$).  The median (mean) difference in $G$ is $0.006$ ($0.005$), or $1.0\%$ ($1.5\%$).  The median (mean) difference in \fgmtwenty\ is $0.03$ ($0.01$), with a median absolute difference of $0.06$.  The sense of this difference is that on average, image pixels are slightly less concentrated ($M_{20}$ is larger) but more unequal ($G$ is larger) under our adaptive $r_*^{16}$ assumption versus the constant $r_*^{\prime}$ assumption.  The magnitude is small enough to have no effect on our conclusions.}

\section{Morphology catalogs} \label{a:morphs}

We make the morphology measurements from this paper available as supplementary tables associated with this article.  These catalogs will also be publicly available on the \illustris\ Project website, illustris-project.org.   Moreover, the code we used to add observational effects to the images of T15 is publicly available at https://bitbucket.org/ptorrey/sunpy.  The ideal synthetic images are already publicly available (T15).  

Table~\ref{tab:morphs} presents example entries of these catalogs.  We imaged each of the 10808 simulated galaxies with $M_* > 10^{9.7} M_{\odot}$ from four viewing directions in four filters.  Since these all lead to different quantitative morphologies, each of these 16 iterations per simulated galaxy is treated as a separate source or line in our catalogs.  We separate our tables by filter, leading to 4 individual catalogs of roughly $43000$ entries.   Filesystem errors cause a very small number ($< 1\%$) of the morphology measurements to fail at random, and so we have cross-matched our catalogs to a uniform set of $42531$ sources per filter.  Finally, we sort by $M_*$ and show the top 20 lines of our unattenuated $g$-band catalog in Table~\ref{tab:morphs}.  

\begin{table*}
\caption{ First 20 entries of our unattenuated $g$-band morphology catalog,
  sorted by $M_*$.  The full catalogs for $u$, $g$, $i$, and
  $H$ will be made available as online-only supplementary material.  The subhalo index is the index into the Subfind FOF catalog arrays for the $z=0$ snapshot (number 135).  $U-B$ is the intrinsic unattenuated AB-system colour of the galaxy's stellar populations.  The camera index is the \sunrise\ camera number.  The remaining quantities are morphology measurements described in Section~\ref{ss:morphology}. \label{tab:morphs}}
\begin{center}
\begin{tabular}{cccccccccc}  
Subhalo index &  $\log_{10} M_*/M_{\odot}$ & SFR ($M_{\odot}$ $\rm yr^{-1}$) & $U-B$ & camera index & $G$ & $M_{20}$ & $C$ & $r_p$ & $R_{1/2}$ \\
\hline
0 & 12.33 & 2.93 & 1.25 & 0 & 0.63 & -2.86 & 5.38 & 170.4 & 51.8\\ 
      &       &       &       & 1 & 0.61 & -2.68 & 4.9 & 109.8 & 37.0\\ 
      &       &       &       & 2 & 0.62 & -2.86 & 5.29 & 139.9 & 45.4\\ 
      &       &       &       & 3 & 0.62 & -2.85 & 5.21 & 140.7 & 45.1\\ 
80734 & 12.16 & 0.05 & 1.32 & 0 & 0.61 & -2.65 & 4.76 & 73.0 & 25.0\\ 
      &       &       &       & 1 & 0.62 & -2.64 & 4.86 & 74.1 & 24.7\\ 
      &       &       &       & 2 & 0.62 & -2.69 & 4.99 & 76.7 & 25.2\\ 
      &       &       &       & 3 & 0.61 & -2.55 & 4.79 & 73.9 & 25.0\\ 
142714 & 12.03 & 0.02 & 1.35 & 0 & 0.59 & -2.42 & 4.21 & 45.1 & 15.8\\ 
      &       &       &       & 1 & 0.59 & -2.44 & 4.24 & 47.0 & 16.1\\ 
      &       &       &       & 2 & 0.59 & -2.42 & 4.2 & 44.9 & 15.8\\ 
      &       &       &       & 3 & 0.59 & -2.38 & 4.24 & 46.1 & 16.1\\ 
99148 & 12.0 & 0.15 & 1.32 & 0 & 0.63 & -2.68 & 5.0 & 51.1 & 16.8\\ 
      &       &       &       & 1 & 0.63 & -2.53 & 4.73 & 57.6 & 18.7\\ 
      &       &       &       & 2 & 0.63 & -2.5 & 4.78 & 56.1 & 18.5\\ 
      &       &       &       & 3 & 0.63 & -2.55 & 4.85 & 56.4 & 18.0\\ 
86186 & 11.96 & 0.1 & 1.28 & 0 & 0.63 & -2.69 & 5.35 & 65.5 & 19.9\\ 
      &       &       &       & 1 & 0.64 & -2.46 & 5.31 & 55.2 & 16.6\\ 
      &       &       &       & 2 & 0.63 & -2.35 & 5.55 & 68.6 & 21.1\\ 
      &       &       &       & 3 & 0.64 & -2.76 & 5.44 & 85.5 & 25.2\\ 
\hline
\end{tabular}

\end{center}
\end{table*}

\bibliographystyle{apj}
\bibliography{$HOME/Dropbox/library}

\begin{thebibliography}{133}
\expandafter\ifx\csname natexlab\endcsname\relax\def\natexlab#1{#1}\fi

\bibitem[{Abadi {et~al.}(2003)Abadi, Navarro, Steinmetz, \& Eke}]{Abadi2003}
Abadi, M.~G., Navarro, J.~F., Steinmetz, M., \& Eke, V.~R. 2003, \apj, 591, 499

\bibitem[{Abraham {et~al.}(1996)Abraham, Tanvir, Santiago, Ellis, Glazebrook,
  \& van~den Bergh}]{Abraham1996}
Abraham, R.~G., Tanvir, N.~R., Santiago, B.~X., Ellis, R.~S., Glazebrook, K.,
  \& van~den Bergh, S. 1996, \mnras, 279

\bibitem[{Abraham {et~al.}(2003)Abraham, van~den Bergh, \& Nair}]{Abraham2003}
Abraham, R.~G., van~den Bergh, S., \& Nair, P. 2003, \apj, 588, 218

\bibitem[{Ahn {et~al.}(2014)Ahn, Alexandroff, {Allende Prieto}, Anders,
  Anderson, Anderton, Andrews, Aubourg, Bailey, Bastien, Bautista, Beers,
  Beifiori, Bender, Berlind, Beutler, Bhardwaj, Bird, Bizyaev, Blake, Blanton,
  Blomqvist, Bochanski, Bolton, Borde, Bovy, Bradley, Brandt, Brauer,
  Brinkmann, Brownstein, Busca, Carithers, Carlberg, Carnero, Carr, Chiappini,
  Chojnowski, Chuang, Comparat, Crepp, Cristiani, Croft, Cuesta, Cunha,
  da~Costa, Dawson, {De Lee}, Dean, Delubac, Deshpande, Dhital, Ealet, Ebelke,
  Edmondson, Eisenstein, Epstein, Escoffier, Esposito, Evans, Fabbian, Fan,
  Favole, {Femen\'{\i}a Castell\'{a}}, {Fern\'{a}ndez Alvar}, Feuillet, {Filiz
  Ak}, Finley, Fleming, Font-Ribera, Frinchaboy, Galbraith-Frew,
  Garc\'{\i}a-Hern\'{a}ndez, {Garc\'{\i}a P\'{e}rez}, Ge, G\'{e}nova-Santos,
  Gillespie, Girardi, {Gonz\'{a}lez Hern\'{a}ndez}, Gott, Gunn, Guo, Halverson,
  Harding, Harris, Hasselquist, Hawley, Hayden, Hearty, {Herrero Dav\'{o}}, Ho,
  Hogg, Holtzman, Honscheid, Huehnerhoff, Ivans, Jackson, Jiang, Johnson,
  Kinemuchi, Kirkby, Klaene, Kneib, Koesterke, Lan, Lang, {Le Goff}, Leauthaud,
  Lee, Lee, Long, Loomis, Lucatello, Lupton, Ma, Mack, Mahadevan, Maia,
  Majewski, Malanushenko, Malanushenko, Manchado, Manera, Maraston, Margala,
  Martell, Masters, McBride, McGreer, McMahon, M\'{e}nard, M\'{e}sz\'{a}ros,
  Miralda-Escud\'{e}, Miyatake, Montero-Dorta, Montesano, More, Morrison, Muna,
  Munn, Myers, Nguyen, Nichol, Nidever, Noterdaeme, Nuza, O'Connell, O'Connell,
  O'Connell, Olmstead, Oravetz, Owen, Padmanabhan, Palanque-Delabrouille, Pan,
  Parejko, Parihar, P\^{a}ris, Pepper, Percival, P\'{e}rez-R\`{a}fols, {Dotto
  Perottoni}, Petitjean, Pieri, Pinsonneault, Prada, Price-Whelan, Raddick,
  Rahman, Rebolo, Reid, Richards, Riffel, Robin, Rocha-Pinto, Rockosi, Roe,
  Ross, Ross, Rossi, Roy, Rubi\~{n}o Martin, Sabiu, S\'{a}nchez, Santiago,
  Sayres, Schiavon, Schlegel, Schlesinger, Schmidt, Schneider, Schultheis,
  Sellgren, Seo, Shen, Shetrone, Shu, Simmons, Skrutskie, Slosar, Smith,
  Snedden, Sobeck, Sobreira, Stassun, Steinmetz, Strauss, Streblyanska, Suzuki,
  Swanson, Terrien, Thakar, Thomas, Thompson, Tinker, Tojeiro, Troup,
  Vandenberg, {Vargas Maga\~{n}a}, Viel, Vogt, Wake, Weaver, Weinberg, Weiner,
  White, White, Wilson, Wisniewski, Wood-Vasey, Y\`{e}che, York, Zamora,
  Zasowski, Zehavi, Zhao, Zheng, \& Zhu}]{Ahn2014}
Ahn, C.~P., {et~al.} 2014, \apjs, 211, 17

\bibitem[{Barro {et~al.}(2013)Barro, Faber, P\'{e}rez-Gonz\'{a}lez, Koo,
  Williams, Kocevski, Trump, Mozena, McGrath, van~der Wel, Wuyts, Bell, Croton,
  Daniel, Dekel, Ashby, Cheung, Ferguson, Fontana, Fang, Giavalisco, Grogin,
  Guo, Hathi, Hopkins, Huang, Koekemoer, Kartaltepe, Lee, Newman, Porter,
  Primack, Ryan, Rosario, Somerville, Salvato, \& Hsu}]{Barro2013}
Barro, G., {et~al.} 2013, \apj, 765, 104

\bibitem[{Bell {et~al.}(2012)Bell, van~der Wel, Papovich, Kocevski, Lotz,
  McIntosh, Kartaltepe, Faber, Ferguson, Koekemoer, Grogin, Wuyts, Cheung,
  Conselice, Dekel, Dunlop, Giavalisco, Herrington, Koo, McGrath, de~Mello,
  Rix, Robaina, \& Williams}]{Bell2012}
Bell, E.~F., {et~al.} 2012, \apj, 753, 167

\bibitem[{Bell {et~al.}(2006)Bell, Gray, \& Szalay}]{Bell2006}
Bell, G., Gray, J., \& Szalay, A. 2006, IEEE Computer, 39, 110

\bibitem[{Bender {et~al.}(1994)Bender, Saglia, \& Gerhard}]{Bender1994}
Bender, R., Saglia, R.~P., \& Gerhard, O.~E. 1994, \mnras, 269

\bibitem[{Bernyk {et~al.}(2014)Bernyk, Croton, Tonini, Hodkinson, Hassan,
  Garel, Duffy, Mutch, \& Poole}]{Bernyk2014}
Bernyk, M., {et~al.} 2014, eprint arXiv:1403.5270

\bibitem[{Bershady {et~al.}(2000)Bershady, Jangren, \&
  Conselice}]{Bershady2000}
Bershady, M.~A., Jangren, A., \& Conselice, C.~J. 2000, \aj, 119, 2645

\bibitem[{Blanton {et~al.}(2003)Blanton, Hogg, Bahcall, Baldry, Brinkmann,
  Csabai, Eisenstein, Fukugita, Gunn, Ivezi\'{c}, Lamb, Lupton, Loveday, Munn,
  Nichol, Okamura, Schlegel, Shimasaku, Strauss, Vogeley, \&
  Weinberg}]{Blanton2003}
Blanton, M.~R., {et~al.} 2003, \apj, 594, 186

\bibitem[{Boggs \& Rogers(1990)}]{Boggs1990}
Boggs, P., \& Rogers, J. 1990, Contemporary Mathematics, 112, 183

\bibitem[{Brammer {et~al.}(2009)Brammer, Whitaker, van Dokkum, Marchesini,
  Labb\'{e}, Franx, Kriek, Quadri, Illingworth, Lee, Muzzin, \&
  Rudnick}]{brammer09}
Brammer, G., {et~al.} 2009, \apjl, 706, L173

\bibitem[{Brennan {et~al.}(2015)Brennan, Pandya, Somerville, Barro, Taylor,
  Wuyts, Bell, Dekel, Ferguson, McIntosh, Papovich, \& Primack}]{Brennan2015}
Brennan, R., {et~al.} 2015, \mnras, 451, 2933

\bibitem[{Bruzual \& Charlot(2003)}]{bc03}
Bruzual, G., \& Charlot, S. 2003, \mnras, 344, 1000

\bibitem[{Bush {et~al.}(2010)Bush, Cox, Hayward, Thilker, Hernquist, \&
  Besla}]{bush10}
Bush, S., Cox, T., Hayward, C., Thilker, D., Hernquist, L., \& Besla, G. 2010,
  \apj, 713, 780

\bibitem[{Cappellari {et~al.}(2007)Cappellari, Emsellem, Bacon, Bureau, Davies,
  {De Zeeuw}, Falcon-Barroso, Krajnovi, Kuntschner, McDermid, Peletier, Sarzi,
  {Van Den Bosch}, \& {Van De Ven}}]{Cappellari2007}
Cappellari, M., {et~al.} 2007, \mnras, 379, 418

\bibitem[{Cappellari {et~al.}(2011{\natexlab{a}})Cappellari, Emsellem,
  Krajnovi\'{c}, McDermid, Scott, {Verdoes Kleijn}, Young, Alatalo, Bacon,
  Blitz, Bois, Bournaud, Bureau, Davies, Davis, de~Zeeuw, Duc, Khochfar,
  Kuntschner, Lablanche, Morganti, Naab, Oosterloo, Sarzi, Serra, \&
  Weijmans}]{Cappellari2011a}
---. 2011{\natexlab{a}}, \mnras, 413, 813

\bibitem[{Cappellari {et~al.}(2011{\natexlab{b}})Cappellari, Emsellem,
  Krajnovi\'{c}, McDermid, Serra, Alatalo, Blitz, Bois, Bournaud, Bureau,
  Davies, Davis, de~Zeeuw, Khochfar, Kuntschner, Lablanche, Morganti, Naab,
  Oosterloo, Sarzi, Scott, Weijmans, \& Young}]{Cappellari2011}
---. 2011{\natexlab{b}}, \mnras, 416, 1680

\bibitem[{Cen(2014)}]{Cen2014}
Cen, R. 2014, \apj, 781, 38

\bibitem[{Ceverino {et~al.}(2014)Ceverino, Klypin, Klimek, Trujillo-Gomez,
  Churchill, Primack, \& Dekel}]{Ceverino2014}
Ceverino, D., Klypin, A., Klimek, E.~S., Trujillo-Gomez, S., Churchill, C.~W.,
  Primack, J., \& Dekel, A. 2014, \mnras, 442, 1545

\bibitem[{Chabrier(2003)}]{chabrier03}
Chabrier, G. 2003, \pasp, 115, 763

\bibitem[{Christensen {et~al.}(2014)Christensen, Brooks, Fisher, Governato,
  McCleary, Quinn, Shen, \& Wadsley}]{Christensen2014}
Christensen, C.~R., Brooks, A.~M., Fisher, D.~B., Governato, F., McCleary, J.,
  Quinn, T.~R., Shen, S., \& Wadsley, J. 2014, \mnrasl, 440, L51

\bibitem[{Conselice {et~al.}(2003)Conselice, Bershady, Dickinson, \&
  Papovich}]{Conselice2003}
Conselice, C.~J., Bershady, M.~A., Dickinson, M., \& Papovich, C. 2003, \aj,
  126, 1183

\bibitem[{Cox {et~al.}(2006)Cox, Dutta, {Di Matteo}, Hernquist, Hopkins,
  Robertson, \& Springel}]{cox06_kinematics}
Cox, T., Dutta, S., {Di Matteo}, T., Hernquist, L., Hopkins, P., Robertson, B.,
  \& Springel, V. 2006, \apj, 650, 791

\bibitem[{Crain {et~al.}(2009)Crain, Theuns, {Dalla Vecchia}, Eke, Frenk,
  Jenkins, Kay, Peacock, Pearce, Schaye, Springel, Thomas, White, \&
  Wiersma}]{Crain2009}
Crain, R.~A., {et~al.} 2009, \mnras, 399, 1773

\bibitem[{Crain {et~al.}(2015)Crain, Schaye, Bower, Furlong, Schaller, Theuns,
  {Dalla Vecchia}, Frenk, McCarthy, Helly, Jenkins, Rosas-Guevara, White, \&
  Trayford}]{Crain2015}
---. 2015, \mnras, 450, 1937

\bibitem[{Croton {et~al.}(2006)Croton, Springel, White, {De Lucia}, Frenk, Gao,
  Jenkins, Kauffmann, Navarro, \& Yoshida}]{croton06}
Croton, D., {et~al.} 2006, \mnras, 365, 11

\bibitem[{Davis {et~al.}(1985)Davis, Efstathiou, Frenk, \& White}]{Davis1985}
Davis, M., Efstathiou, G., Frenk, C.~S., \& White, S. D.~M. 1985, \apj, 292,
  371

\bibitem[{{De Lucia} \& Blaizot(2007)}]{DeLucia2007}
{De Lucia}, G., \& Blaizot, J. 2007, \mnras, 375, 2

\bibitem[{Dekel {et~al.}(2009)Dekel, Sari, \& Ceverino}]{Dekel2009}
Dekel, A., Sari, R., \& Ceverino, D. 2009, \apj, 703, 785

\bibitem[{Dolag {et~al.}(2009)Dolag, Borgani, Murante, \& Springel}]{Dolag2009}
Dolag, K., Borgani, S., Murante, G., \& Springel, V. 2009, \mnras, 399, 497

\bibitem[{Dressler(1980)}]{dressler80}
Dressler, A. 1980, \apj, 236, 351

\bibitem[{Dubois {et~al.}(2014)Dubois, Pichon, Welker, {Le Borgne}, Devriendt,
  Laigle, Codis, Pogosyan, Arnouts, Benabed, Bertin, Blaizot, Bouchet, Cardoso,
  Colombi, de~Lapparent, Desjacques, Gavazzi, Kassin, Kimm, McCracken,
  Milliard, Peirani, Prunet, Rouberol, Silk, Slyz, Sousbie, Teyssier, Tresse,
  Treyer, Vibert, \& Volonteri}]{Dubois2014}
Dubois, Y., {et~al.} 2014, \mnras, 444, 1453

\bibitem[{Dugan {et~al.}(2014)Dugan, Bryan, Gaibler, Silk, \& Haas}]{Dugan2014}
Dugan, Z., Bryan, S., Gaibler, V., Silk, J., \& Haas, M. 2014, \apj, 796, 113

\bibitem[{Emsellem {et~al.}(2007)Emsellem, Cappellari, Krajnovi, {Van De Ven},
  Bacon, Bureau, Davies, {De Zeeuw}, Falcon-Barroso, Kuntschner, McDermid,
  Peletier, \& Sarzi}]{Emsellem2007}
Emsellem, E., {et~al.} 2007, \mnras, 379, 401

\bibitem[{Fall \& Romanowsky(2013)}]{Fall2013}
Fall, S.~M., \& Romanowsky, A.~J. 2013, \apjl, 769, L26

\bibitem[{Franx {et~al.}(2008)Franx, van Dokkum, Schreiber, Wuyts, Labb\'{e},
  \& Toft}]{Franx2008}
Franx, M., van Dokkum, P.~G., Schreiber, N. M.~F., Wuyts, S., Labb\'{e}, I., \&
  Toft, S. 2008, \apj, 688, 770

\bibitem[{Freeman {et~al.}(2013)Freeman, Izbicki, Lee, Newman, Conselice,
  Koekemoer, Lotz, \& Mozena}]{Freeman2013}
Freeman, P.~E., Izbicki, R., Lee, A.~B., Newman, J.~A., Conselice, C.~J.,
  Koekemoer, A.~M., Lotz, J.~M., \& Mozena, M. 2013, \mnras, 434, 282

\bibitem[{Genel {et~al.}(2015)Genel, Fall, Hernquist, Vogelsberger, Snyder,
  Rodriguez-Gomez, Sijacki, \& Springel}]{Genel2015}
Genel, S., Fall, S.~M., Hernquist, L., Vogelsberger, M., Snyder, G.~F.,
  Rodriguez-Gomez, V., Sijacki, D., \& Springel, V. 2015, \apj, 804, L40

\bibitem[{Genel {et~al.}(2014)Genel, Vogelsberger, Springel, Sijacki, Nelson,
  Snyder, Rodriguez-Gomez, Torrey, \& Hernquist}]{Genel2014}
Genel, S., {et~al.} 2014, \mnras, 445, 175

\bibitem[{Glasser(1962)}]{Glasser1962}
Glasser, G.~J. 1962, Journal of the American Statistical Association, 57, 648

\bibitem[{Guedes {et~al.}(2011)Guedes, Callegari, Madau, \& Mayer}]{Guedes2011}
Guedes, J., Callegari, S., Madau, P., \& Mayer, L. 2011, \apj, 742, 76

\bibitem[{Gunn {et~al.}(1981)Gunn, Stryker, \& Tinsley}]{Gunn1981}
Gunn, J.~E., Stryker, L.~L., \& Tinsley, B.~M. 1981, \apj, 249, 48

\bibitem[{Guo {et~al.}(2012)Guo, White, Angulo, Henriques, Lemson,
  Boylan-Kolchin, Thomas, \& Short}]{Guo2012a}
Guo, Q., White, S., Angulo, R.~E., Henriques, B., Lemson, G., Boylan-Kolchin,
  M., Thomas, P., \& Short, C. 2012, \mnras, 428, 1351

\bibitem[{Guo \& White(2009)}]{Guo2009}
Guo, Q., \& White, S. D.~M. 2009, \mnras, 396, 39

\bibitem[{Guo {et~al.}(2011)Guo, White, Boylan-Kolchin, {De Lucia}, Kauffmann,
  Lemson, Li, Springel, \& Weinmann}]{Guo2011}
Guo, Q., {et~al.} 2011, \mnras, 413, 101

\bibitem[{Hanisch {et~al.}(2001)Hanisch, Farris, Greisen, Pence, Schlesinger,
  Teuben, Thompson, \& Warnock}]{Hanisch2001}
Hanisch, R.~J., Farris, A., Greisen, E.~W., Pence, W.~D., Schlesinger, B.~M.,
  Teuben, P.~J., Thompson, R.~W., \& Warnock, A. 2001, \aap, 376, 359

\bibitem[{Henriques {et~al.}(2012)Henriques, White, Lemson, Thomas, Guo,
  Marleau, \& Overzier}]{Henriques2012}
Henriques, B. M.~B., White, S. D.~M., Lemson, G., Thomas, P.~A., Guo, Q.,
  Marleau, G.-D., \& Overzier, R.~A. 2012, \mnras, 421, 2904

\bibitem[{Henriques {et~al.}(2015)Henriques, White, Thomas, Angulo, Guo,
  Lemson, Springel, \& Overzier}]{Henriques2015}
Henriques, B. M.~B., White, S. D.~M., Thomas, P.~A., Angulo, R., Guo, Q.,
  Lemson, G., Springel, V., \& Overzier, R. 2015, \mnras, 451, 2663

\bibitem[{Hinshaw {et~al.}(2013)Hinshaw, Larson, Komatsu, Spergel, Bennett,
  Dunkley, Nolta, Halpern, Hill, Odegard, Page, Smith, Weiland, Gold, Jarosik,
  Kogut, Limon, Meyer, Tucker, Wollack, \& Wright}]{Hinshaw2013}
Hinshaw, G., {et~al.} 2013, \apjs, 208, 19

\bibitem[{Hopkins {et~al.}(2007)Hopkins, Hernquist, Cox, Robertson, \&
  Krause}]{hopkinsfp_07b}
Hopkins, P., Hernquist, L., Cox, T., Robertson, B., \& Krause, E. 2007, \apj,
  669, 67

\bibitem[{Hopkins {et~al.}(2014)Hopkins, Keres, Onorbe, Faucher-Giguere,
  Quataert, Murray, \& Bullock}]{Hopkins2014}
Hopkins, P.~F., Keres, D., Onorbe, J., Faucher-Giguere, C.-A., Quataert, E.,
  Murray, N., \& Bullock, J.~S. 2014, \mnras, 445, 581

\bibitem[{Hunter(2007)}]{Hunter:2007}
Hunter, J.~D. 2007, Computing In Science \& Engineering, 9, 90

\bibitem[{Jones {et~al.}(2001)Jones, Oliphant, Peterson, \& Others}]{SCIPY}
Jones, E., Oliphant, T., Peterson, P., \& Others. 2001, {{SciPy}: Open source
  scientific tools for {Python}, www.scipy.org}

\bibitem[{Jonsson(2006)}]{jonsson06}
Jonsson, P. 2006, \mnras, 372, 2

\bibitem[{Jonsson {et~al.}(2010)Jonsson, Groves, \& Cox}]{jonsson09}
Jonsson, P., Groves, B., \& Cox, T. 2010, \mnras, 403, 17

\bibitem[{Jonsson \& Primack(2010)}]{Jonsson:2010gpu}
Jonsson, P., \& Primack, J.~R. 2010, New Astron., 15, 509

\bibitem[{Kassin {et~al.}(2014)Kassin, Brooks, Governato, Weiner, \&
  Gardner}]{Kassin2014}
Kassin, S.~A., Brooks, A., Governato, F., Weiner, B.~J., \& Gardner, J.~P.
  2014, \apj, 790, 89

\bibitem[{Kauffmann {et~al.}(2003)Kauffmann, Heckman, White, Charlot, Tremonti,
  Peng, Seibert, Brinkmann, Nichol, SubbaRao, \& York}]{Kauffmann2003}
Kauffmann, G., {et~al.} 2003, \mnras, 341, 54

\bibitem[{Kere\v{s} {et~al.}(2012)Kere\v{s}, Vogelsberger, Sijacki, Springel,
  \& Hernquist}]{Keres2012}
Kere\v{s}, D., Vogelsberger, M., Sijacki, D., Springel, V., \& Hernquist, L.
  2012, \mnras, 425, 2027

\bibitem[{Khandai {et~al.}(2015)Khandai, {Di Matteo}, Croft, Wilkins, Feng,
  Tucker, DeGraf, \& Liu}]{Khandai2015}
Khandai, N., {Di Matteo}, T., Croft, R., Wilkins, S., Feng, Y., Tucker, E.,
  DeGraf, C., \& Liu, M.-S. 2015, \mnras, 450, 1349

\bibitem[{Kitzbichler \& White(2007)}]{Kitzbichler2007}
Kitzbichler, M.~G., \& White, S. D.~M. 2007, \mnras, 376, 2

\bibitem[{Kormendy \& Bender(1996)}]{Kormendy1996}
Kormendy, J., \& Bender, R. 1996, \apj, 464, L119

\bibitem[{Krajnovi\'{c} {et~al.}(2008)Krajnovi\'{c}, Bacon, Cappellari, Davies,
  de~Zeeuw, Emsellem, Falc\'{o}n-Barroso, Kuntschner, McDermid, Peletier,
  Sarzi, van~den Bosch, \& van~de Ven}]{Krajnovic2008}
Krajnovi\'{c}, D., {et~al.} 2008, \mnras, 390, 93

\bibitem[{Laureijs {et~al.}(2011)Laureijs, Amiaux, Arduini, Augu\`{e}res,
  Brinchmann, Cole, Cropper, Dabin, Duvet, Ealet, Garilli, Gondoin, Guzzo,
  Hoar, Hoekstra, Holmes, Kitching, Maciaszek, Mellier, Pasian, Percival,
  Rhodes, {Saavedra Criado}, Sauvage, Scaramella, Valenziano, Warren, Bender,
  Castander, Cimatti, {Le F\`{e}vre}, Kurki-Suonio, Levi, Lilje, Meylan,
  Nichol, Pedersen, Popa, {Rebolo Lopez}, Rix, Rottgering, Zeilinger, Grupp,
  Hudelot, Massey, Meneghetti, Miller, Paltani, Paulin-Henriksson, Pires,
  Saxton, Schrabback, Seidel, Walsh, Aghanim, Amendola, Bartlett, Baccigalupi,
  Beaulieu, Benabed, Cuby, Elbaz, Fosalba, Gavazzi, Helmi, Hook, Irwin, Kneib,
  Kunz, Mannucci, Moscardini, Tao, Teyssier, Weller, Zamorani, {Zapatero
  Osorio}, Boulade, Foumond, {Di Giorgio}, Guttridge, James, Kemp, Martignac,
  Spencer, Walton, Bl\"{u}mchen, Bonoli, Bortoletto, Cerna, Corcione, Fabron,
  Jahnke, Ligori, Madrid, Martin, Morgante, Pamplona, Prieto, Riva, Toledo,
  Trifoglio, Zerbi, Abdalla, Douspis, Grenet, Borgani, Bouwens, Courbin,
  Delouis, Dubath, Fontana, Frailis, Grazian, Koppenh\"{o}fer, Mansutti,
  Melchior, Mignoli, Mohr, Neissner, Noddle, Poncet, Scodeggio, Serrano, Shane,
  Starck, Surace, Taylor, Verdoes-Kleijn, Vuerli, Williams, Zacchei, Altieri,
  {Escudero Sanz}, Kohley, Oosterbroek, Astier, Bacon, Bardelli, Baugh,
  Bellagamba, Benoist, Bianchi, Biviano, Branchini, Carbone, Cardone, Clements,
  Colombi, Conselice, Cresci, Deacon, Dunlop, Fedeli, Fontanot, Franzetti,
  Giocoli, Garcia-Bellido, Gow, Heavens, Hewett, Heymans, Holland, Huang,
  Ilbert, Joachimi, Jennins, Kerins, Kiessling, Kirk, Kotak, Krause, Lahav, van
  Leeuwen, Lesgourgues, Lombardi, Magliocchetti, Maguire, Majerotto, Maoli,
  Marulli, Maurogordato, McCracken, McLure, Melchiorri, Merson, Moresco,
  Nonino, Norberg, Peacock, Pello, Penny, Pettorino, {Di Porto}, Pozzetti,
  Quercellini, Radovich, Rassat, Roche, Ronayette, Rossetti, Sartoris,
  Schneider, Semboloni, Serjeant, Simpson, Skordis, Smadja, Smartt, Spano,
  Spiro, Sullivan, Tilquin, Trotta, Verde, Wang, Williger, Zhao, Zoubian, \&
  Zucca}]{Laureijs2011}
Laureijs, R., {et~al.} 2011, eprint arXiv:1110.3193

\bibitem[{Lemson(2006)}]{Lemson2006}
Lemson, G. 2006, ArXiv astro-ph/0608019

\bibitem[{Lintott {et~al.}(2008)Lintott, Schawinski, Slosar, Land, Bamford,
  Thomas, Raddick, Nichol, Szalay, Andreescu, Murray, \&
  Vandenberg}]{Lintott2008}
Lintott, C.~J., {et~al.} 2008, \mnras, 389, 1179

\bibitem[{Lotz {et~al.}(2008{\natexlab{a}})Lotz, Jonsson, Cox, \&
  Primack}]{lotz08}
Lotz, J., Jonsson, P., Cox, T., \& Primack, J. 2008{\natexlab{a}}, \mnras, 391,
  1137

\bibitem[{Lotz {et~al.}(2010{\natexlab{a}})Lotz, Jonsson, Cox, \&
  Primack}]{lotz10}
---. 2010{\natexlab{a}}, \mnras, 404, 575

\bibitem[{Lotz {et~al.}(2008{\natexlab{b}})Lotz, Davis, Faber, Guhathakurta,
  Gwyn, Huang, Koo, {Le Floc'h}, Lin, Newman, Noeske, Papovich, Willmer, Coil,
  Conselice, Cooper, Hopkins, Metevier, Primack, Rieke, \& Weiner}]{lotz08_hst}
Lotz, J., {et~al.} 2008{\natexlab{b}}, \apj, 672, 177

\bibitem[{Lotz {et~al.}(2011)Lotz, Jonsson, Cox, Croton, Primack, Somerville,
  \& Stewart}]{Lotz2011}
Lotz, J.~M., Jonsson, P., Cox, T.~J., Croton, D., Primack, J.~R., Somerville,
  R.~S., \& Stewart, K. 2011, \apj, 742, 103

\bibitem[{Lotz {et~al.}(2010{\natexlab{b}})Lotz, Jonsson, Cox, \&
  Primack}]{Lotz2010}
Lotz, J.~M., Jonsson, P., Cox, T.~J., \& Primack, J.~R. 2010{\natexlab{b}},
  \mnras, 404, 590

\bibitem[{Lotz {et~al.}(2004)Lotz, Primack, \& Madau}]{Lotz2004}
Lotz, J.~M., Primack, J., \& Madau, P. 2004, \aj, 128, 163

\bibitem[{Lu {et~al.}(2014)Lu, Wechsler, Somerville, Croton, Porter, Primack,
  Behroozi, Ferguson, Koo, Guo, Safarzadeh, Finlator, Castellano, White,
  Sommariva, \& Moody}]{Lu2014}
Lu, Y., {et~al.} 2014, \apj, 795, 123

\bibitem[{Mandelker {et~al.}(2014)Mandelker, Dekel, Ceverino, Tweed, Moody, \&
  Primack}]{Mandelker2014}
Mandelker, N., Dekel, A., Ceverino, D., Tweed, D., Moody, C.~E., \& Primack, J.
  2014, \mnras, 443, 3675

\bibitem[{Marinacci {et~al.}(2013)Marinacci, Pakmor, \&
  Springel}]{Marinacci2013}
Marinacci, F., Pakmor, R., \& Springel, V. 2013, \mnras, 437, 1750

\bibitem[{Moody {et~al.}(2014{\natexlab{a}})Moody, Guo, Mandelker, Ceverino,
  Mozena, Koo, Dekel, \& Primack}]{Moody2014}
Moody, C.~E., Guo, Y., Mandelker, N., Ceverino, D., Mozena, M., Koo, D.~C.,
  Dekel, A., \& Primack, J. 2014{\natexlab{a}}, \mnras, 444, 1389

\bibitem[{Moody {et~al.}(2014{\natexlab{b}})Moody, Romanowsky, Cox, Novak, \&
  Primack}]{Moody2014a}
Moody, C.~E., Romanowsky, A.~J., Cox, T.~J., Novak, G.~S., \& Primack, J.~R.
  2014{\natexlab{b}}, \mnras, 444, 1475

\bibitem[{Munshi {et~al.}(2013)Munshi, Governato, Brooks, Christensen, Shen,
  Loebman, Moster, Quinn, \& Wadsley}]{Munshi2013}
Munshi, F., {et~al.} 2013, \apj, 766, 56

\bibitem[{Muratov {et~al.}(2015)Muratov, Keres, Faucher-Giguere, Hopkins,
  Quataert, \& Murray}]{Muratov2015}
Muratov, A.~L., Keres, D., Faucher-Giguere, C.-A., Hopkins, P.~F., Quataert,
  E., \& Murray, N. 2015, eprint arXiv:1501.03155

\bibitem[{Naab {et~al.}(2014)Naab, Oser, Emsellem, Cappellari, Krajnovi,
  McDermid, Alatalo, Bayet, Blitz, Bois, Bournaud, Bureau, Crocker, Davies,
  Davis, de~Zeeuw, Duc, Hirschmann, Johansson, Khochfar, Kuntschner, Morganti,
  Oosterloo, Sarzi, Scott, Serra, Ven, Weijmans, \& Young}]{Naab2014}
Naab, T., {et~al.} 2014, \mnras, 444, 3357

\bibitem[{Nelson {et~al.}(2015{\natexlab{a}})Nelson, Genel, Vogelsberger,
  Springel, Sijacki, Torrey, \& Hernquist}]{Nelson2015}
Nelson, D., Genel, S., Vogelsberger, M., Springel, V., Sijacki, D., Torrey, P.,
  \& Hernquist, L. 2015{\natexlab{a}}, \mnras, 448, 59

\bibitem[{Nelson {et~al.}(2015{\natexlab{b}})Nelson, Pillepich, Genel,
  Vogelsberger, Springel, Torrey, Rodriguez-Gomez, Sijacki, Snyder, Griffen,
  Marinacci, Blecha, Sales, Xu, \& Hernquist}]{Nelson2015a}
Nelson, D., {et~al.} 2015{\natexlab{b}}, eprint arXiv:1504.00362

\bibitem[{Noeske {et~al.}(2007)Noeske, Weiner, Faber, Papovich, Koo,
  Somerville, Bundy, Conselice, Newman, Schiminovich, Floc'h, Coil, Rieke,
  Lotz, Primack, Barmby, Cooper, Davis, Ellis, Fazio, Guhathakurta, Huang,
  Kassin, Martin, Phillips, Rich, Small, Willmer, \& Wilson}]{Noeske:2007a}
Noeske, K.~G., {et~al.} 2007, \apjl, 660, L43

\bibitem[{Omand {et~al.}(2014)Omand, Balogh, \& Poggianti}]{Omand2014}
Omand, C. M.~B., Balogh, M.~L., \& Poggianti, B.~M. 2014, \mnras, 440, 843

\bibitem[{Overzier {et~al.}(2013)Overzier, Lemson, Angulo, Bertin, Blaizot,
  Henriques, Marleau, \& White}]{Overzier2012}
Overzier, R., Lemson, G., Angulo, R.~E., Bertin, E., Blaizot, J., Henriques, B.
  M.~B., Marleau, G.-D., \& White, S. D.~M. 2013, \mnras, 428, 778

\bibitem[{Pedrosa {et~al.}(2014)Pedrosa, Tissera, \& {De Rossi}}]{Pedrosa2014}
Pedrosa, S.~E., Tissera, P.~B., \& {De Rossi}, M.~E. 2014, \aap, 567, A47

\bibitem[{Peng {et~al.}(2010)Peng, Lilly, Kova\v{c}, Bolzonella, Pozzetti,
  Renzini, Zamorani, Ilbert, Knobel, Iovino, Maier, Cucciati, Tasca, Carollo,
  Silverman, Kampczyk, de~Ravel, Sanders, Scoville, Contini, Mainieri,
  Scodeggio, Kneib, {Le F\`{e}vre}, Bardelli, Bongiorno, Caputi, Coppa, de~la
  Torre, Franzetti, Garilli, Lamareille, {Le Borgne}, {Le Brun}, Mignoli,
  Montero, Pello, Ricciardelli, Tanaka, Tresse, Vergani, Welikala, Zucca,
  Oesch, Abbas, Barnes, Bordoloi, Bottini, Cappi, Cassata, Cimatti, Fumana,
  Hasinger, Koekemoer, Leauthaud, Maccagni, Marinoni, McCracken, Memeo, Meneux,
  Nair, Porciani, Presotto, \& Scaramella}]{Peng2010}
Peng, Y.-j., {et~al.} 2010, \apj, 721, 193

\bibitem[{Pillepich {et~al.}(2014)Pillepich, Vogelsberger, Deason,
  Rodriguez-Gomez, Genel, Nelson, Torrey, Sales, Marinacci, Springel, Sijacki,
  \& Hernquist}]{Pillepich2014}
Pillepich, A., {et~al.} 2014, \mnras, 444, 237

\bibitem[{Porter {et~al.}(2014)Porter, Somerville, Primack, \&
  Johansson}]{Porter2014}
Porter, L.~A., Somerville, R.~S., Primack, J.~R., \& Johansson, P.~H. 2014,
  \mnras, 444, 942

\bibitem[{Robitaille {et~al.}(2013)Robitaille, Tollerud, Greenfield,
  Droettboom, Bray, Aldcroft, Davis, Ginsburg, Price-Whelan, Kerzendorf,
  Conley, Crighton, Barbary, Muna, Ferguson, Grollier, Parikh, Nair,
  G\"{u}nther, Deil, Woillez, Conseil, Kramer, Turner, Singer, Fox, Weaver,
  Zabalza, Edwards, {Azalee Bostroem}, Burke, Casey, Crawford, Dencheva, Ely,
  Jenness, Labrie, Lim, Pierfederici, Pontzen, Ptak, Refsdal, Servillat, \&
  Streicher}]{Robitaille2013}
Robitaille, T.~P., {et~al.} 2013, \aap, 558, A33

\bibitem[{Rodriguez-Gomez {et~al.}(2015)Rodriguez-Gomez, Genel, Vogelsberger,
  Sijacki, Pillepich, Sales, Torrey, Snyder, Nelson, Springel, Ma, \&
  Hernquist}]{Rodriguez-Gomez2015}
Rodriguez-Gomez, V., {et~al.} 2015, \mnras, 449, 49

\bibitem[{Romanowsky \& Fall(2012)}]{Romanowsky2012}
Romanowsky, A.~J., \& Fall, S.~M. 2012, \apjs, 203, 17

\bibitem[{Sales {et~al.}(2012)Sales, Navarro, Theuns, Schaye, White, Frenk,
  Crain, \& {Dalla Vecchia}}]{Sales2012}
Sales, L.~V., Navarro, J.~F., Theuns, T., Schaye, J., White, S. D.~M., Frenk,
  C.~S., Crain, R.~A., \& {Dalla Vecchia}, C. 2012, \mnras, 423, 1544

\bibitem[{Sales {et~al.}(2014)Sales, Vogelsberger, Genel, Torrey, Nelson,
  Rodriguez-Gomez, Wang, Pillepich, Sijacki, Springel, \&
  Hernquist}]{Sales2014}
Sales, L.~V., {et~al.} 2014, \mnrasl, 447, L6

\bibitem[{Scannapieco {et~al.}(2010)Scannapieco, Gadotti, Jonsson, \&
  White}]{scannapieco10}
Scannapieco, C., Gadotti, D., Jonsson, P., \& White, S. 2010, \mnras, L99

\bibitem[{Scarlata {et~al.}(2007)Scarlata, Carollo, Lilly, Sargent, Feldmann,
  Kampczyk, Porciani, Koekemoer, Scoville, Kneib, Leauthaud, Massey, Rhodes,
  Tasca, Capak, Maier, McCracken, Mobasher, Renzini, Taniguchi, Thompson,
  Sheth, Ajiki, Aussel, Murayama, Sanders, Sasaki, Shioya, \&
  Takahashi}]{Scarlata2007}
Scarlata, C., {et~al.} 2007, \apjs, 172, 406

\bibitem[{Schaye {et~al.}(2010)Schaye, Vecchia, Booth, Wiersma, Theuns, Haas,
  Bertone, Duffy, McCarthy, \& van~de Voort}]{Schaye2010}
Schaye, J., {et~al.} 2010, \mnras, 402, 1536

\bibitem[{Schaye {et~al.}(2014)Schaye, Crain, Bower, Furlong, Schaller, Theuns,
  {Dalla Vecchia}, Frenk, McCarthy, Helly, Jenkins, Rosas-Guevara, White, Baes,
  Booth, Camps, Navarro, Qu, Rahmati, Sawala, Thomas, \& Trayford}]{Schaye2014}
---. 2014, \mnras, 446, 521

\bibitem[{Scoville {et~al.}(2007)Scoville, Aussel, Brusa, Capak, Carollo,
  Elvis, Giavalisco, Guzzo, Hasinger, Impey, Kneib, LeFevre, Lilly, Mobasher,
  Renzini, Rich, Sanders, Schinnerer, Schminovich, Shopbell, Taniguchi, \&
  Tyson}]{Scoville2007}
Scoville, N., {et~al.} 2007, \apjs, 172, 1

\bibitem[{Sijacki {et~al.}(2015)Sijacki, Vogelsberger, Genel, Springel, Torrey,
  Snyder, Nelson, \& Hernquist}]{Sijacki2015}
Sijacki, D., Vogelsberger, M., Genel, S., Springel, V., Torrey, P., Snyder,
  G.~F., Nelson, D., \& Hernquist, L. 2015, \mnras, 452, 575

\bibitem[{Sijacki {et~al.}(2012)Sijacki, Vogelsberger, Kere\v{s}, Springel, \&
  Hernquist}]{Sijacki2012}
Sijacki, D., Vogelsberger, M., Kere\v{s}, D., Springel, V., \& Hernquist, L.
  2012, \mnras, 424, 2999

\bibitem[{Snyder {et~al.}(2011)Snyder, Cox, Hayward, Hernquist, \&
  Jonsson}]{snyder11a}
Snyder, G., Cox, T., Hayward, C., Hernquist, L., \& Jonsson, P. 2011, \apj,
  741, 77

\bibitem[{Snyder {et~al.}(2015)Snyder, Lotz, Moody, Peth, Freeman, Ceverino,
  Primack, \& Dekel}]{Snyder2015a}
Snyder, G.~F., Lotz, J., Moody, C., Peth, M., Freeman, P., Ceverino, D.,
  Primack, J., \& Dekel, A. 2015, \mnras, 451, 4290

\bibitem[{Somerville {et~al.}(2008)Somerville, Hopkins, Cox, Robertson, \&
  Hernquist}]{Somerville2008}
Somerville, R.~S., Hopkins, P.~F., Cox, T.~J., Robertson, B.~E., \& Hernquist,
  L. 2008, \mnras, 391, 481

\bibitem[{Sparre {et~al.}(2015)Sparre, Hayward, Springel, Vogelsberger, Genel,
  Torrey, Nelson, Sijacki, \& Hernquist}]{Sparre2015}
Sparre, M., {et~al.} 2015, \mnras, 447, 3548

\bibitem[{Spergel {et~al.}(2015)Spergel, Gehrels, Baltay, Bennett,
  Breckinridge, Donahue, Dressler, Gaudi, Greene, Guyon, Hirata, Kalirai,
  Kasdin, Macintosh, Moos, Perlmutter, Postman, Rauscher, Rhodes, Wang,
  Weinberg, Benford, Hudson, Jeong, Mellier, Traub, Yamada, Capak, Colbert,
  Masters, Penny, Savransky, Stern, Zimmerman, Barry, Bartusek, Carpenter,
  Cheng, Content, Dekens, Demers, Grady, Jackson, Kuan, Kruk, Melton, Nemati,
  Parvin, Poberezhskiy, Peddie, Ruffa, Wallace, Whipple, Wollack, \&
  Zhao}]{Spergel2015}
Spergel, D., {et~al.} 2015, eprint arXiv:1503.03757

\bibitem[{Springel(2010)}]{Springel2010}
Springel, V. 2010, \mnras, 401, 791

\bibitem[{Springel \& Hernquist(2003)}]{springel03}
Springel, V., \& Hernquist, L. 2003, \mnras, 339, 289

\bibitem[{Springel {et~al.}(2001)Springel, White, Tormen, \&
  Kauffmann}]{Springel2001}
Springel, V., White, S. D.~M., Tormen, G., \& Kauffmann, G. 2001, \mnras, 328,
  726

\bibitem[{Strateva {et~al.}(2001)Strateva, Ivezi\'{c}, Knapp, Narayanan,
  Strauss, Gunn, Lupton, Schlegel, Bahcall, Brinkmann, Brunner, Budav\'{a}ri,
  Csabai, Castander, Doi, Fukugita, Gy\H~ory, Hamabe, Hennessy, Ichikawa,
  Kunszt, Lamb, McKay, Okamura, Racusin, Sekiguchi, Schneider, Shimasaku, \&
  York}]{strateva01_gv}
Strateva, I., {et~al.} 2001, \aj, 122, 1861

\bibitem[{Strauss {et~al.}(2002)Strauss, Weinberg, Lupton, Narayanan, Annis,
  Bernardi, Blanton, Burles, Connolly, Dalcanton, Doi, Eisenstein, Frieman,
  Fukugita, Gunn, Ivezi\'{c}, Kent, Kim, Knapp, Kron, Munn, Newberg, Nichol,
  Okamura, Quinn, Richmond, Schlegel, Shimasaku, SubbaRao, Szalay, {Vanden
  Berk}, Vogeley, Yanny, Yasuda, York, \& Zehavi}]{Strauss2002}
Strauss, M.~A., {et~al.} 2002, \aj, 124, 1810

\bibitem[{Taghizadeh-Popp {et~al.}(2015)Taghizadeh-Popp, Fall, White, \&
  Szalay}]{Taghizadeh-Popp2015}
Taghizadeh-Popp, M., Fall, S.~M., White, R.~L., \& Szalay, A.~S. 2015, \apj,
  801, 14

\bibitem[{Tinsley(1968)}]{tinsley68}
Tinsley, B. 1968, \apj, 151, 547

\bibitem[{Tinsley \& Gunn(1976)}]{Tinsley1976}
Tinsley, B.~M., \& Gunn, J.~E. 1976, \apj, 203, 52

\bibitem[{Tinsley \& Larson(1978)}]{Tinsley1978}
Tinsley, B.~M., \& Larson, R.~B. 1978, \apj, 221, 554

\bibitem[{Torrey {et~al.}(2014)Torrey, Vogelsberger, Genel, Sijacki, Springel,
  \& Hernquist}]{Torrey2014}
Torrey, P., Vogelsberger, M., Genel, S., Sijacki, D., Springel, V., \&
  Hernquist, L. 2014, \mnras, 438, 1985

\bibitem[{Torrey {et~al.}(2015)Torrey, Snyder, Vogelsberger, Hayward, Genel,
  Sijacki, Springel, Hernquist, Nelson, Kriek, Pillepich, Sales, \&
  McBride}]{Torrey2015}
Torrey, P., {et~al.} 2015, \mnras, 447, 2753

\bibitem[{van~den Bergh(1976)}]{VanDenBergh1976}
van~den Bergh, S. 1976, \apj, 206, 883

\bibitem[{Vogelsberger {et~al.}(2013)Vogelsberger, Genel, Sijacki, Torrey,
  Springel, \& Hernquist}]{Vogelsberger2013}
Vogelsberger, M., Genel, S., Sijacki, D., Torrey, P., Springel, V., \&
  Hernquist, L. 2013, \mnras, 436, 3031

\bibitem[{Vogelsberger {et~al.}(2012)Vogelsberger, Sijacki, Kere\v{s},
  Springel, \& Hernquist}]{Vogelsberger2012}
Vogelsberger, M., Sijacki, D., Kere\v{s}, D., Springel, V., \& Hernquist, L.
  2012, \mnras, 425, 3024

\bibitem[{Vogelsberger {et~al.}(2014{\natexlab{a}})Vogelsberger, Genel,
  Springel, Torrey, Sijacki, Xu, Snyder, Nelson, \&
  Hernquist}]{Vogelsberger2014b}
Vogelsberger, M., {et~al.} 2014{\natexlab{a}}, \mnras, 444, 1518

\bibitem[{Vogelsberger {et~al.}(2014{\natexlab{b}})Vogelsberger, Genel,
  Springel, Torrey, Sijacki, Xu, Snyder, Bird, Nelson, \&
  Hernquist}]{Vogelsberger2014a}
---. 2014{\natexlab{b}}, Nature, 509, 177

\bibitem[{Wellons {et~al.}(2015)Wellons, Torrey, Ma, Rodriguez-Gomez,
  Vogelsberger, Kriek, van Dokkum, Nelson, Genel, Pillepich, Springel, Sijacki,
  Snyder, Nelson, Sales, \& Hernquist}]{Wellons2015}
Wellons, S., {et~al.} 2015, \mnras, 449, 361

\bibitem[{Wells {et~al.}(1981)Wells, Greisen, \& Harten}]{Wells1981}
Wells, D.~C., Greisen, E.~W., \& Harten, R.~H. 1981, \aaps, 44, 363

\bibitem[{Whitaker {et~al.}(2012)Whitaker, Kriek, van Dokkum, Bezanson,
  Brammer, Franx, \& Labb\'{e}}]{Whitaker2012}
Whitaker, K.~E., Kriek, M., van Dokkum, P.~G., Bezanson, R., Brammer, G.,
  Franx, M., \& Labb\'{e}, I. 2012, \apj, 745, 179

\bibitem[{White {et~al.}(2015)White, Somerville, \& Ferguson}]{White2015}
White, C.~E., Somerville, R.~S., \& Ferguson, H.~C. 2015, \apj, 799, 201

\bibitem[{Williams {et~al.}(2010)Williams, Quadri, Franx, van Dokkum, Toft,
  Kriek, \& Labb\'{e}}]{Williams2010}
Williams, R.~J., Quadri, R.~F., Franx, M., van Dokkum, P., Toft, S., Kriek, M.,
  \& Labb\'{e}, I. 2010, \apj, 713, 738

\bibitem[{Woo {et~al.}(2015)Woo, Dekel, Faber, \& Koo}]{Woo2015}
Woo, J., Dekel, A., Faber, S.~M., \& Koo, D.~C. 2015, \mnras, 448, 237

\bibitem[{Woo {et~al.}(2012)Woo, Dekel, Faber, Noeske, Koo, Gerke, Cooper,
  Salim, Dutton, Newman, Weiner, Bundy, Willmer, Davis, \& Yan}]{Woo2012}
Woo, J., {et~al.} 2012, \mnras, 428, 3306

\bibitem[{Wuyts {et~al.}(2011)Wuyts, {F\"{o}rster Schreiber}, van~der Wel,
  Magnelli, Guo, Genzel, Lutz, Aussel, Barro, Berta, Cava, Graci\'{a}-Carpio,
  Hathi, Huang, Kocevski, Koekemoer, Lee, {Le Floc'h}, McGrath, Nordon,
  Popesso, Pozzi, Riguccini, Rodighiero, Saintonge, \& Tacconi}]{Wuyts2011}
Wuyts, S., {et~al.} 2011, \apj, 742, 96

\bibitem[{Zolotov {et~al.}(2015)Zolotov, Dekel, Mandelker, Tweed, Inoue,
  DeGraf, Ceverino, Primack, Barro, \& Faber}]{Zolotov2015}
Zolotov, A., {et~al.} 2015, \mnras, 450, 2327

\end{thebibliography}
\bsp


\end{document}